\documentclass[english,a4paper,12pt,twoside,openany]{book}
\usepackage[utf8]{inputenc}
\usepackage[english]{babel}

\usepackage{notoccite}

\usepackage{amsmath}
\usepackage{amsfonts}
\usepackage{amssymb}
\usepackage{amsthm}
\usepackage{slashed}
\usepackage{graphics}
\usepackage{graphicx}
\usepackage{fancyhdr}
\usepackage{multirow}
\usepackage{ragged2e}
\usepackage{hyperref}
\usepackage{fancyhdr}
\usepackage{bm}
\usepackage[left=2.5cm,top=2cm,bottom=2cm,right=2.5cm,includehead]{geometry}

\usepackage{xcolor}
\hypersetup{
	colorlinks,
	linkcolor={red!50!black},
	citecolor={blue!50!black},
	urlcolor={blue!80!black}
}

\usepackage{multicol}

\usepackage[titletoc]{appendix}
\usepackage{youngtab}
\usepackage{dsfont}
\usepackage{tikz}
\usetikzlibrary{shapes,arrows}
\usepackage{bbm}

\usepackage{verbatim}
\usepackage{indentfirst}

\usepackage{framed}
\usepackage[normalem]{ulem}
\usepackage{enumerate}
\usepackage{cancel}
\usepackage{physics}
\usepackage{mathtools}

\usepackage{feynmf,subfigure}

\usepackage[numbers,sort&compress]{natbib}

\usepackage{etoolbox}

\makeatletter
\patchcmd{\@makechapterhead}{\vspace*{50\p@}}{}{}{}
\patchcmd{\@makeschapterhead}{\vspace*{50\p@}}{}{}{}
\makeatother

\theoremstyle{definition}

\newcommand{\diff}{\mathrm{d}}
\newcommand{\lag}{\mathcal{L}}

\fancypagestyle{plain}{%
	\fancyhf{} 
	\fancyfoot{} 
	
	}

\renewcommand{\chaptermark}[1]%
{\markboth{{\thechapter.\ #1}}{}}
\renewcommand{\sectionmark}[1]%
{\markright{{\thesection.\ #1}}}

\begin{document}
\begin{fmffile}{diag}

\title{The information loss paradox}
\author{Francisco Mart\'inez L\'opez}
\date{\today}

\setcounter{page}{0}
\pagenumbering{roman}

\thispagestyle{empty}

\begin{picture}(50,50)
\put(-85,-725){\hbox{\includegraphics[scale=1]{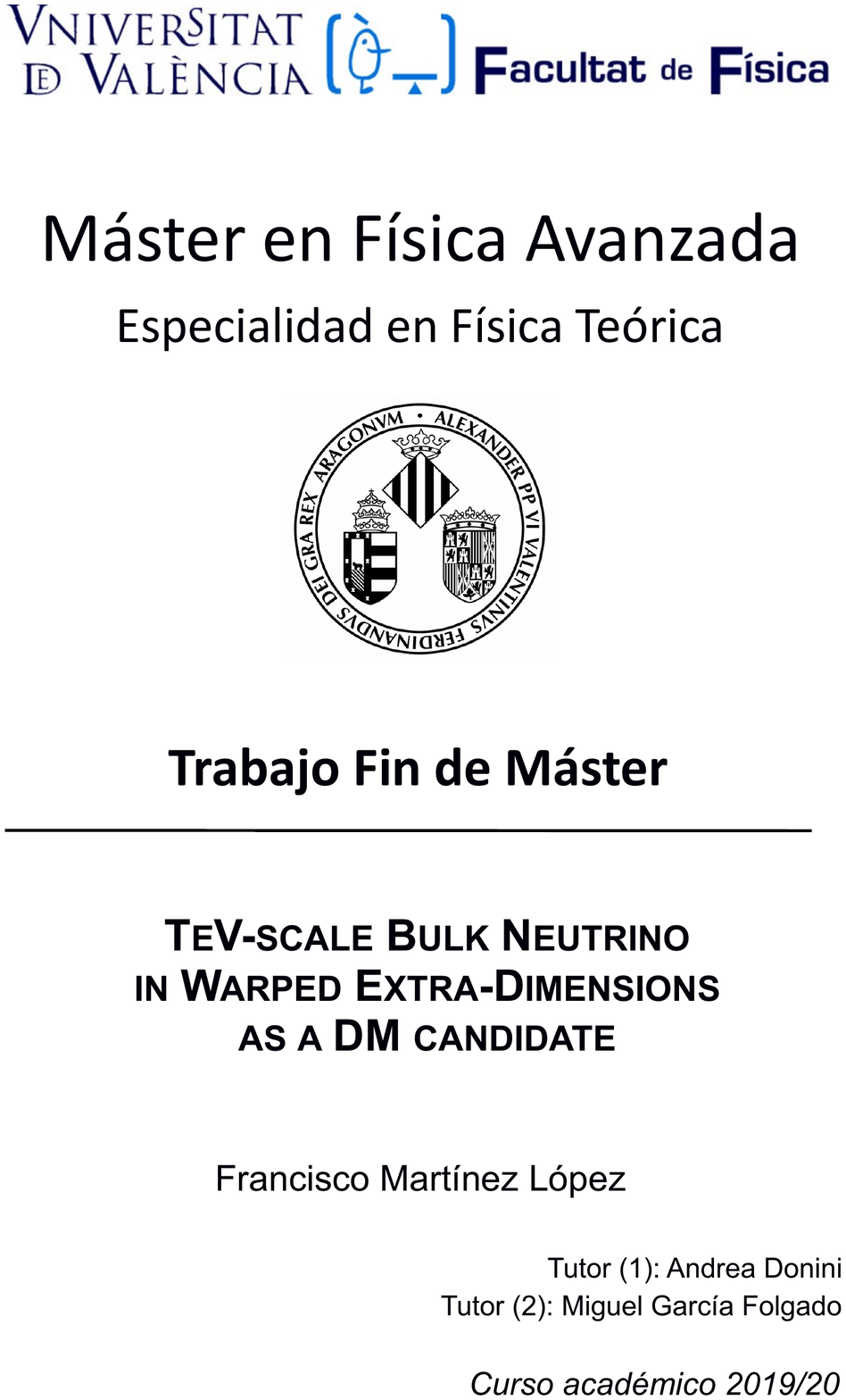}}}
\end{picture}

\clearpage


		
		
		
		
		
		
		
		
		

		

		
		
		
		
		
	

\newpage

\begin{center}
	\textbf{\small{Resumen}}
		\justify
		En esta tesis hemos estudiado la posibilidad de identificar uno de los neutrinos dextrógiros que participan en el mecanismo del balancín de tipo I como un posible candidato a ser la Materia Oscura (MO), en el paradigma de las dimensiones enrolladas extra. Como un primer paso, hemos revisado la literatura existente sobre los escenarios de Randall-Sundrum y el mecanismo de estabilización de Goldberger-Wise. Hemos discutido una posible realización del balancín de tipo I, implementado por compactificación de la dimensión extra. Aprovechando los resultados de estudios recientes, exploramos como la MO térmica puede ser incluida en estos escenarios. También resumimos los límites experimentales presentes en las búsquedas de MO y neutrinos Majorana pesados, así como las ligaduras teóricas que se imponen. Nos centramos en generar una jerarquía específica entre los neutrinos dextrógiros, que permita la correcta implementación del balancín a la vez que tiene un neutrino estéril en la escala de los eV (que podría ser de ayuda para explicar o suavizar las llamadas ``anomalías de neutrinos''), otro en el rango de masas apropiado para ser la MO y el último de ellos suficientemente pesado para desacoplarse. En nuestra búsqueda en el espacio de parámetros hemos encontrado importantes regiones donde este patrón es posible. Además, hemos identificado algunos de los problemas que ponen en riesgo la integridad del modelo: este no da cuenta de las masas y las mezclas medidas en el sector de los neutrinos ligeros y por otro lado existen decaimientos peligrosos que hacen que nuestro candidato a MO no sea lo suficientemente estable. Por último, discutimos ambos problemas y proponemos soluciones razonables a los mismos.

\end{center}

\vspace*{0.5cm}

\begin{center}
	\textbf{\small{Abstract}}
		\justify
		In this thesis, we have studied the possibility to identify one of the right-handed neutrinos entering the type-I seesaw mechanism as a Dark Matter (DM) candidate, within the warped extra-dimensions paradigm. As a first step, we have reviewed the existing literature on the Randall-Sundrum scenarios and the Goldberger-Wise stabilization mechanism. We have discussed a feasible realization of the type-I seesaw mechanism, implemented by warped compactification of an extra dimension. Taking advantage of recent studies, we have explored how the thermal DM can be included in the warped scenario. We have also outlined the current experimental bounds on the searches for DM and heavy Majorana neutrinos, as well as the theoretical constraints imposed. We have focused on generating a specific hierarchy between the right-handed neutrinos, that allows a correct realization of the seesaw while having an eV-scale sterile neutrino (that may be helpful to explain or soften the so-called ``neutrino anomalies''), another one lying in the appropriate mass range for the DM and the last one sufficiently heavy to decouple. In our search of the parameter space, we found significant regions where this pattern can be realized. Moreover, we identified some problems that compromise the integrity of the model: it does not account for the measured masses and mixing of the light neutrino sector and, moreover, there exist some dangerous decay channels that prevents the DM candidate to be sufficiently stable. We discussed both problems and propose some reasonable solutions.

\end{center}

\newpage

\tableofcontents

\mainmatter
\setcounter{page}{1}

\chapter{Introduction}\label{Chapter 1}


The Standard Model of Fundamental Interactions (SM) have provided a deep understanding of the electromagnetic, weak and strong interactions, and over the past decades it has passed all kind of precision tests with astonishing robustness. The discovery of the Higgs boson at LHC \cite{Collaboration2012} has strengthened the theory even further. However, the SM by itself can not explain certain observed phenomena, such as the baryon asymmetry of the Universe, the existence of Dark Matter and Dark Energy or the origin of neutrino masses.


Within the SM, neutrinos are massless. This prediction is in direct conflict with the results of neutrino oscillation experiments \cite{Fukuda1998}, which necessarily require tiny but non-zero masses for them. Since the masses of all SM charged fermions come from their interaction with the SM Higgs, one can wonder if adding new Higgs-like particles or new fermions can contribute to understand these small neutrino masses. A trivial way to accommodate them is the introduction of right-handed neutrinos, which will generate a Dirac mass term for neutrinos. However, their smallness will require an unnaturally small (according to the 't Hooft definition of {\em naturalness} \cite{Hooft1980}) value for the Yukawa couplings. An alternative possibility is adding a Majorana mass term for the right-handed neutrinos, possibly related to some spontaneous symmetry breaking. This constitutes the so-called type-I seesaw mechanism \cite{Minkowski1977,GellMann2013,Yanagida:1979as,Glashow1979,Mohapatra1980,Schechter1980}, where large Majorana masses can account for small masses of active neutrinos.


Another puzzle to be solved in order to have a complete picture of the Universe is the nature of Dark Matter (DM). From astrophysical observations (see, for example, Ref. \cite{Bertone2004} and refs. therein), we are aware of the existence of some unknown matter which only interacts gravitationally with other particles. Usually, extensions of the SM include feasible DM candidates. These are usually very stable, heavy particles with small interactions (if any) with SM particles. These states are known as weakly interacting massive particles (WIMPs) \cite{Dimopoulos1981,Appelquist2000,Arun2017}. LHC data have put stringent bounds on the masses of these particles, pushing them towards the TeV-scale. Experiments looking for DM through its direct interaction with matter or its decays into SM particles have set the interaction cross-section between DM and SM particles to be very small for DM masses below $1 \ \mathrm{TeV}$.


One interesting possibility is that the interaction between DM and SM particles can be enhanced as a consequence of gravity feeling more than the usual $3+1$ space-time dimensions. This kind of scenarios involving extra-dimensions have been proposed in the literature to solve the hierarchy problem between the electroweak and Planck scales. In these models, the fundamental scale of gravity is denoted as $M_{D}$ whose relation to $M_{P}$ depends on the geometry of space-time \cite{ArkaniHamed1998,Antoniadis1998,ArkaniHamed1999}. In the case of warped extra-dimensions (also known as Randall-Sundrum models) the difference between $M_{P}$ and $M_{D}$ is not so large, but they introduce an exponential suppression to all physical masses \cite{Randall1999,Randall1999a,Gogberashvili1998,Gogberashvili1998a,Gogberashvili1999}. Therefore the fundamental mass scales are set to be order $M_{P}$, but non-fundamental scales such as the electroweak scale may be much smaller.


In the context of warped extra-dimensions, the implementation of a type-I seesaw-like mechanism can be traced back to Ref. \cite{Neubert2000}. In order to explain the so-called neutrino anomalies (experimental results that could be explained adding one or more ``sterile'' neutrino states to the SM neutrino spectrum \cite{Aguilar2001,Collaboration2018a}), an interesting extension can be found in Ref. \cite{Kawai2019}, where the neutrino mass matrix is built in such a way so as to obtain some light neutrinos plus one eV-scale and one or more heavy ``sterile'' states. The possibility of DM particles having an enhanced gravitational interaction has been studied previously, first advanced in Refs. \cite{Lee2014,Lee2014a} and later discussed in Refs. \cite{Folgado2020a,Folgado2020b}. In this work we explore the possibility of combining both ideas, trying to identify one of the right-handed neutrinos entering the seesaw as a fermionic DM candidate, letting this particle to penetrate in the bulk. We explore the allowed parameter space, finding a significant region compatible with our hypothesis and the experimental bounds. In adition, we also study the stability of this DM candidate, noticing some tension between our minimal choice and experimental data. We propose a possible extension of the model, based on the addition of an extra Higgs doublet (see Ref. \cite{Branco2011} and refs. therein), which can stabilize the DM. This possibility will be, however, further explored in future work beyond the limit of this thesis.


The thesis is organized as follows: in Chap. \ref{Chapter 2} we review the basic features of the warped extra-dimensions scenario and introduce the Goldberger-Wise mechanism to stabilize the size of the extra dimension; in Chap. \ref{Chapter 3} we outline a possible realization of the type-I seesaw mechanism within the framework of RS scenarios; in Chap. \ref{Chapter 4} we explore how the DM can be included in the warped scenario and show our results for the annihilation cross-section of DM particles into SM particles and on-shell KK-gravitons; in Chap. \ref{Chapter 5} we review the experimental bounds on the model, both the ones on the masses of the DM and KK-gravitons and the corresponding ones on the masses and couplings of the heavy Majorana neutrinos; in Chap. \ref{Chapter 6} we explore the allowed parameter space; in Chap. \ref{Chapter 7} we review some of the difficulties of the model and outline possible solutions that we are working on; and eventually in Chap. \ref{Chapter 8} we conclude. In the Appendices we give some mathematical expressions used throughout the thesis: in App. \ref{Appendix A} we give the formul$\mathrm{\ae}$ for the annihilation cross-section of Majorana fermion DM particles into SM particles and KK-gravitons; and in App. \ref{Appendix B} we review the formalism of neutrino mixing.

\chapter{Warped Extra-Dimensions}\label{Chapter 2}

If space-time is higher dimensional, with $D\equiv4+n$ space-time dimensions, then there would be a fundamental $(4+n)$-dimensional Planck scale, $M_{D}$. In that case, the usual 4-dimensional Planck scale, $M_{P}$, should be given by this fundamental scale and the geometry of the extra space-time dimensions. For models where the higher dimensional space-time is the product of the usual 4-dimensional space-time and a $n$-dimensional compact space that relation reads:
\begin{equation}
M_{P}^2=M_{D}^{2+n} V_n,
\end{equation}
where $V_n$ is the volume of the compact space. Assuming that the compact space is large, the hierarchy problem can be addressed \cite{ArkaniHamed1998,Antoniadis1998,ArkaniHamed1999}. This type of scenarios are known as large extra-dimensions (LED).

While the scenario of LED does eliminate the hierarchy between the weak scale $m_{EW}$ and the Planck scale $M_{P}$, it introduces a new hierarchy between the compactification scale $\mu_c=1/R$ and $m_{EW}$. As an alternative approach to generate the hierarchy, we can consider that the metric is not factorizable, but rather the four-dimensional metric is multiplied by a warp factor which is a rapidly changing function of an additional dimension \cite{Randall1999,Randall1999a,Gogberashvili1998,Gogberashvili1998a,Gogberashvili1999}.

Parametrizing the fifth dimension with the angular coordinate $-\pi\leq\phi\leq\pi$, the metric results:
\begin{equation}\label{3}
\diff s^2=\mathrm{e}^{-2\sigma(\phi)} \eta_{\mu\nu} \diff x^\mu \diff x^\nu- r_c^2 \diff \phi^2,
\end{equation}
where $\sigma(\phi)=k r_c |\phi|$. The signature of the metric is $\left(+,-,-,-,-\right)$, $1/k$ is the curvature radius of the 5th-dimension and $r_c$ is the proportionality constant between $\phi$ and the proper distance in the extra dimension (thus related to its size).

We specify as a boundary condition periodicity in $\phi$, then identifying $(x,\phi)$ with $(x,\phi+2\pi)$. Moreover, we impose the reflectivity condition $\phi=-\phi$. The resulting quotient space $S_{1}/\mathbb{Z}_{2}$ is an orbifold, therefore, the metric is completely specified in $0\leq\phi\leq\pi$. At the two orbifold fixed points $\phi=0,\pi$ we locate two 3-branes, extending in the $x^\mu$-directions (named UV-brane and IR-brane, respectively). The resulting 5-dimensional space-time is a slice of anti-de Sitter space ($AdS_{5}$).

The 5-dimensional action is written as:
\begin{equation}\label{4}
S=S_{bulk}+S_{IR}+S_{UV}.
\end{equation}

We explicitly write these terms as:
\begin{equation}\label{5}
\begin{split}
S_{bulk}&=\frac{16\pi}{M_{5}^3}\int \diff^4x\int^{\pi}_{0} r_c \diff \phi \sqrt{\vphantom{G}^{(5)}G} \left[\vphantom{R}^{(5)}R-\Lambda_{5}\right],\\
S_{IR}&=\int \diff^4x \sqrt{-g} \left[\lag_{SM}+\lag_{DM}-f_{IR}^4\right],\\
S_{UV}&=\int \diff^4x \sqrt{-g} \left[-f_{UV}^4+...\right],
\end{split}
\end{equation}
where $M_{5}$ is the fundamental scale of gravity, $\vphantom{G}^{(5)}G$ and $\vphantom{R}^{(5)}R$ are the 5-dimensional metric and Ricci scalar, respectively, $\Lambda_{5}$ a 5-dimensional cosmological constant in the bulk and $f_{IR}$, $f_{UV}$ the corresponding tensions (or vacuum energies) of both branes. We have placed all SM and DM fields in the IR-brane, whereas we have assumed that the UV-brane is populated just by Planck-suppressed physics.

If our metric is a classical solution to Einstein field equations, the brane tensions have to be chosen to cancel the 5-dimensional cosmological term:
\begin{equation}\label{6}
f_{IR}^4=-f_{UV}^4=\sqrt{-24M_5^3\Lambda_5}.
\end{equation}

The relation between the Planck mass $M_{P}$ and the fundamental mass scale $M_5$ can be obtained by identification, when considering fluctuations about Minkowski space in (\ref{5}):
\begin{equation}\label{7}
\bar{M}_{P}^2=\frac{M_5^2}{k} \left(1-\mathrm{e}^{-2k\pi r_{c}}\right),
\end{equation}
where $\bar{M}_{P}\equiv\frac{M_{P}}{\sqrt{8\pi}}$ is the reduced Planck mass.

\section{Physical implications}

As the size of the fifth dimension $r_{c}$ is assumed to be small, the extra dimension can not be proven in present experiments. Therefore, we can make use of a 4-dimensional effective field theory for our description. Expanding the 4-dimensional component of the metric around Minkowski space-time gives, at first order:
\begin{equation}
G_{\mu\nu}=\mathrm{e}^{-2\sigma}\left(\eta_{\mu\nu}+\frac{2}{M_{5}^{2/3}} h_{\mu\nu}\right).
\end{equation}

The graviton field $h_{\mu\nu}$ represents the metric fluctuations around flat space-time. We can decompose the metric fluctuations in terms of a Kaluza-Klein tower of 4-dimensional fields:
\begin{equation}
h_{\mu\nu}\left(x,\phi\right)=\sum_{n=0}^{\infty} h_{\mu\nu}^{(n)}\left(x\right) \frac{\chi^{(n)}\left(\phi\right)}{\sqrt{r_c}},
\end{equation}
where $h_{\mu\nu}^{(n)}\left(x\right)$ are the Kaluza-Klein modes of the graviton and $\chi^{(n)}\left(\phi\right)$ the corresponding wavefunctions. In the transverse-traceless gauge, the equations of motion for the modes are \cite{Davoudiasl2000}:
\begin{equation}
\left(\eta^{\alpha\beta}\partial_{\alpha}\partial_{\beta}+m_n^2\right)h_{\mu\nu}^{(n)}\left(x\right)=0.
\end{equation}
where $m_n>0$ are their masses. Using Einstein equations and the equations of motion, we obtain a second-order differential equation for the wavefunctions:
\begin{equation}
\frac{-1}{r_c^2}\frac{\diff}{\diff \phi} \left(\mathrm{e}^{-4\sigma}\frac{\diff \chi^{(n)}\left(\phi\right)}{\diff \phi}\right)=m_n^2\mathrm{e}^{-2\sigma}\chi^{(n)}\left(\phi\right).
\end{equation}

Its solution is found to be a linear combination of Bessel functions of order 2:
\begin{equation}
\chi^{(n)}\left(\phi\right)=\frac{\mathrm{e}^{2\sigma}}{N_n}\left[J_{2}(z_n)+\alpha_nY_{2}(z_n)\right],
\end{equation}
where $z_n(\phi)=m_n\mathrm{e}^{\sigma(\phi)}/k$, $N_n$ are normalization factors and $\alpha_n$ some coefficients. In the limit $m_n/k\ll1$ and $\mathrm{e}^{k\pi r_c}\gg1$, requiring the first derivative of $\chi^{(n)}\left(\phi\right)$ to be continuous at the orbifold fixed points yields:
\begin{equation}
\alpha_n\sim x_n^2 \mathrm{e}^{-2k \pi r_c},
\end{equation}
where $x_n$ is the $n$th-root of the Bessel function of first order $J_1$. The masses of the KK-gravitons depend on these roots, and therefore are not equally spaced:
\begin{equation}
m_n=kx_ne^{-k \pi r_c}.
\end{equation}

The normalization factors are found after imposing the orthonormality condition:
\begin{equation}
\int_{-\pi}^{\pi}\diff \phi \ \mathrm{e}^{-2\sigma}\chi^{(m)}\left(\phi\right)\chi^{(n)}\left(\phi\right)=\delta_{mn}.
\end{equation}
Working in the same approximation as before, we find:
\begin{equation}
\begin{split}
N_0&=-\frac{1}{\sqrt{k r_c}},\\
N_n&=\frac{1}{\sqrt{2k r_c}} \mathrm{e}^{k\pi r_c} J_2(x_n); \ n>0.
\end{split}
\end{equation}

At the IR-brane, imposing the constraint $\phi=\pi$ in the 5-dimensional action, we find an effective interaction Lagrangian of the form:

\begin{equation}
\lag_{eff}=-\frac{1}{M_{5}^{3/2}} T^{\mu\nu} h_{\mu\nu}\left(x,\phi=\pi\right)=-\frac{1}{M_{5}^{3/2}} T^{\mu\nu}\sum_{n=0}^{\infty} h_{\mu\nu}^{(n)}\left(x\right) \frac{\chi^{(n)}\left(\phi=\pi\right)}{\sqrt{r_c}}.
\end{equation}

The wavefunctions evaluated at the IR-brane take the form:
\begin{equation}
\begin{split}
\chi^{(0)}\left(\phi=\pi\right)&=\sqrt{kr_c}\left(1-\mathrm{e}^{-2k\pi r_c}\right)=-\sqrt{r_c} \ \frac{M_5^{3/2}}{\bar{M}_{P}},\\
\chi^{(n)}\left(\phi=\pi\right)&=\sqrt{kr_c}\mathrm{e}^{k\pi r_c}=\sqrt{r_c} \ \mathrm{e}^{k\pi r_c} \frac{M_5^{3/2}}{\bar{M}_{P}}
\end{split}
\end{equation}
therefore yielding an expansion of the effective Lagrangian:
\begin{equation}
\lag_{eff}=-\frac{1}{M_{5}^{3/2}} T^{\mu\nu} h_{\mu\nu}\left(x,\phi=\pi\right)=-\frac{1}{\bar{M}_{P}} T^{\mu\nu} h_{\mu\nu}^{(0)}\left(x\right)-\frac{\mathrm{e}^{k\pi r_c}}{\bar{M}_{P}} T^{\mu\nu}\sum_{n=1}^{\infty} h_{\mu\nu}^{(n)}\left(x\right).
\end{equation}

We can clearly see that the zero-mode couples to matter with the usual gravitational strength $\bar{M}_{P}$, however, the massive KK-excitations are only suppressed by $\Lambda\equiv\bar{M}_{P} \ \mathrm{e}^{-k\pi r_c}$. The effective gravitational coupling is, thus, enhanced due to the rescaling factor $\sqrt{\vphantom{G}^{(5)}G}/\sqrt{-g}$. For moderate choices of $\sigma$, like $kr_c\sim10$, we see that $\Lambda\ll M_{P}$ and therefore the hierarchy problem is correctly addressed. In the following, we will assume $\Lambda=\mathcal{O}(1 \ \mathrm{TeV})$.

In the expansion of the 5-dimensional metric, we do not consider the contribution of the graviphoton $G_{\mu5}$ and the graviscalar $G_{55}$. Due to the breaking of 5-dimensional translational invariance induced by the presence of the branes, the towers of KK-modes of the graviphoton and of the graviscalar are absorbed by the KK-gravitons, in the unitary gauge, to get massive spin-2 KK-modes. The zero-mode of the graviphoton does not play any role in the phenomenology of the model (see, e.g., Ref. \cite{Randall1999}). Eventually, the graviscalar will play a role in the stabilization of the fifth dimension, and will be discussed below.

\section{Introducing the radion}

In Randall-Sundrum scenarios, stabilizing the size of the extra-dimension to be $y=\pi r_c$ can be achieved through the following mechanism. If we add a scalar bulk field $S$ with a potential $V(S)$ and some localized potential terms $\delta\left(y=0\right) V_{\mathrm{UV}}(S)$ and $\delta\left(y=\pi r_c\right) V_{\mathrm{IR}}(S)$, we can generate an effective potential $V(\varphi)$ for the 4-dimensional field:
\begin{equation}
\varphi=f \ \mathrm{e}^{-\pi k T},
\end{equation}
where $f=\sqrt{-24M_5^3/k}$ and $\expval{T}=r_c$. The minimum of this potential can give the desired value of $k r_c$, without extreme fine-tuning \cite{Goldberger1999,Goldberger1999a}.

The field $S$ will have also a KK-tower, but we do not take it into account as it is expected to be heavy \cite{Goldberger1999b}. The only light field in the spectrum is a combination of the zero-modes of the graviscalar and $S$, the so-called radion, $r$. Its mass can be obtained from the effective potential, and is given by:
\begin{equation}
m_r^2=\frac{k^2 v_v^2}{3 M_5^3} \left(\frac{m^2}{4k^2}\right)^2 \ \mathrm{e}^{-2\pi k r_c},
\end{equation}
where $v_v$ is the value of $S$ at the IR-brane and $m$ is its mass parameter. In general $m^2/4k^2\ll1$, and therefore the radion mass is much smaller than the mass of the first KK-graviton mode. This is known as the Goldberger-Wise stabilization mechanism \cite{Goldberger1999}.

The radion interacts both with DM and SM particles. It couples to them through the trace of the energy-momentum tensor. Although the massless gauge bosons do not contribute to the trace, the radion couple to them by mean of quark and $W$ boson loops and the trace anomaly. The radion Lagrangian takes the form:
\begin{equation}
\lag_{r}=\frac{1}{2} \partial_{\mu} r \ \partial^{\mu} r - \frac{1}{2} m_{r}^2 r^2 + \frac{1}{\sqrt{6}\Lambda} r T + \frac{\alpha_{EM} C_{EM}}{8 \pi \sqrt{6}\Lambda} r F_{\mu\nu} F^{\mu\nu} + \frac{\alpha_{S} C_{3}}{8 \pi \sqrt{6}\Lambda} r \sum_{a} F_{\mu\nu}^{a} F^{a\mu\nu},
\end{equation}
where $F_{\mu\nu}$ and $F_{\mu\nu}^{a}$ are the Yang-Mills tensors of $U(1)_{EM}$ and $SU(3)_{c}$, respectively. The coefficients $C_{EM}$ and $C_{3}$ encode the information of the massless gauge bosons contribution and can be found in App. B of Ref. \cite{Folgado2020a}.
\chapter{Sterile neutrinos from an extra dimension}\label{Chapter 3}

Within the SM, neutrinos are massless. The neutrino oscillation experiments constitute the first indication of physics Beyond the Standard Model (BSM), as they are an unambiguous indicative of neutrinos having small but non-zero masses \cite{Fukuda1998}. The possibility of generate these masses from a Yukawa term of the form $\bar{L}_{L}\tilde{\Phi}N_{R}$, where we added three right-handed neutrino singlets, would give ``unnaturally'' small values for the couplings. A reasonable alternative consists on adding a Majorana mass term for these right-handed neutrinos in addition to the Yukawa coupling with the light species. The diagonalization of this mass sector can explain the smallness of neutrino masses, if the right-handed ones are sufficiently heavy. This constitutes the so-called type-I seesaw mechanism \cite{Minkowski1977,GellMann2013,Yanagida:1979as,Glashow1979,Mohapatra1980,Schechter1980}.

The aim of the present Chapter is to accommodate a sterile neutrino in the realization of the seesaw. To do so, we work within the extra-dimensional Randall-Sundrum scenario (See Refs. \cite{Randall1999,Randall1999a,Davoudiasl2000,Davoudiasl2001}).

\section{Motivation}

The results of the MiniBooNE Collaboration \cite{Collaboration2018a} combined with the ones of the Liquid Scintillator Neutrino Detector (LSND) \cite{Aguilar2001} report a $6.0 \ \sigma$ excess in the $\nu_e$ and $\bar{\nu}_e$ appearance experiments. This can be interpreted as a signal for the existence of sterile neutrinos of order $\sim 1 \ \mathrm{eV}$.

It is reasonable to accommodate eV-scale sterile neutrinos, which are SM singlets, working in the framework of a type-I seesaw mechanism \cite{Minkowski1977,GellMann2013,Yanagida:1979as,Glashow1979,Mohapatra1980,Schechter1980}. In order to successfully accomplish this task, we have to introduce at least two heavy singlet neutrinos (see Ref. \cite{Donini2011}). The Majorana masses of the right-handed neutrinos can be dynamically motivated by the spontaneous breaking of the $U(1)_{\mathrm{B-L}}$ symmetry group. As the minimal $U(1)_{\mathrm{B-L}}$-extended SM is anomaly free only if there are three right-handed neutrinos, we can realize the seesaw spontaneously breaking $U(1)_{\mathrm{B-L}}$.

In this 3+1+2 neutrino model, one of the right-handed neutrinos is eV-scale and the remaining two are much heavier (seesaw scale). The mass matrix results:
\begin{equation}\label{3.1}
\left(\begin{array}{cc}M_{\nu}^{4\times 4}&m_D^{4\times 2}\\ \left(m_D^{4\times 2}\right)^{T}&M^{2\times 2}\end{array}\right),
\end{equation}
where
\begin{equation}\label{3.2}
M_{\nu}^{4\times 4}=\left(\begin{array}{cccc}0&0&0&*\\ 0&0&0&*\\0&0&0&*\\ *&*&*&M_1\end{array}\right).
\end{equation}

The stars represent non-zero elements, and $M_1\sim1 \ \mathrm{eV}$ as discussed before. The resulting seesaw formula for the 3+1 neutrinos is:
\begin{equation}\label{3.3}
m_{\nu}^{4\times 4}\approx M_{\nu}^{4\times 4} - m_D^{4\times 2} \left(M^{2\times 2}\right)^{-1} \left(m_D^{4\times 2}\right)^{T}.
\end{equation}

A simple compactification scenario giving rise to the desired structure can be built within the Randall-Sundrum paradigm.

\section{Construction of the model}\label{Section3.2}

Consider the minimal $U(1)_{\mathrm{B-L}}$-extended Standard Model, embedded in a slice of $AdS_{5}$ (the Randall-Sundrum background) \cite{Randall1999,Randall1999a,Gogberashvili1998,Gogberashvili1998a,Gogberashvili1999}. We assume that the SM fields and the $U(1)_{\mathrm{B-L}}$ Higgs field $\Phi_{\mathrm{BL}}$ are localized on the IR-brane ($\phi=\pi$), whereas the $U(1)_{\mathrm{B-L}}$ gauge field and some right-handed neutrino fields extend over the 5th-dimension. As before, at the UV-brane we just locate a brane tension. In the following, we will follow mainly the discussions of Refs. \cite{Kawai2019,Davoudiasl2001}.

In this model, the right-handed neutrinos $N_i^c$ are identified as the KK zero-modes of the 5-dimensional fields $\Psi(x,\phi)$ \cite{Huber2003,Perez2009,Csaki2008,Fong2011,Iyer2013}. Starting from the 5-dimensional action of the bulk fermions:
\begin{equation}\label{3.4}
S_{bf}=\int \diff^4 x \int_{-\pi}^{\pi} r_c \diff \phi \sqrt{\vphantom{G}^{(5)}G} \left[e^{\mu}_{a}\left(\frac{i}{2}\bar{\Psi}\gamma^a \partial_{\mu}\Psi + h.c.+\frac{\omega_{bc\mu}}{8}\bar{\Psi}\left\{\gamma^a,\sigma^{bc}\right\}\Psi\right)-\mathrm{sgn}(\phi)m\bar{\Psi}\Psi\right],
\end{equation}
where $e^{\mu}_{a}=\mathrm{diag}\left(\mathrm{e}^{\sigma},\mathrm{e}^{\sigma},\mathrm{e}^{\sigma},\mathrm{e}^{\sigma},1\right)$ is the inverse vielbein, $\omega_{bc\mu}$ the spin connection and $\gamma^a=\left(\gamma^{\mu},i\gamma_5\right)$ the 4-dimensional representation of the 5-dimensional gamma matrices. The $\mathrm{sgn}(\phi)$ term is introduced in order to conserve $\phi$-parity, as required by the $\mathbb{Z}_2$ orbifold symmetry. Because our metric is diagonal, the only non-vanishing terms of the spin connection have $b=\mu$ or $c=\mu$, giving no contribution to the action.

We can decompose the bulk fermion field in its chiral components:
\begin{equation}\label{3.5}
\Psi_{\mathrm{L,R}}=\frac{1}{2}\left(1\mp\gamma_5\right)\Psi,
\end{equation}
and these into its KK-excitations:
\begin{equation}\label{3.6}
\Psi_{\mathrm{L,R}}=\sum_{n=0}^{\infty}\mathrm{e}^{-2\sigma} \psi_{\mathrm{L,R}}^{(n)}(x) \frac{f_{\mathrm{L,R}}^{(n)}\left(\phi\right)}{\sqrt{r_c}}.
\end{equation}

Implementing this reduction in the action gives the bulk Dirac equations:
\begin{equation}\label{3.7}
\left(\pm \frac{1}{r_c} \partial_{\phi}+m\right)f_{\mathrm{R,L}}^{(n)}\left(\phi\right)=m_n \mathrm{e}^\sigma f_{\mathrm{L,R}}^{(n)}\left(\phi\right),
\end{equation}
and the orthonormality relations:
\begin{equation}\label{3.8}
\int_{-\pi}^{\pi}\diff \phi \ \mathrm{e}^{\sigma} f_{\mathrm{R}}^{(m)*}\left(\phi\right) f_{\mathrm{R}}^{(n)}\left(\phi\right)=\int_{-\pi}^{\pi}\diff \phi \ \mathrm{e}^{\sigma} f_{\mathrm{L}}^{(m)*}\left(\phi\right) f_{\mathrm{L}}^{(n)}\left(\phi\right)=\delta_{mn}.
\end{equation}

The general solution to Eq. (\ref{3.7}) for $n\neq 0$ is given by:
\begin{equation}\label{3.9a}
f_{\mathrm{L,R}}^{(n)}=\frac{\mathrm{e}^{\sigma/2}}{N^{\mathrm{L,R}}_{n}}\left[J_{\frac{1}{2}\mp \nu}\left(z^{\mathrm{L,R}}_{n}\right)+\beta^{\mathrm{L,R}}_{n} Y_{\frac{1}{2}\mp \nu}\left(z^{\mathrm{L,R}}_{n}\right)\right],
\end{equation}
where $z^{\mathrm{L,R}}_{n}(\phi)=m^{\mathrm{L,R}}_{n}\mathrm{e}^{\sigma(\phi)}/k$, $m^{\mathrm{L,R}}_{n}$ is the mass of the n-th KK-excitation $\psi_{\mathrm{L,R}}^{(n)}(x)$, $\nu=m/k$, $N^{\mathrm{L,R}}_{n}$ the normalization factors and the coefficients $\beta^{\mathrm{L,R}}_{n}$ are constants. The zero-mode yields:
\begin{equation}\label{3.10a}
f_{\mathrm{L,R}}^{(0)}=\frac{\mathrm{e}^{\nu\sigma}}{N^{\mathrm{L,R}}_{0}}.
\end{equation}

As the action has to be $\mathbb{Z}_2$-symmetric, the left and right modes must have opposite $\mathbb{Z}_2$-parity. If we choose $f_{\mathrm{L}}^{(n)}$ to be $\mathbb{Z}_2$-even, therefore $f_{\mathrm{R}}^{(n)}$ is $\mathbb{Z}_2$-odd. With this choice, we must impose the boundary conditions:
\begin{equation}\label{3.11a}
\begin{split}
\left(\frac{\diff}{\diff \phi}-mr_c\right)f_{\mathrm{L}}^{(n)}|_{\phi=0,\pi}&=0,\\
f_{\mathrm{R}}^{(n)}|_{\phi=0,-\pi}&=0.\\
\end{split}
\end{equation}

In the case of the left-handed wavefunctions, we obtain:
\begin{equation}\label{3.12a}
\begin{split}
&\beta^{\mathrm{L}}_{n}=-\frac{J_{-\left(\frac{1}{2}+ \nu\right)}\left(m^{\mathrm{L}}_{n}/k\right)}{Y_{-\left(\frac{1}{2}+ \nu\right)}\left(m^{\mathrm{L}}_{n}/k\right)},\\
&J_{-\left(\frac{1}{2}+ \nu\right)}\left(x^{\mathrm{L}}_{n}\right)+\beta^{\mathrm{L}}_{n}Y_{-\left(\frac{1}{2}+ \nu\right)}\left(x^{\mathrm{L}}_{n}\right)=0.\\
\end{split}
\end{equation}
where $x_{n}^{L}$ is the n-th zero of this equation. Similarly, for the right-handed ones:
\begin{equation}\label{3.13a}
\begin{split}
&\beta^{\mathrm{R}}_{n}=-\frac{J_{\frac{1}{2}+ \nu}\left(m^{\mathrm{R}}_{n}/k\right)}{Y_{\frac{1}{2}+ \nu}\left(m^{\mathrm{R}}_{n}/k\right)},\\
&J_{\frac{1}{2}+ \nu}\left(x^{\mathrm{R}}_{n}\right)+\beta^{\mathrm{R}}_{n}Y_{\frac{1}{2}+ \nu}\left(x^{\mathrm{R}}_{n}\right)=0.\\
\end{split}
\end{equation}

From these relations, the coefficients and the masses of the KK-modes can be obtained, by numerically solving the appropriate Bessel function roots.

The orthonormality relations (\ref{3.8}) give us the expression for the normalization constants for $n\neq 0$:
\begin{equation}\label{3.14a}
N^{\mathrm{L,R}}_{n}=\left(\frac{\mathrm{e}^{\pi k r_c}}{x_n^{\mathrm{L,R}} \sqrt{kr_c}}\right)\sqrt{\left\{\left(z^{\mathrm{L,R}}_{n}\right)^2\left[J_{\frac{1}{2}\mp \nu}\left(z^{\mathrm{L,R}}_{n}\right)+\beta^{\mathrm{L,R}}_{n} Y_{\frac{1}{2}\mp \nu}\left(z^{\mathrm{L,R}}_{n}\right)\right]^2\right\}^{z^{\mathrm{L,R}}_{n}(\phi=\pi)}_{z^{\mathrm{L,R}}_{n}(\phi=0)}},
\end{equation}
and for the zero-mode:
\begin{equation}\label{3.15a}
N^{\mathrm{L,R}}_{0}=\sqrt{\frac{2\left[\mathrm{e}^{(1+2\nu)\pi k r_c}-1\right]}{(1+2\nu)kr_c}}.
\end{equation}

We are only interested in the zero modes. Following our choice for the $\mathbb{Z}_2$-parity, $f_{\mathrm{R}}^{(0)}$ turns out to be trivial and, as the zero-modes have vanishing mass, we obtain a expression for the left zero-mode:
\begin{equation}\label{3.9}
f_{\mathrm{L}}^{(0)}=\sqrt{\frac{\left(1+2\nu\right)k r_c}{\mathrm{e}^{\left(1+2\nu\right)\pi k r_c}-1}} \ \mathrm{e}^{\nu\sigma}.
\end{equation}
Considering three generations of 5-dimensional fields $\Psi_i$ with different masses, their left-handed component can be decomposed as:
\begin{equation}\label{3.10}
\Psi_{i,\mathrm{L}}=\psi_{i,\mathrm{L}}^{(0)}(x) \ \sqrt{\frac{\left(1+2\nu_i\right)k}{\mathrm{e}^{\left(1+2\nu_i\right)\pi k r_c}-1}} \ \mathrm{e}^{(2+\nu_i)\sigma}+...
\end{equation}

The left-handed zero-mode fields $\psi_{i,\mathrm{L}}^{(0)}(x)$ are identified as the conjugate of the right-handed sterile neutrino fields $N_i^c$ \cite{Grossman1999,Chang1999,Davoudiasl2001}. The 5-dimensional mass parameters $m_i$ are of the same order as the curvature $k$, but their sign can be either positive or negative.

\section{Yukawa couplings}

The action of the $U(1)_{\mathrm{B-L}}$ scalar sector on the IR-brane is:
\begin{equation}\label{3.11}
S_{\Phi_{\mathrm{BL}}}=-\int \diff^4 x \int_{-\pi}^{\pi} r_c \diff \phi \sqrt{\vphantom{G}^{(5)}G} \ \delta\left(\phi-\pi\right) \left\{(D_{\mu}\tilde{\Phi}_{\mathrm{BL}})^{\dagger}D^{\mu}\tilde{\Phi}_{\mathrm{BL}}+\lambda\left(\tilde{\Phi}_{\mathrm{BL}}^{\dagger}\tilde{\Phi}_{\mathrm{BL}}-\frac{1}{2}\tilde{v}_{\mathrm{BL}}^2\right)^2\right\}.
\end{equation}

The rescaling $\Phi_{\mathrm{BL}}=\mathrm{e}^{-\pi k r_c} \ \tilde{\Phi}_{\mathrm{BL}}$ and $v_{\mathrm{BL}}=\mathrm{e}^{-\pi k r_c} \ \tilde{v}_{\mathrm{BL}}$ reduce the action (\ref{3.11}) to a canonically normalized action in 4-dimensions, corresponding to a scalar $\Phi$ spontaneously breaking the $U(1)_{\mathrm{B-L}}$ symmetry.

The Majorana masses of the right-handed neutrinos will be generated through the $U(1)_{\mathrm{B-L}}$ scalar field expectation value on the IR-brane. The resulting Yukawa term is written as:
\begin{equation}\label{3.12}
S_{MY}=-\int \diff^4 x \int_{-\pi}^{\pi} r_c \diff \phi \sqrt{\vphantom{G}^{(5)}G} \ \delta\left(\phi-\pi\right) \frac{\lambda_{ij}}{M_5} \tilde{\Phi}_{\mathrm{BL}}\bar{\Psi}_i\Psi_j.
\end{equation}

Using KK-reduction (\ref{3.10}) and the rescaling of the scalar, the action becomes:
\begin{equation}\label{3.13}
S_{MY}=-\int \diff^4 x \lambda_{ij}^{\mathrm{eff}} \Phi_{\mathrm{BL}}\bar{N}_i^cN_j + \text{KK modes},
\end{equation}
where the effective 4-dimensional Majorana Yukawa couplings are given by:
\begin{equation}\label{3.14}
\lambda_{ij}^{\mathrm{eff}} =\sqrt{\frac{\left(1+2\nu_i\right)\left(1+2\nu_j\right)}{\left(\mathrm{e}^{\left(1+2\nu_i\right)\pi k r_c}-1\right)\left(\mathrm{e}^{\left(1+2\nu_j\right)\pi k r_c}-1\right)}} \ \mathrm{e}^{(1+\nu_i+\nu_j)\pi k r_c} \ \lambda_{ij}.
\end{equation}

Defining the quantity:
\begin{equation}\label{3.15}
\omega_i\equiv\sqrt{\frac{\left(1+2\nu_i\right)\mathrm{e}^{(1+2\nu_i)\pi k r_c}}{\mathrm{e}^{\left(1+2\nu_i\right)\pi k r_c}-1}},
\end{equation}
we can compactly write $\lambda_{ij}^{\mathrm{eff}}=\omega_i\omega_j\lambda_{ij}$.

Once singlet fermions are added to the Lagrangian, it is unavoidable to include a Dirac Yukawa term in the action:
\begin{equation}\label{3.16}
S_{DY}=-\int \diff^4 x \int_{-\pi}^{\pi} r_c \diff \phi \sqrt{\vphantom{G}^{(5)}G} \ \delta\left(\phi-\pi\right) \left\{\frac{y_{\alpha i}}{\sqrt{M_5}} \ \bar{l}^{\alpha}H\Psi_i + \mathrm{h.c.}\right\}.
\end{equation}

After spontaneous electroweak breaking, this term will induce a Dirac mass term that mixes the left-handed neutrinos with the new singlet fermions (the right-handed neutrinos). Upon KK-reduction and rescaling the Higgs and lepton doublets as $\Phi=\mathrm{e}^{-\pi k r_c} \ H$ and $L^{\alpha}=\mathrm{e}^{-\frac{3}{2}\pi k r_c} \ l^{\alpha}$, this term becomes:
\begin{equation}\label{3.17}
S_{DY}=-\int \diff^4 x \left\{y_{\alpha i}^{\mathrm{eff}} \bar{L}^{\alpha}\Phi N_i^c + \mathrm{h.c.}\right\}.
\end{equation}

The effective Dirac Yukawa couplings are:
\begin{equation}\label{3.18}
y_{\alpha i}^{\mathrm{eff}}=\sqrt{\frac{\left(1+2\nu_i\right)}{\mathrm{e}^{\left(1+2\nu_i\right)\pi k r_c}-1}} \ \mathrm{e}^{(\frac{1}{2}+\nu_i)\pi k r_c} \ y_{\alpha i}=\omega_i y_{\alpha i}.
\end{equation}

In the end, after $U(1)_{\mathrm{B-L}}$ and electroweak symmetries are spontaneously broken, we are left with the following mass terms for the neutrino sector:
\begin{equation}\label{3.19}
S_{\nu}=-\int \diff^4 x \left\{m_{\alpha i} \bar{\nu}^{\alpha}N_i + \mathrm{h.c.}+ M_{ij} \bar{N}_i^cN_j\right\}.
\end{equation}

The matrix $M_{ij}$ is real and symmetric and can be diagonalized without loss of generality \cite{Donini2011,Santamaria1993}.

\section{Seesaw realization}

From the previous developments, the entries of the mass matrix (\ref{3.1}) are given by:
\begin{equation}\label{3.27}
\begin{split}
M_{ij}&\equiv\lambda_{ij}^{\mathrm{eff}} v_{\mathrm{BL}},\\
m_{\alpha i}&\equiv y_{\alpha i}^{\mathrm{eff}} v_{\mathrm{EW}},
\end{split}
\end{equation}
and therefore reads:
\begin{equation}\label{3.28}
\left(\begin{array}{cc}0&y^{\mathrm{eff}} v_{\mathrm{EW}}\\\left(y^{\mathrm{eff}}\right)^{T} v_{\mathrm{EW}}&\lambda^{\mathrm{eff}} v_{\mathrm{BL}}\end{array}\right).
\end{equation}

If we are to evaluate this matrix, we have to make some assumptions about the values of the dimensionless mass parameters $\nu_i$. The mild assumption:
\begin{equation}\label{3.29}
1+2\nu_1<0; \ 1+2\nu_2>0; \ 1+2\nu_3>0,
\end{equation}
along with assuming all the Yukawa couplings to be $\mathcal{O}(1)$ and the $U(1)_{\mathrm{B-L}}$ breaking scale $v_{\mathrm{BL}}\sim10^{13} \ \mathrm{GeV}$ result in the correct combination:
\begin{equation}\label{3.30}
\left\{\begin{array}{ccc}M_{\nu}^{4\times 4}&\lesssim &1 \ \mathrm{eV},\\m_{D}^{2\times 2}&\sim& 100 \ \mathrm{GeV},\\M^{2\times 2}&\sim &10^{13} \ \mathrm{GeV}.\end{array}\right.
\end{equation}

This simple warped compactification scenario can address the large mass hierarchy between the singlet neutrinos. Just with small fluctuations in the mass parameters of the 5-dimensional bulk fermions, an eV-scale sterile neutrino can arise in the 4-dimensional effective theory.

\chapter{Gravity-mediated DM in Warped Extra-Dimensions}\label{Chapter 4}

One of the main motivations for physics Beyond the Standard Model (BSM) is to address the nature of Dark Matter (DM). While astrophysical data points towards the existence of this non-baryonic matter, high-energy collider experiments have not yet observed any candidate to fill this rôle. Searches at the LHC have set the lower bound on the masses of the candidates to the TeV range, a region of the parameter space difficult to test.

\section{Experimental evidence for DM}

Our present knowledge reveals that ordinary, baryonic matter does not dominate the composition of the Universe. Astrophysical observations point towards the existence of an unknown type of matter which is approximately five times more abundant than this. That so-called ``Dark Matter'' constitutes one of the main puzzles of modern Physics.

\begin{figure}
	\centering
	\includegraphics[width=1\linewidth]{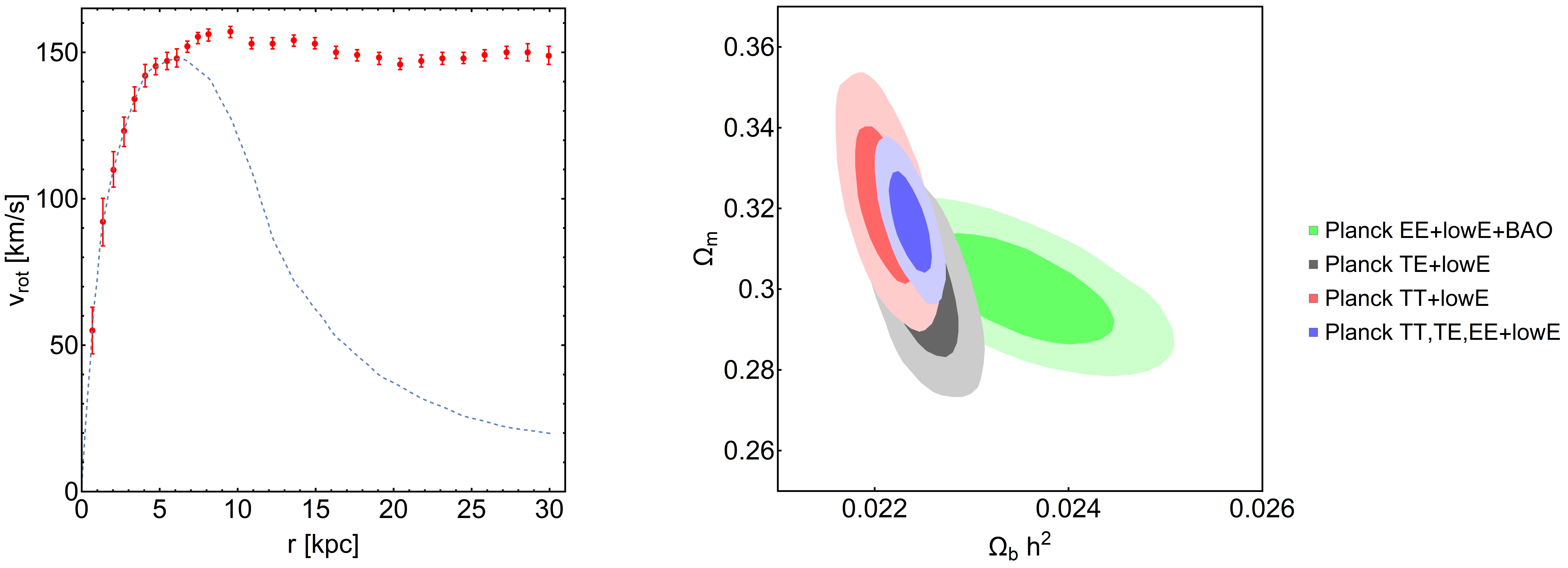}
	\caption{\textit{Left panel: Rotational velocities of HI regions in NGC 3198 (red points), taken from Ref. \cite{Begeman1989}, compared to the expected Keplerian behaviour (dashed blue line). Right panel: Constraints on the $\Omega_{b} h^2$, $\Omega_{m}$ plane, comparing different combinations of Planck EE, TE, TT, lowE and BAO (taken from Ref. \cite{Collaboration2018}).}}
	\label{fig:dm_evidence}
\end{figure}

The first evidence for the existence of DM come from observations of velocity distributions along spiral galaxies. The rotational velocity profiles do not match the expected Keplerian dependence $v \propto 1/\sqrt{r}$, being $r$ the distance to the luminous matter distribution. On the contrary, the velocity remains almost constant from a certain point \cite{Rubin1970}. In Fig. \ref{fig:dm_evidence} (left panel) we show the measured rotational velocities of HI regions in NGC 3198 \cite{Begeman1989}, comparing them with the expected Keplerian behaviour. The most feasible explanation for these plain profiles is that it should exist some missing non-luminous mass in the regions with low luminous matter density. Gravitational lensing offers another hint for the existence of DM. The estimations for galactic cluster masses coming from lensing measurements of bright objects yield much larger masses than the inferred from cluster's luminosity estimates \cite{Bergmann1990}. In order to account for these excesses, large amounts of DM are needed. Cosmological observations reveal the need of an electrically neutral matter existing long before recombination to explain the structure formation, as CMB fluctuations are to small too account for them \cite{Smoot1992}. From CMB anisotropies the total and baryonic matter densities can be estimated. In Fig. \ref{fig:dm_evidence} (right panel) we show the constraints in these parameters impossed by PLANCK \cite{Collaboration2018}. The best fit of the data collected gives:
\begin{equation}
\left\{\begin{array}{ccc}\Omega_{m}h^2&=&0.1430\pm 0.0011,\\\Omega_{b}h^2&=&0.02237 \pm 0.00015,\end{array}\right.
\end{equation}
where the subscripts $m$ and $b$ refers to the total and baryonic matter densities respectively. From this result we can clearly see that baryonic matter is not the only type of matter in the Universe. Thus, DM accounts for approximately the $85 \%$ of matter.

\section{The DM Relic Abundance in the Freeze-Out scenario}

Within the cosmological ``standard model'', DM is assumed to be constituted by stable and heavy particles (i.e. long-lived and non-relativistic). In the so-called thermal freeze-out scenario, DM is in thermal equilibrium with the rest of the particle content of the Early Universe. The time evolution of DM density in this case is given by the Boltzmann equation:
\begin{equation}\label{4.1}
\frac{\diff n_{\mathrm{DM}}}{\diff t}=-3 H(t) n_{\mathrm{DM}}- \expval{\sigma v} \left[ n_{\mathrm{DM}}^2-\left( n_{\mathrm{DM}}^{eq}\right)^2\right],
\end{equation}
where $T$ is the temperature, $H(T)$ the Hubble parameter and $ n_{\mathrm{DM}}^{eq}$ the DM density in equilibrium. As the Universe expanded, it cooled down. Then, below some temperature $T_{\mathrm{FO}}$ the density $n_{\mathrm{DM}}$ ``freezes out'' and it is no longer modified by thermal fluctuations. This occurs when the thermally-averaged cross-section $\expval{\sigma v}$ times the number density falls below $H(T)$. At this time DM decouples from the rest of the particles, leaving a constant density in its co-moving frame, the relic abundance.

The relic abundance can be computed from experimental data, starting from the DM density. Solving (\ref{4.1}), one can found that the thermally-averaged cross-section at freeze-out is $\expval{\sigma_{\mathrm{FO}} v}=2.2\times 10^{-26} \ \mathrm{cm^3/s}$ \cite{Steigman2012}. For $m_{\mathrm{DM}}>10 \ \mathrm{GeV}$ the relic abundance is insensitive to $m_{\mathrm{DM}}$ and therefore $\sigma_{\mathrm{FO}}$ is not a function of the DM mass. Usually, $\expval{\sigma v}$ is computed in the so-called velocity expansion, assuming small relative velocity between the DM particles. However, as this expansion can fail in the neighbourhood of resonances, the exact value of $\expval{\sigma v}$ can be calculated as \cite{Gondolo1991}:
\begin{equation}
\expval{\sigma v}=\frac{1}{8 m_{\mathrm{DM}}^4TK_{2}^{2}\left(m_{\mathrm{DM}}/T\right)}\int_{4m_{\mathrm{DM}}^2}^{\infty} \diff s \ \sigma(s) \left(s-4m_{\mathrm{DM}}^2\right) \sqrt{s} K_1\left(\sqrt{s}/T\right),
\end{equation}
where $K_1$ and $K_2$ are the modified Bessel functions. DM particles within this scenario are usually known as Weakly Interacting Massive Particles (WIMP's).

Ongoing Direct and Indirect Detection searches for massive DM particles \cite{Aprile2017,FermiLAT2016,Aguilar2014} that could give the observed relic abundance within the freeze-out scenario have not been successful to date. Bounds on non-gravitational interactions of DM particles with masses below 1 TeV are currently stronger than $10 \ \mathrm{pb}$, pointing out that DM particles are either heavier than expected, or with sub-weak interactions.

A possibility previously explored in the literature is based on the enhancement of the gravitational interaction between SM particles and DM, as a result of gravity feeling more than the usual 3+1 space-time dimensions \cite{Lee2014,Lee2014a,Folgado2020a}. These extra-dimensional models have been proposed to address the hierarchy problem between the electroweak scale and the Planck scale \cite{Randall1999,Randall1999a}. Even though these models do not say anything about the nature of DM, they can be used to find out if it is possible to obtain the correct DM relic abundance and through which channels.

The so-called Weakly Interacting Massive Particle (WIMP) paradigm is one of the main candidate to explain DM evolution. These heavy particles would be thermally produced during an early epoch by the relativistic plasma. As the Universe expands and cools down, their production is exponentially suppressed. At some temperature, the DM particles become so rare that their annihilation rate drops and since then its number density does not change, leaving a relic abundance of DM behind. Thus, it is said that the density of DM ``freezes-out'' at certain temperature $T_{\mathrm{FO}}< m_{\mathrm{DM}}$. This possibility has been studied within the context of warped extra-dimensions in Refs. \cite{Folgado2020a,Folgado2020b}, where the relic abundance is achieved thorough the enhancement of the gravitational interaction due to the presence of the extra space-time dimensions. Here, we are going to mainly follow their reasoning.

\section{Contributions to $\expval{\sigma v}$ in the RS model}

If we want to check that some model reproduces the correct relic abundance at freeze-out, the parameter we have to compute is $\expval{\sigma v}$. To do so, first we should calculate the total DM annihilation cross-section:
\begin{equation}\label{4.2}
\begin{split}
\sigma&=\sum_{\mathrm{SM}} \sigma_{\mathrm{ve}}\left(\mathrm{DM} \ \mathrm{DM}\rightarrow \mathrm{SM} \ \mathrm{SM}\right)+\sum_{n=1}\sum_{m=1} \sigma\left(\mathrm{DM} \ \mathrm{DM}\rightarrow G_n \ G_m\right)\\
&+\sigma\left(\mathrm{DM} \ \mathrm{DM}\rightarrow r \ r\right)+\sum_{n=1} \sigma\left(\mathrm{DM} \ \mathrm{DM}\rightarrow G_n \ r\right).
\end{split}
\end{equation}

The first term accounts for the annihilation to any SM particle, through virtual exchange (``ve'') of KK-gravitons or the radion. The second one correspond to DM annihilation into two KK-gravitons, $G_n$ and $G_m$, where we can have $n$ different from $m$ as KK-number is not conserved (as a consequence of the explicit breaking of momentum conservation in the 5th-dimension due to the presence of the branes). The third term corresponds to annihilation into a pair of radions. Eventually, the fourth term accounts for the annihilation into a KK-graviton and a radion.

If $m_{\mathrm{DM}}<m_{G_1}, \ m_{r}$, then only the first channel exists. This process can occur through the two mediators mentioned previously. The dominance of these channels will depend on the particular values of the radion mass and the KK-graviton masses. The second channel to open is the DM annihilation into radions, as within the Goldberger-Wise stabilization mechanism the radion is expected to be lighter than the first KK-graviton. Then, for $m_{\mathrm{DM}}>\left(m_{G_1}+m_{r}\right)/2$, the decay channel into a KK-graviton and the radion opens. Eventually, for $m_{\mathrm{DM}}>m_{G_1}$ annihilation of DM particles into two on-shell KK-gravitons becomes possible. As KK-number is not conserved, we have to sum over all possible modes fulfilling the condition $2m_{\mathrm{DM}}>m_{G_n}+m_{G_m}$. We do not take into account annihilation into KK-graviton zero-modes as these channels are Planck suppressed.

\section{Fermionic DM annihilation cross-section in RS}

\begin{figure}
	\centering
	\includegraphics[width=1\linewidth]{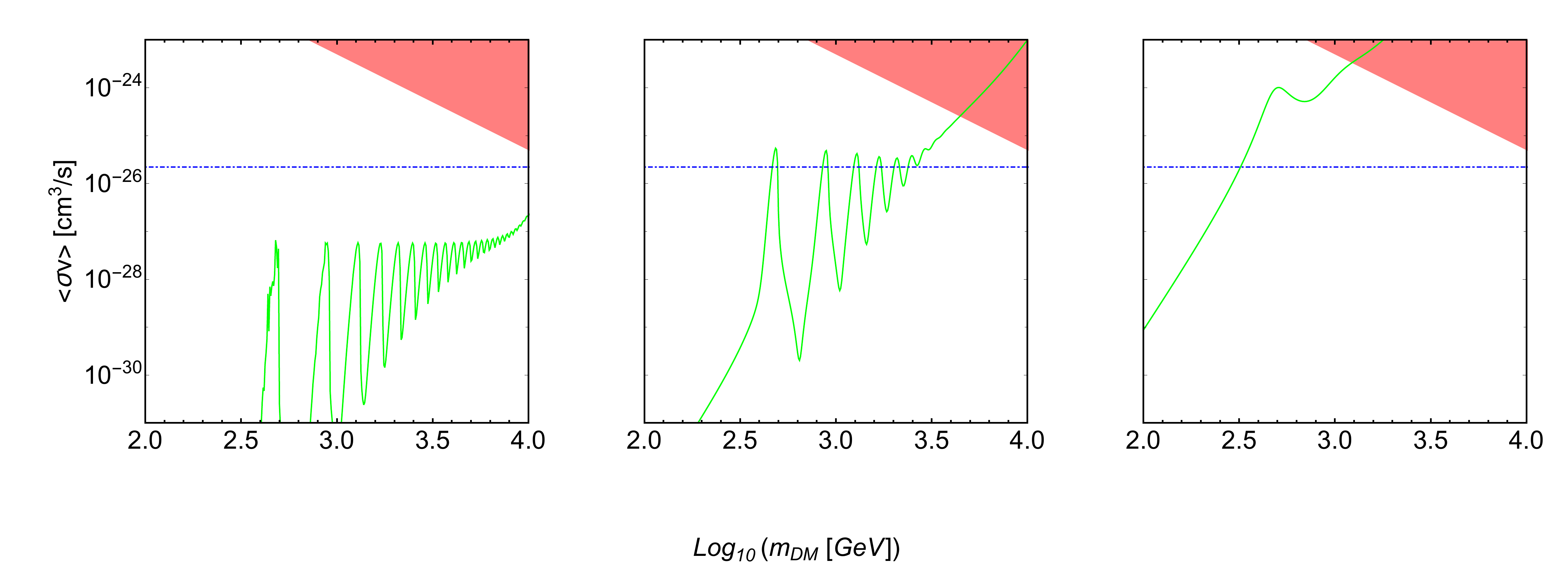}
	\caption{\textit{DM thermal average total annihilation cross section for the fermionic case (solid green line).  The blue dash-dotted line represent $\expval{\sigma_{\mathrm{FO}} v}=2.2\times 10^{-26} \ \mathrm{cm^3/s}$. The red-shaded area represents the theoretical unitarity bound $\expval{\sigma v} \geq 1/s$, where we can no longer trust the theory. Left panel: $m_{G_1} = 1 \ \mathrm{TeV}$, $\Lambda = 100 \ \mathrm{TeV}$; middle panel: $m_{G_1} = 1 \ \mathrm{TeV}$, $\Lambda = 10 \ \mathrm{TeV}$; right panel: $m_{G_1} = 1 \ \mathrm{TeV}$, $\Lambda = 1 \ \mathrm{TeV}$.}}
	\label{fig:relic}
\end{figure}

In Fig. \ref{fig:relic} we represent the total thermally-averaged cross-section of the fermionic DM as a function of its mass (solid green line). The three panels in the figure have been obtained for the choices of parameters $m_{G_1} = 1 \ \mathrm{TeV}$, $\Lambda = 100 \ \mathrm{TeV}$ (left panel), $m_{G_1} = 1 \ \mathrm{TeV}$, $\Lambda = 10 \ \mathrm{TeV}$ (middle panel) and $m_{G_1} = 1 \ \mathrm{TeV}$, $\Lambda = 1 \ \mathrm{TeV}$ (right panel). In all panels, the dash-dotted blue line represents the value of the thermally-averaged cross-section at freeze-out, $\expval{\sigma_{\mathrm{FO}} v}$. We notice that for large values of $\Lambda$, the resonant production of KK-graviton dominates in the low mass range. The direct production of two on-shell KK-gravitons becomes kinematically allowed for $m_{\mathrm{DM}}\geq m_{G_{1}}$. For a wide range of parameters, this channel allows to achieve the correct relic abundance. The red-shaded area in the upper-right corner represents the region where we trespassed the theoretical unitarity bound $\expval{\sigma v} \geq 1/s$, and therefore the effective theory is no longer trustable.

\chapter{Experimental bounds and theoretical constraints}\label{Chapter 5}

Even though the parameter space of the model is huge, it is strongly constrained by both experimental and theoretical bounds. In this Chapter, we analyse the different constraints that are relevant for our study. In the first section, we will focus on the bounds involving the DM particles and KK-graviton modes. Then, we will move on to summarize the current limits of heavy Majorana neutrinos searches.

As we are going to discuss in the next Chapter, we are interested in generate a specific hierarchy between the right-handed neutrinos. We want to identify one of them with the DM particle, therefore, it should lie in the appropriate mass range and fulfill its constraints. We also look for a light sterile neutrino, thus, it must be consistent with data from direct Majorana neutrino detection.

\section{DM constraints}

In this section we review the bounds from resonance searches at LHC, direct and indirect DM searches. These resonance searches strongly constrain the plane $\left(m_{G_1},\Lambda\right)$. We also discuss some constraints imposed by the limitations of the theory itself. For a detailed discussion see Refs. \cite{Folgado2020,Folgado2020a,Folgado2020b}.

\subsection{LHC bounds}

Resonance searches at LHC give strong bounds to the model. In the spectrum of the theory, the particles that may be resonantly produced at the LHC are the KK-graviton modes and the radion. First, one have to compute the production cross-section of these particles at the LHC, in order to estimate the consequences of LHC data in the parameter space later.

KK-graviton modes can be produced from $gg$ and $q\bar{q}$. The production cross-section of the KK-graviton $n$th-mode at the LHC is given by \cite{Giudice2018}:
\begin{equation}
\sigma_{pp\rightarrow G_{n}}\left(m_{G_n}\right)=\frac{\pi}{48\Lambda^2}\left[3 \lag_{gg}\left(m_{G_n}^2\right)+4\sum_{q} \lag_{q\bar{q}}\left(m_{G_n}^2\right)\right],
\end{equation}
where
\begin{equation}
\lag_{ij}\left(\hat{s}\right)=\frac{\hat{s}}{s}\int_{\hat{s}/s}^{1} \frac{\diff x}{x} f_{i}(x)f_j\left(\frac{\hat{s}}{xs}\right),
\end{equation}

where $f_{i}$ are the Parton Distribution Functions (PDF's) and the hats are introduced to distinguish between the variable and the $s$ parameter of the accelerator. In order to perform the calculations, the leading-order MSTW2008 Parton Distribution Functions (PDF's) $f_i(x)$ at $Q^2=m_{G_1}^2$ \cite{Martin2009} are used.

In the case of the radion, as the $q\bar{q}r$ vertex is proportional to the quark mass, its production cross-section in the LHC is dominated by gluon fusion. This interaction between gluons and the radion is similar to that between gluons and the SM Higgs, so one can use those well-known results for Higgs production at the LHC \cite{Spira1995}. Properly rescaling the Lagrangian, the cross-section at LHC results:
\begin{equation}
\sigma_{pp\rightarrow r}=\frac{\alpha_s^2 C_3^2}{1536\pi\Lambda^2}\lag_{gg}\left(m_r^2\right).
\end{equation}

As the trace of the stress-energy tensor of the gluon vanishes, the cross-section has to be calculated at one-loop. The explicit expression of $C_3$ can be found in App. B of Ref. \cite{Folgado2020a}.

\begin{figure}[t]
	\centering
	\includegraphics[width=0.9\linewidth]{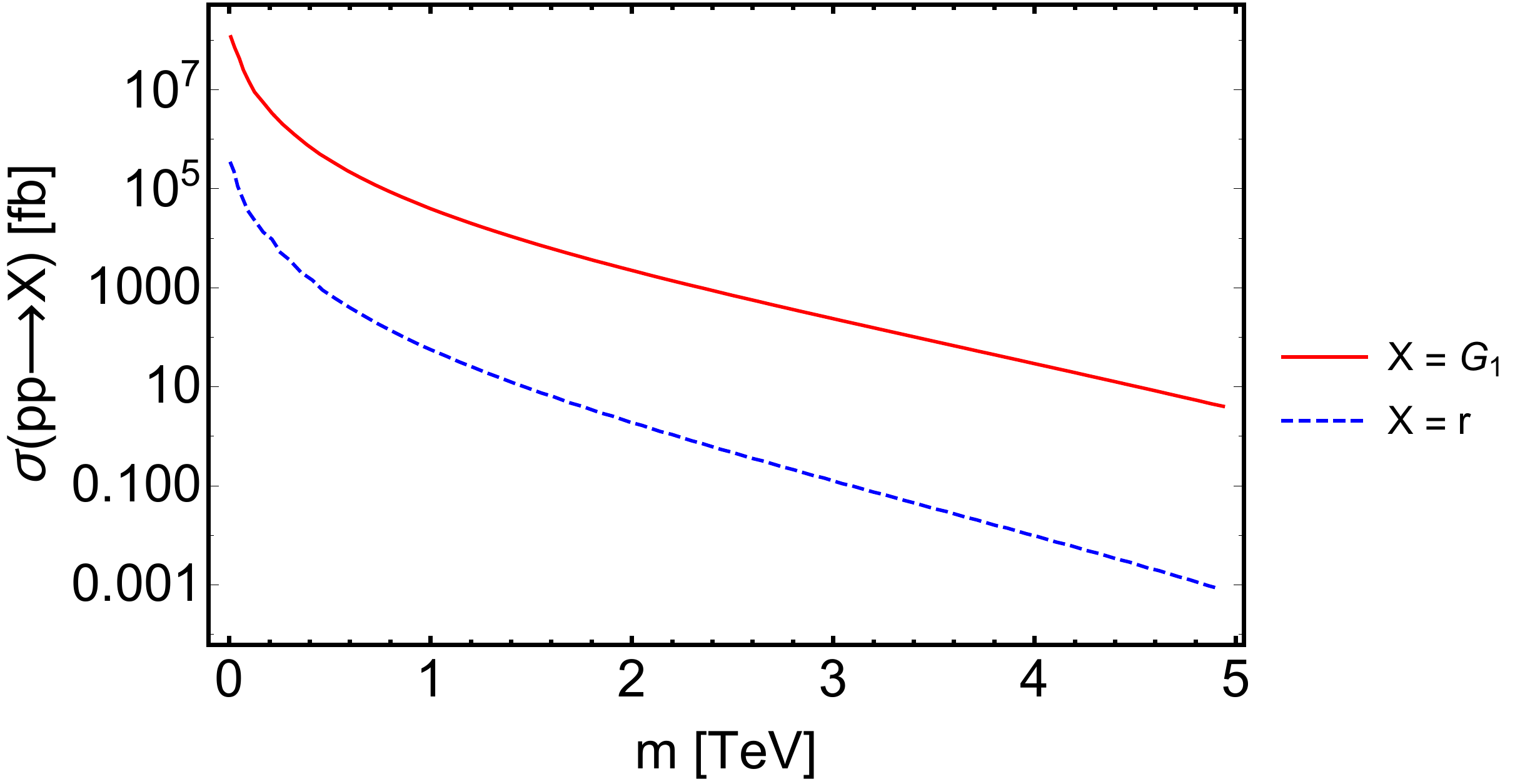}
	\caption[Caption]{\textit{Estimated first KK-graviton mode and radion production cross-section at LHC with $\sqrt{s}=13 \ \mathrm{TeV}$, for $\Lambda=5 \ \mathrm{TeV}$.}}
	\label{fig:lhccross}
\end{figure}

\begin{figure}[h!]
	\centering
	\includegraphics[width=0.9\linewidth]{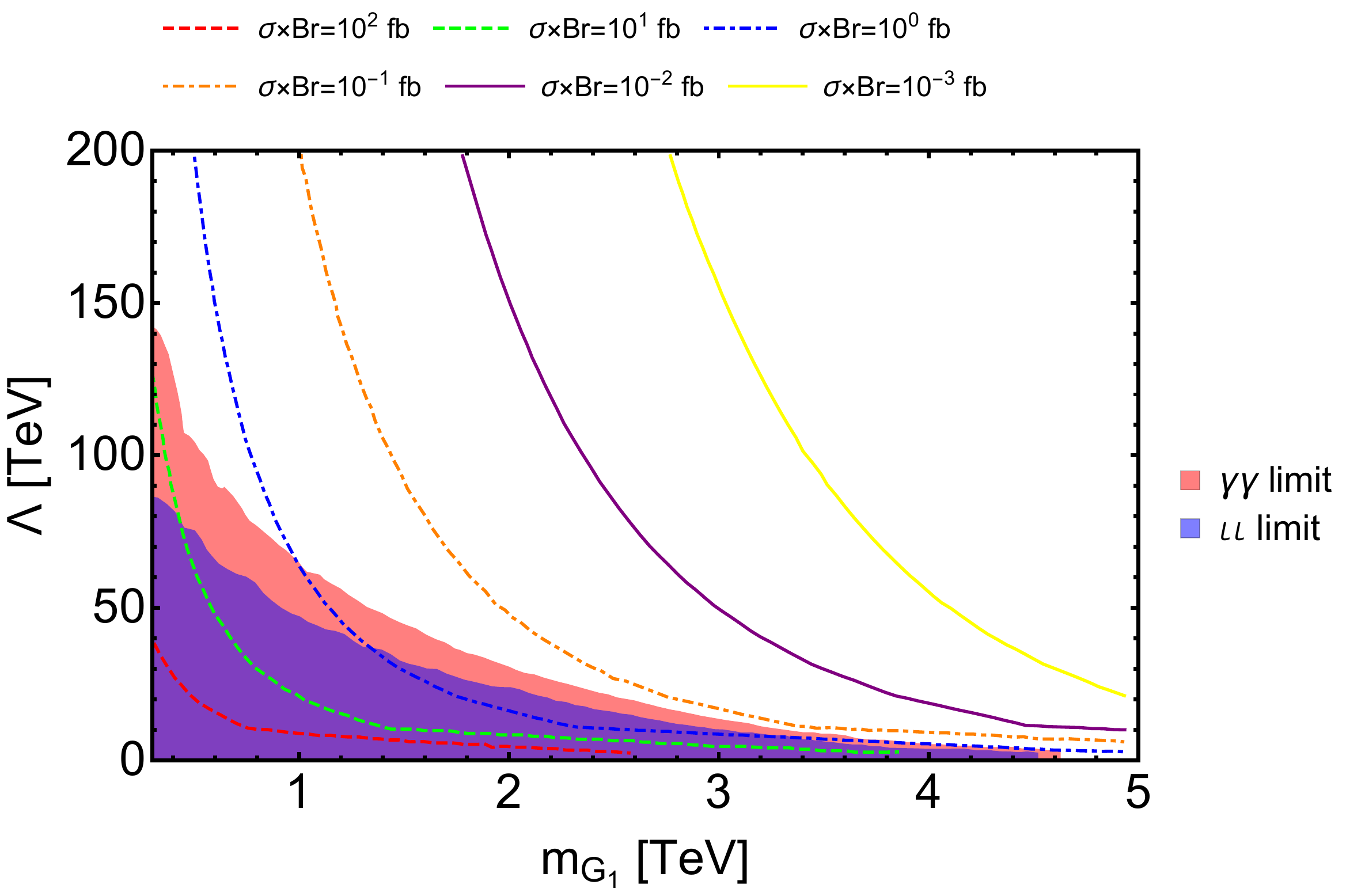}
	\caption[Caption]{\textit{Excluded regions of the  $(m_{G_1},\Lambda)$ plane at the LHC Run II with $\sqrt{s}=13 \ \mathrm{TeV}$ and $36 \ \mathrm{fb^{-1}}$, through resonant production of KK-graviton modes decaying into leptons (light blue area) and photons (light red area), extracted from \cite{Collaboration2017} and \cite{Collaboration2017a}.}}
	\label{fig:lhcbounds}
\end{figure}

In Fig. \ref{fig:lhccross} we represent the two production cross-sections at LHC, the solid red line stands for $pp\rightarrow G_1$ and the dashed blue line for $pp\rightarrow r$, with $\sqrt{s}=13 \ \mathrm{TeV}$ and $\Lambda=5 \ \mathrm{TeV}$. The dependence on $\Lambda$ is just a scaling factor, and therefore generalization to other values of $\Lambda$ is straightforward. One can notice that the radion cross-section is smaller than the one of the graviton in all the plotting range, with a difference of at least a couple of orders of magnitude. For this reason, the constraints on the model are dominated by KK-graviton searches.

In Fig. \ref{fig:lhcbounds} we represent the experimental bounds on $\sigma(pp\rightarrow\gamma\gamma)$ (blue-shaded area) and on $\sigma(pp\rightarrow\ell\ell)$ (red-shaded area). Comparing these with the theoretical expectations $\sigma\times\mathrm{BR}\left(G_1\rightarrow\gamma\gamma\right)$ and $\sigma\times\mathrm{BR}\left(G_1\rightarrow\ell\ell\right)$, ranging from $10^2 \ \mathrm{fb}$ (bottom red-dashed line) to $10^{-3} \ \mathrm{fb}$ (top yellow-solid line), we can set a excluded region in the plane $\left(m_{G_1},\Lambda\right)$. One can notice that the strongest bounds are given by $pp\rightarrow G_1 \rightarrow \gamma\gamma$.

\subsection{Direct and indirect DM searches}

Independently of its spin, the total cross-section for elastic scattering between DM and nuclei is given by \cite{CarrilloMonteverde2018}:
\begin{equation}
\sigma_{p}=\left[\frac{m_{DM}m_{p}}{A \pi (m_{DM}+m_{p})}\right]^2 \left[A f_{p}^{DM}+(A-Z)f_{n}^{DM}\right]^2,
\end{equation}
where $Z$ and $A$ are the atomic and mass number respectively, $m_{DM}$ the mass of the DM particle and $m_{p}$ the mass of the proton. The form factors can be written as:
\begin{equation}
\begin{split}
f_{p}^{DM}&=\frac{m_{DM}m_{p}}{4 m_{G_1}^2\Lambda^2} \left\{\sum_{q=u,d,s,c,b} 3\left[q(2)+\bar{q}(2)\right]+\sum_{q=u,d,s}\frac{1}{3}f_{Tq}^{p}\right\},\\
f_{n}^{DM}&=\frac{m_{DM}m_{p}}{4 m_{G_1}^2\Lambda^2} \left\{\sum_{q=u,d,s,c,b} 3\left[q(2)+\bar{q}(2)\right]+\sum_{q=u,d,s}\frac{1}{3}f_{Tq}^{n}\right\},\\
\end{split}
\end{equation}
where $q(2)$ and $\bar{q}(2)$ are the second moment of the PDF's:
\begin{equation}
q(2)=\int_{0}^{1} \diff x x f_{q}(x),
\end{equation}
and $f_{Tq}^{p,n}$ the mass fractions of the light quarks in nucleons, which are displayed in Table \ref{tab:massfraction}. The second moments of the PDF's can be computed using Ref. \cite{Martin2009}.

\begin{table}[t]
	\centering
	\begin{tabular}{|c|c|c|c|}
		\hline
		$f_{Tq}^{N}$ & $u$ & $d$ & $s$ \\
		\hline
		$p$ & 0.023 & 0.032 & 0.020 \\
		\hline
		$n$ & 0.017 & 0.041 & 0.020 \\
		\hline
	\end{tabular}
	\caption{\textit{Mass fraction of the light quarks in a nucleon, as given in \cite{Hisano2010}}.}
	\label{tab:massfraction}
\end{table}

\begin{figure}[t]
	\centering
	\includegraphics[width=1.0\linewidth]{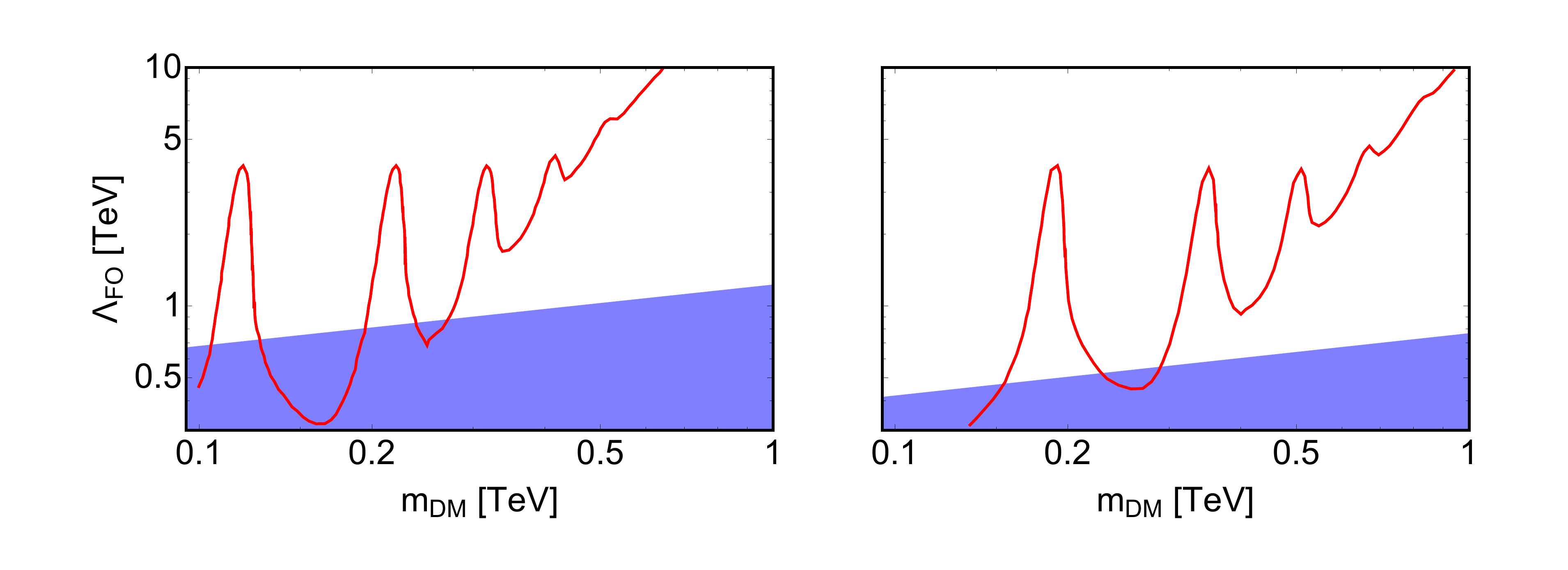}
	\caption[Caption]{\textit{Values of $\Lambda$ for which the observed DM relic abundance, $\Lambda_{FO}$, is achieved as a function of the DM particle mass $m_{DM}$, for various values of $m_{G_1}$. The bounds form DD DM searches in the $\left(m_{DM},\Lambda\right)$ plane are represented by the blue-shaded area. Left panel: $m_{G_1}=250 \ \mathrm{GeV}$; right panel: $m_{G_1}=400 \ \mathrm{GeV}$.}}
	\label{fig:DDDM}
\end{figure}

The strongest bounds from DM Direct Detection (DD) searches comes from XENON1T, its exclusion curve set certain constraints on the space $\left(m_{DM},m_{G_1},\Lambda\right)$ \cite{Aprile2017}. In Fig. \ref{fig:DDDM} we show the bounds in the $\left(m_{DM},\Lambda\right)$ plane, for the values $m_{G_1}=250 \ \mathrm{GeV}$ (left panel) and $m_{G_1}=400 \ \mathrm{GeV}$ (right panel), represented by the blue-shaded area. The red solid line represent the values of $\Lambda$ required to achieve the observed DM relic abundance, $\Lambda_{FO}$, as a function of the DM particle mass. One can notice the resonant behaviour of $\Lambda_{FO}$ for $m_{DM}<2m_{G_1}$, where the DM annihilation cross-section is dominated by virtual KK-graviton exchange. As the value of $m_{DM}$ increases, eventually the production of two on-shell KK-gravitons dominates and $\Lambda_{FO}$ grows smoothly. In the excluded region, the relic abundance is achieved through KK-graviton resonances, and therefore the exclusion bounds will have a striped pattern. In any case, bounds from DD DM searches result weaker than those imposed by the LHC.

Bounds from DM Indirect Detection (ID) searches also arises in models where a continuum spectra of SM particles are generated, as it is our case. Besides DM annihilation through exchange of virtual KK-gravitons into SM particles is $d$-wave suppressed, the channels into on-shell KK-gravitons or radions can produce observable signals as they later decay into SM. The Fermi-LAT collaboration has analysed gamma-ray from Dwarf spheroidal galaxies \cite{FermiLAT2016} and the galactic center \cite{Collaboration2015,Collaboration2017b}. The AMS Collaboration has reported positron \cite{Aguilar2014} and anti-proton \cite{Aguilar2016} fluxes coming from the galactic center. However, these data constraint DM particles with masses below $\sim 100 \ \mathrm{GeV}$, and therefore in our case of heavy DM ($\sim 1 \ \mathrm{TeV}$) the bounds have no significant impact in our parameter space.

\subsection{Unitarity limit}

We are performing a tree-level computation, and therefore one should worry about unitarity issues. The annihilation cross-section of DM into a pair of on-shell KK-gravitons diverges as $m_{\mathrm{DM}}^6/\left(m_{G_n}^2m_{G_m}^2\right)$. When $m_{\mathrm{DM}}\gg m_{G_n}, \ m_{G_m}$, one have to check if the effective theory is still unitary \cite{Kahlhoefer2016}. If we want the cross-section to be bounded, we must exclude the region $\sigma > 1/s \simeq 1/m_{\mathrm{DM}}^{2}$. When we combine this requirement with the achievement of the DM relic abundance we obtain directly an upper bound for the DM mass $m_{\mathrm{DM}}\lesssim1/\sqrt{\sigma_{\mathrm{FO}}}$. This bound is independent of the geometry of the space-time.

Moreover, we have to pay attention to the self-consistency of the theory framework. In RS scenarios, at energies larger than $\Lambda$ the KK-gravitons result strongly coupled and therefore the effective theory is no longer valid. Thus, we must impose $m_{G_1}<\Lambda$ to trust our results.

\subsection{Successful freeze-out in the RS scenario}

\begin{figure}[t]
	\centering
	\includegraphics[width=0.5\linewidth]{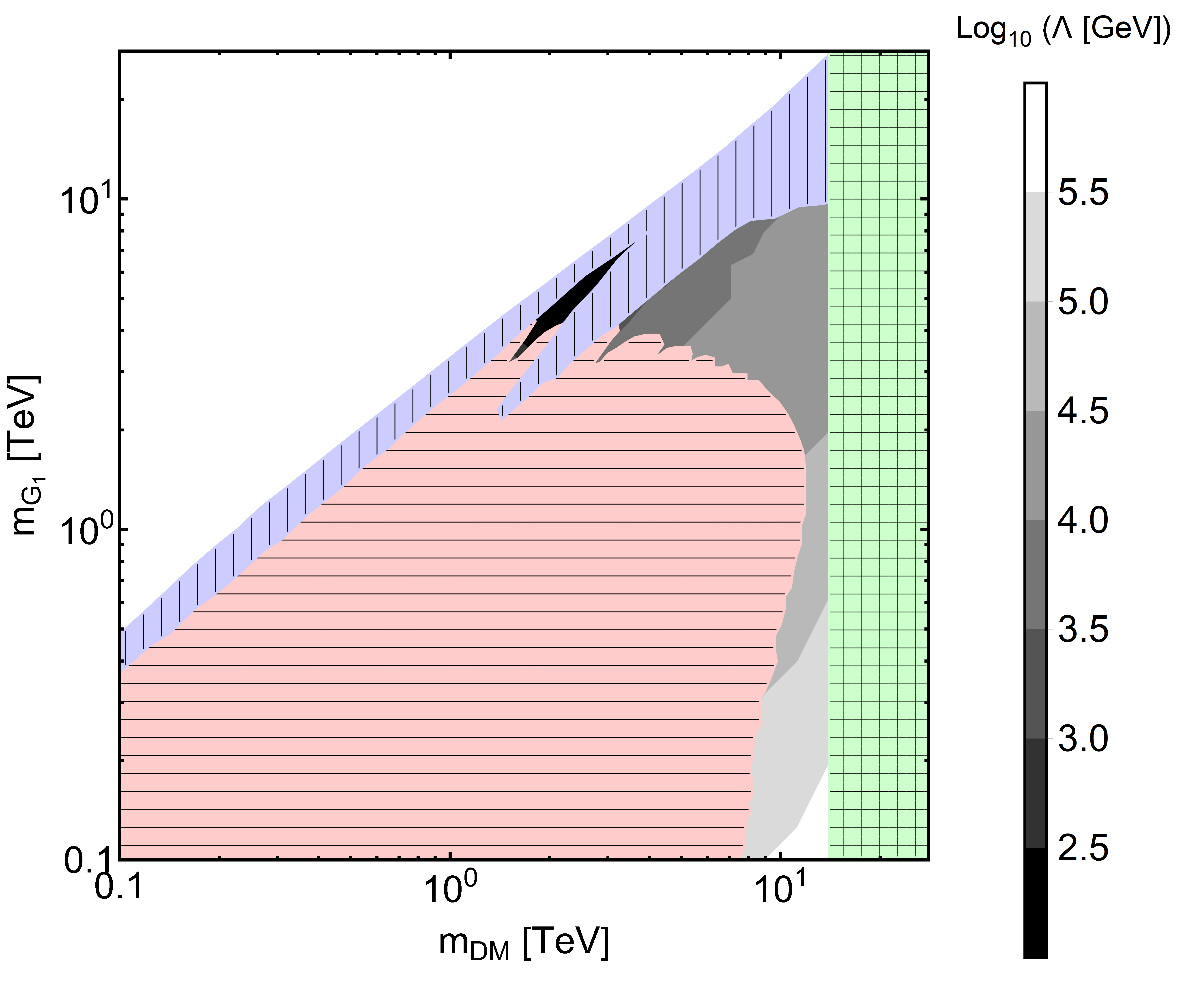}
	\caption[Caption]{\textit{Region of the $\left(m_{\mathrm{DM}},m_{G_1}\right)$ plane for which the DM relic abundance is achieved. The white region corresponds to the part of the parameter space where one can not recover the correct value of the relic abundance. The blue-shaded (vertical-meshed) area represent the region where the 4-dimensional effective theory is untrustable, $\Lambda<m_{G_1}$. The green-shaded (meshed) area represents the region where unitarity conditions are violated. The red-shaded (horizontal-meshed) area represents the excluded region by resonant KK-graviton searches at LHC, with $36 \ \mathrm{fb}^{-1}$ at $\sqrt{s}=13 \ \mathrm{TeV}$. The rest of the colours represent the values of $\Lambda$ required to obtain the correct DM relic abundance.}}
	\label{fig:DMBounds}
\end{figure}

In Fig. \ref{fig:DMBounds} we present the allowed region of the parameter space for which the $\expval{\sigma_{\mathrm{FO}}v}$ can be achieved and the exclusion limits from DD searches, LHC resonance searches and unitarity bounds, as explained in the caption (from Ref. \cite{Folgado2020b}).

\section{Heavy Majorana Neutrino bounds}

In the present section we outline the different bounds that apply to the parameters of the heavy Majorana neutrino sector. First we review the cosmological constraints from BBN and CMB data. Then we focus on the limits imposed by direct searches of heavy neutrinos in accelerator experiments (see Refs. \cite{Drewes2015,Atre2009}).

\subsection{Cosmological constraints}

If sterile neutrinos are present in the primordial plasma during or after BBN, sterile neutrinos can affect the number of effective relativistic degrees of freedom $N_{eff}$ of the plasma. This quantity is constrained at the time of photon decoupling and BBN, from CMB \cite{Collaboration2013} and light elements observation \cite{Steigman2010} respectively. They can affect directly the value of $N_{eff}$ if they come to thermal equilibrium while being relativistic. If they decay before photon decoupling this process leads to a deviation from thermal equilibrium, affecting the SM prediction for $N_{eff}$. If they decay during BBN, their decay products can dissociate nuclei that have already formed. Sterile neutrinos equilibration can be avoided if their mixing is sufficiently small, but if one wants to explain active neutrino masses through seesaw mechanism some of the mixing have to be large enough.

For the three masses $m_{N} \gtrsim 100 \ \mathrm{MeV}$, its contribution to the energy density is expected to be significantly suppressed, either because they decay sufficiently before BBN and/or they become non-relativistic at their decoupling temperature, and therefore get Boltzmann suppressed \cite{Hernandez2014a}.

\subsection{Direct searches for heavy neutrinos}

\begin{figure}[t]
	\centering
	\includegraphics[width=1.0\linewidth]{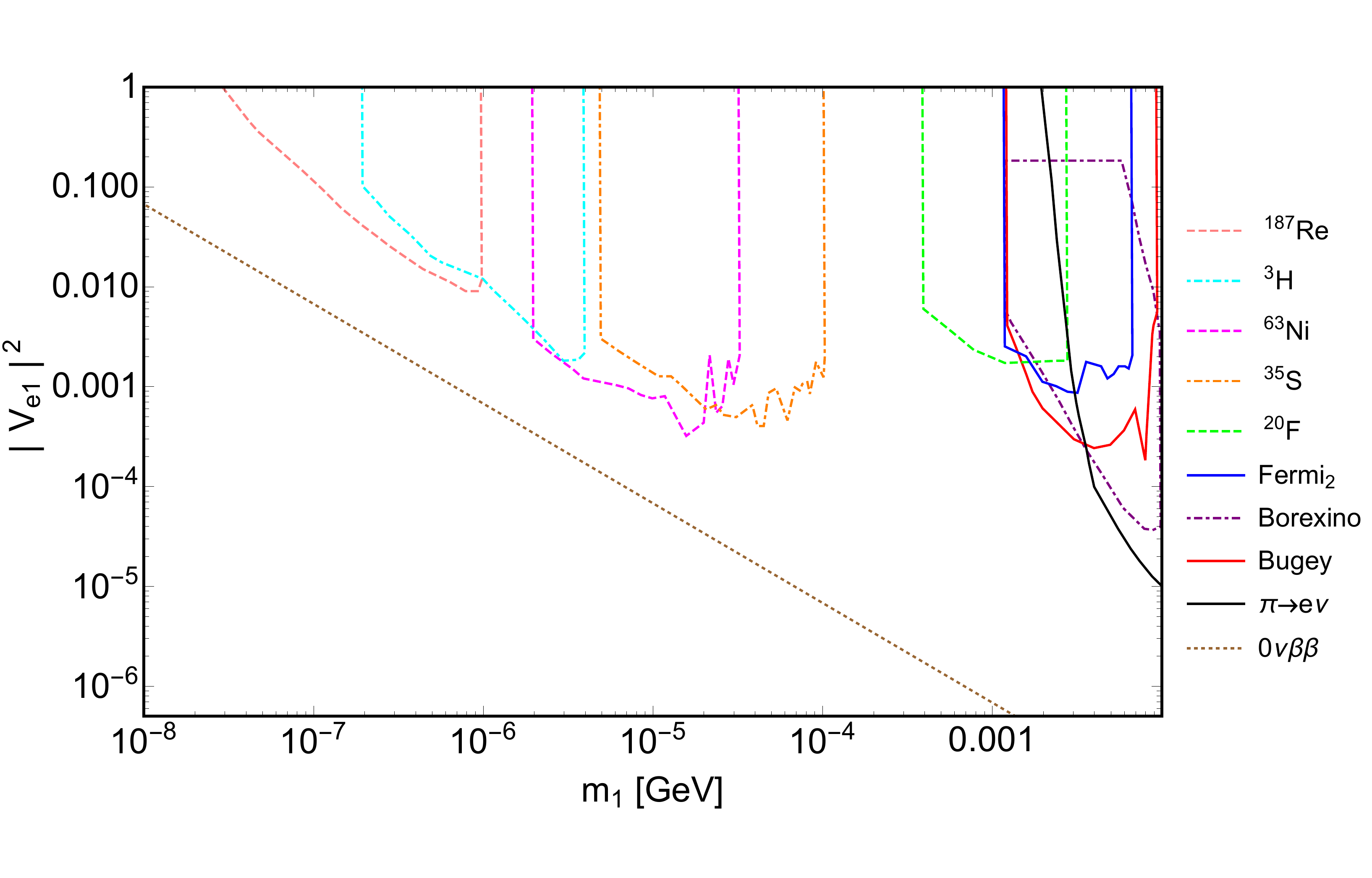}
	\caption[Caption]{\textit{Bounds on the plane $\left(m_1,|V_{e1}|^2\right)$. Kink searches provide the bounds at $95\% \ \mathrm{C.L.}$ from $\hphantom{a}^{187}Re$ (dashed pink line) \cite{Galeazzi2001}, $\hphantom{a}^{3}H$ (dash-dotted cyan line) \cite{Hiddemann1995}, $\hphantom{a}^{63}Ni$ (dashed magenta) \cite{Holzschuh1999}, $\hphantom{a}^{35}S$ (dash-dotted orange line) \cite{Holzschuh2000} decays and at $90\% \ \mathrm{C.L.}$ from $\hphantom{a}^{20}F$ and $\mathrm{Fermi}_2$ \cite{Deutsch1990}. Reactor and solar neutrino data give the limits at $90\% \ \mathrm{C.L.}$ from Borexino \cite{Back2003} and Bugey \cite{Hagner1995} are represented by the dash-dotted purple and solid red lines respectively. Peak searches \cite{Britton1992} exclude the region with solid black contour ($\pi\rightarrow e \nu$). Neutrinoless double beta decay data \cite{Benes2005} gives the limit represented by a dotted brown line.}}
	\label{fig:NuESmallBounds}
\end{figure}

\begin{figure}[t]
	\centering
	\includegraphics[width=1.0\linewidth]{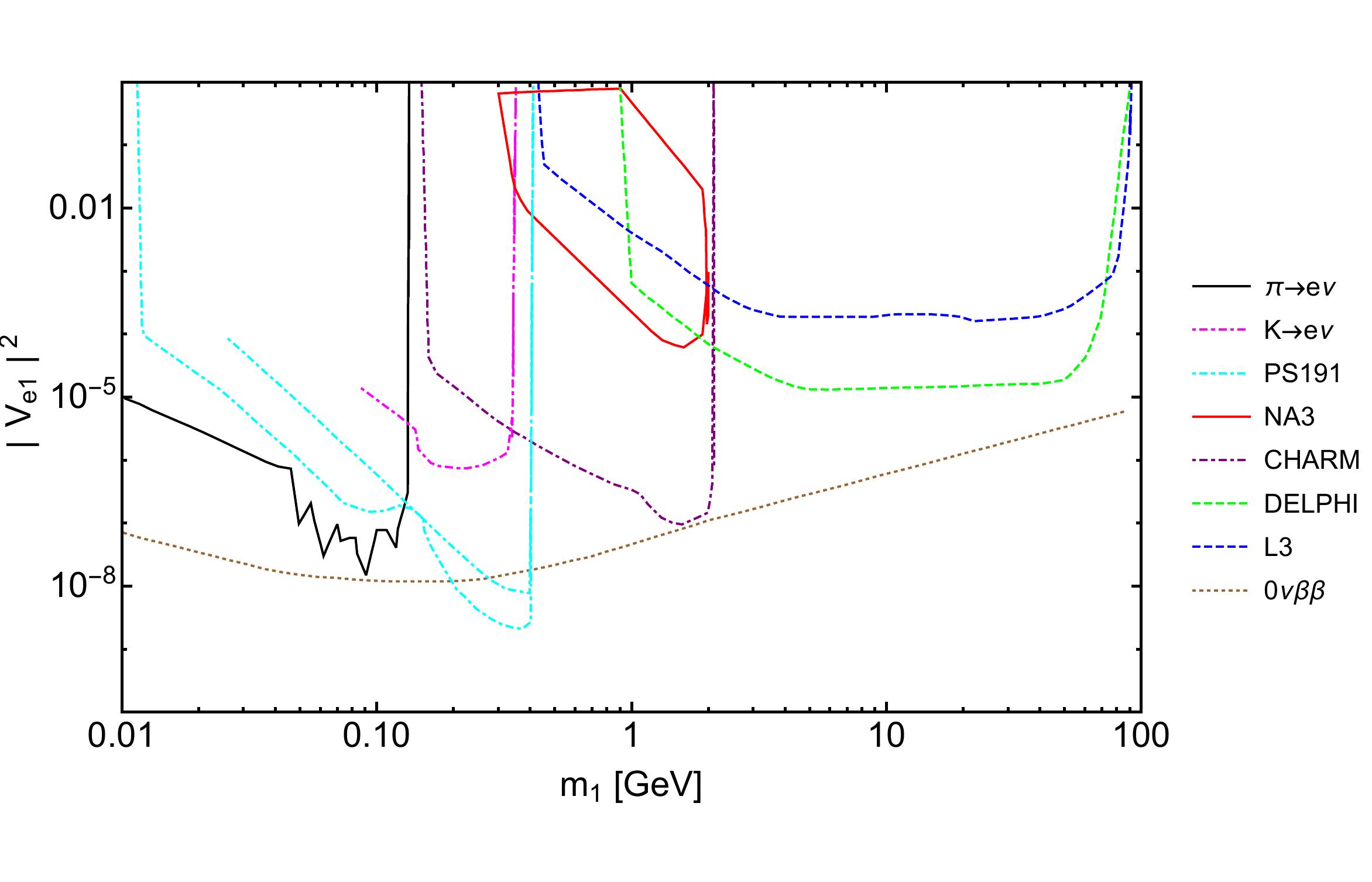}
	\caption[Caption]{\textit{Bounds on the plane $\left(m_1,|V_{e1}|^2\right)$. The peak searches \cite{Britton1992,Berghofer:1981ty,Yamazaki:1984de} exclude the regions with solid black contour ($\pi\rightarrow e \nu$) and dot-dashed magenta contour ($K\rightarrow e \nu$). From beam-dump experiments one obtain the limits at $90\% \ \mathrm{C.L.}$ taken from Ref. \cite{Bernardi1988} (PS191), Ref. \cite{Badier1986} (NA3) and Ref. \cite{Dorenbosch1986} (CHARM). The limits at $95\% \ \mathrm{C.L.}$ represented by the dashed green and blue lines correspond to Ref. \cite{Abreu1997} (DELPHI) and Ref. \cite{Adriani1992} (L3) respectively. The dotted brown line represent the boundary of the excluded region obtained from a reanalysis of neutrinoless double beta decay data \cite{Benes2005}.}}
	\label{fig:NuEBounds}
\end{figure}

Peak searches in leptonic decays of mesons constitute a powerful probe of heavy neutrino mixing with both $\nu_e$ and $\nu_\mu$ \cite{Shrock1980}. When in one of these decays a heavy neutrino is produced, $M^{+}\rightarrow \ell^+ N_1$, the resulting lepton spectrum will show a line:
\begin{equation}
E_{\ell}=\frac{m_{M}^2+m_{\ell}^2-m_{1}^2}{2 m_{M}},
\end{equation}
where $m_{M}$ is the meson mass, $E_{\ell}$ and $m_{\ell}$ the energy and mass of the lepton respectively and $m_1$ the mass of the heavy neutrino. The relative branching fraction of the decay is controlled by the mixing angle between heavy and light neutrinos:
\begin{equation}
\frac{\Gamma\left(M^{+} \rightarrow \ell^{+} N_{1}\right)}{\Gamma\left(M^{+} \rightarrow \ell^{+} \nu_{\ell}\right)} \approx \rho |V_{\ell 1}|^{2},
\end{equation}
where the kinematical factor $\rho$ is written as \cite{Shrock1980}:
\begin{equation}
\rho=\frac{\sqrt{1+\mu_{\ell}^2+\mu_{1}^2-2\left(\mu_{\ell}+\mu_{1}+\mu_{\ell}\mu_{1}\right)}\left[\mu_{\ell}+\mu_{1}-\left(\mu_{\ell}-\mu_{1}\right)^2\right]}{\mu_{\ell}\left(1-\mu_{\ell}\right)^2},
\end{equation}
with $\mu_{i}=m_{i}^2/m_{M}^2$. The helicity suppression of the SM decay $M^{\pm}\rightarrow \ell^{\pm} N_{1}$ gives an enhancement to $\rho$, as $M^{\pm} \rightarrow \ell^{\pm} N_{1}$ can be 4-5 orders of magnitude larger than the former. The bounds from these decays are very strong, as they only rely in the existence of a heavy neutrino which mixes with $\nu_e$ and $\nu_\mu$.

One can also look for decay products of these heavy neutrinos. In every process which emit light neutrinos, $N_1$ will be also produced if kinematically allowed. The branching fraction for the emission of $N_1$ will be proportional to the mixing $|V_{\ell 1}|^{2}$. The heavy neutrino would then decay via Charged and Neutral Current interactions, giving as decay product neutrinos and other SM ``visible'' particles. In performed beam-dump experiments, they searched for this ``visible'' decay products of heavy neutrinos produced in meson decays. Production of heavy neutrinos from $Z$ decays have to be also considered.

Neutrinoless double beta decay also set constraints on the mixing parameter $|V_{e1}|$ for a wide range of heavy neutrino masses. For heavy neutrinos with $m_{n}\gg 1 \ \mathrm{GeV}$, this bound results \cite{Benes2005,Belanger1995}:
\begin{equation}
\sum_{n} \frac{|V_{e1}|^2}{m_n}<5 \times 10^{-5} \ \mathrm{TeV}^{-1}.
\end{equation}

There are also searches for kinks in the $\beta$-spectrum of nuclei \cite{Shrock1980}. This bounds are significant only for right-handed neutrino masses below $\lesssim 10^{-2} \ \mathrm{GeV}$. For some nuclear species, the decays into heavy neutrinos also contribute to their Kurie plot, producing a kink at the endpoint electron energy:
\begin{equation}
E_e=\frac{M_i^2+m_e^2-\left(M_f+m_1\right)^2}{2 M_i},
\end{equation}
where $M_i$ and $M_f$ are the initial and final masses of the nuclei respectively. In Fig. \ref{fig:NuESmallBounds} we show the excluded regions given by different nuclei up to $0.01 \ \mathrm{GeV}$, reported in Refs. \cite{Galeazzi2001,Hiddemann1995,Holzschuh1999,Holzschuh2000,Deutsch1990}. From reactor and solar neutrino experiments we report the limits from Borexino \cite{Back2003} (dash-dotted purple line) and Bugey (solid red line) \cite{Hagner1995}, which looks for decays into $e^{-}e^{+}$ pairs. The solid black line represents the excluded region by peak searches from pion decays \cite{Britton1992}. The dotted brown line represents the limit from neutrinoless double beta decay \cite{Benes2005}.

In Fig. \ref{fig:NuEBounds} we plot the bounds on $|V_{e1}|^2$ in the mass range between $0.01 \ \mathrm{GeV}$ to $100 \ \mathrm{GeV}$. For sufficiently high masses, pion decays give the bound, $90\% \ \mathrm{C.L.}$, represented by the solid black line \cite{Britton1992}. For heavier masses, the bound from $\beta$-spectrum of kaon decay appears, indicated by the dash-dotted magenta line \cite{Berghofer:1981ty,Yamazaki:1984de}. The bounds at $90\% \ \mathrm{C.L.}$ of the beam-dump experiments from Refs. \cite{Bernardi1988,Badier1986,Bergmann1998} (cyan dash-dotted, solid red and purple dash-dotted lines respectively) only assume production of $N_1$ in meson decays and look for its visible decay products. In Refs. \cite{Abreu1997,Adriani1992} from data analysis of DELPHI and L3 detectors one obtain the limits at $95\% \ \mathrm{C.L.}$ represented by the dashed green and blue lines. The dotted brown line represents the limit from neutrinoless double beta decay experiments \cite{Benes2005}, in the case that heavy neutrinos are Majorana fermions.

\begin{figure}[t]
	\centering
	\includegraphics[width=1.0\linewidth]{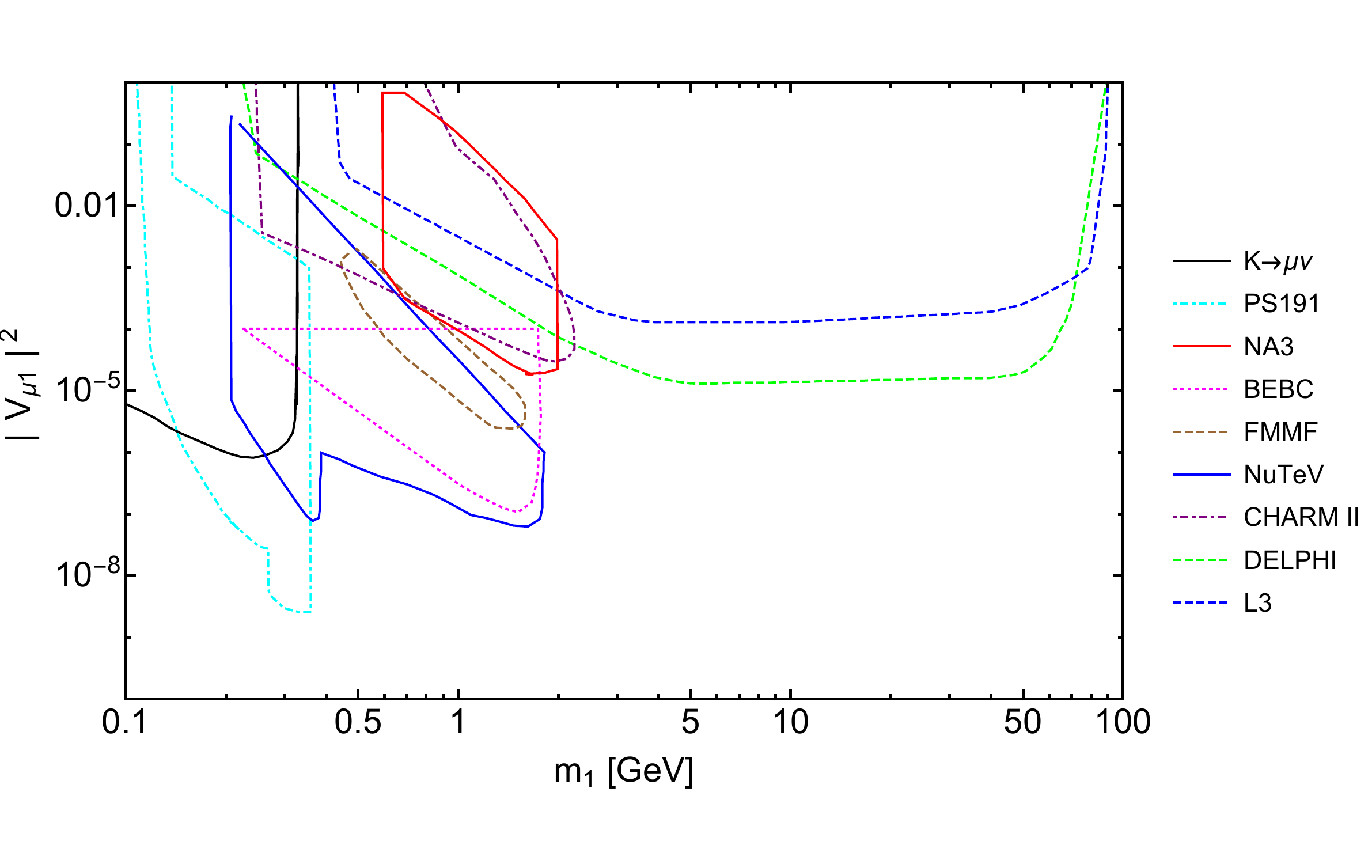}
	\caption[Caption]{\textit{Bounds on the plane $\left(m_1,|V_{\mu 1}|^2\right)$. The peak searches \cite{Kusenko2004} exclude the regions with solid black contour ($K\rightarrow \mu \nu$). The limits at $90\% \ \mathrm{C.L.}$ are taken from Ref. \cite{Bernardi1988} (PS191), Ref. \cite{Badier1986} (NA3), Ref. \cite{CooperSarkar1985} (BEBC), Ref. \cite{Gallas1995} (FMMF), Ref. \cite{Vaitaitis1999} (NuTeV) and Ref. \cite{Vilain1995} (CHARM II). The limits at $95\% \ \mathrm{C.L.}$ represented by the dashed green and blue lines correspond to Ref. \cite{Abreu1997} (DELPHI) and Ref. \cite{Adriani1992} (L3) respectively.}}
	\label{fig:NuMuBounds}
\end{figure}

\begin{figure}[h!]
	\centering
	\includegraphics[width=1.0\linewidth]{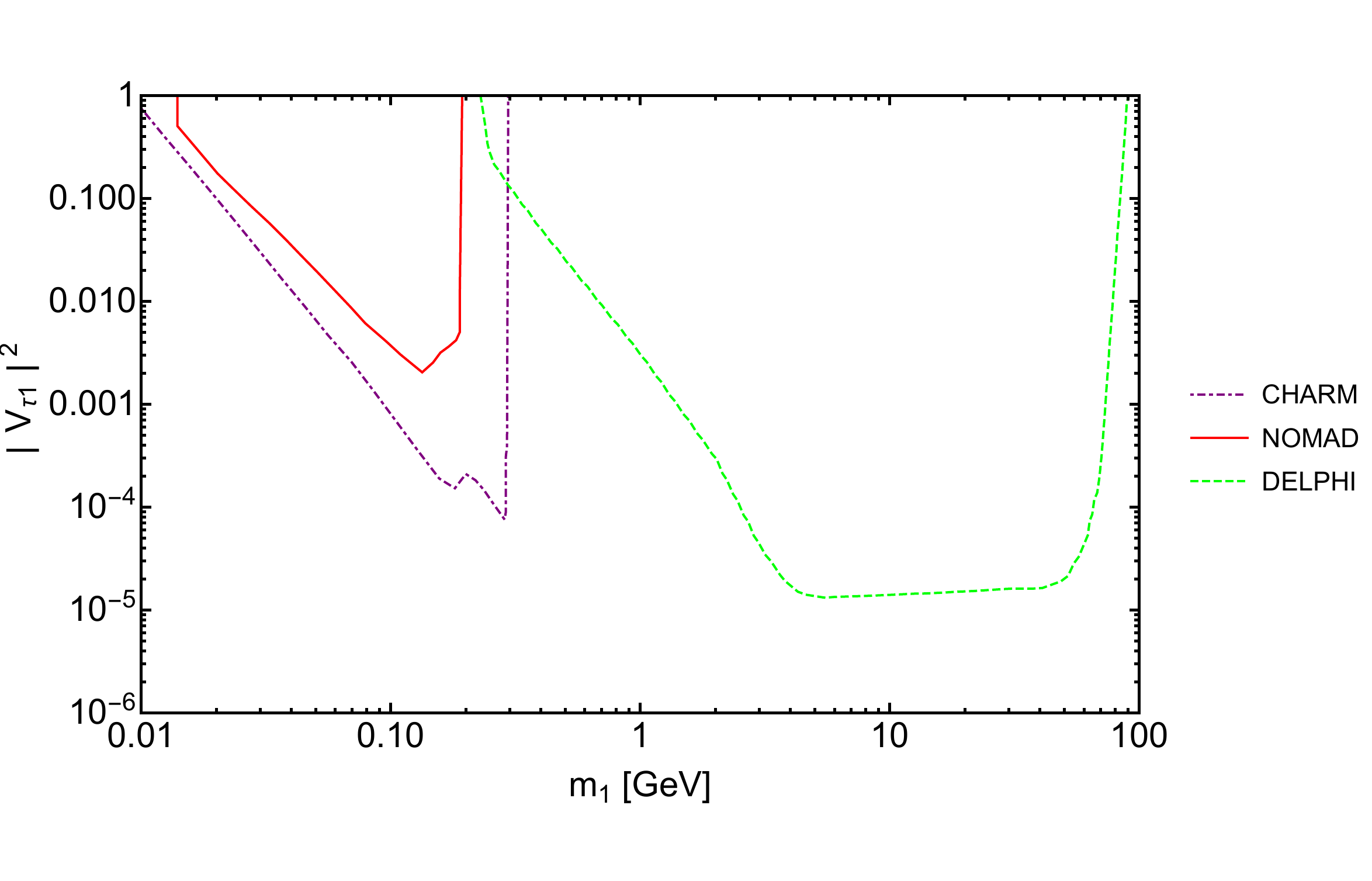}
	\caption[Caption]{\textit{Bounds on the plane $\left(m_1,|V_{\tau 1}|^2\right)$ from heavy neutrino decay searches, taken from Ref. \cite{Orloff2002} (CHARM) and Ref. \cite{Collaboration2001a} (NOMAD) at $90\% \ \mathrm{C.L.}$ and from Ref. \cite{Abreu1997} (DELPHI) at $95\% \ \mathrm{C.L.}$.}}
	\label{fig:NuTauBounds}
\end{figure}

In Fig. \ref{fig:NuMuBounds} the bounds on $|V_{\mu1}|^2$ are reported, for masses larger than $100 \ \mathrm{MeV}$. As in the case of the mixing with $\nu_e$, kaon decays provide very stringent bounds \cite{Kusenko2004}, represented by the solid black contour. Limits from decay searches found in beam-dump experiments \cite{Bernardi1988,Badier1986,CooperSarkar1985,Gallas1995,Vaitaitis1999} and from direct production of heavy neutrinos in DELPHI \cite{Abreu1997} (dashed green line), L3 \cite{Adriani1992} (dashed blue line) and CHARM II \cite{Vilain1995} (dash-dotted purple line) detectors are also represented.

In Fig. \ref{fig:NuTauBounds} we have represented the bounds on $|V_{\tau1}|^2$. The bounds from CHARM \cite{Orloff2002} and NOMAD \cite{Collaboration2001a} detectors at $90\% \ \mathrm{C.L.}$ (dash-dotted purple and solid red lines respectively) assume heavy neutrino production via $D$ and $\tau$ decays. The bound at $95\% \ \mathrm{C.L.}$ from DELPHI \cite{Abreu1997} (dashed green line) assumes $N_1$ production from $Z^0$ decays. With respect to the bounds on $|V_{e1}|^2$ and $|V_{\mu1}|^2$ from DELPHI, there is a kinematical suppression of $\tau$ production which weakens the constraints in the range $m_1 \sim 2-3 \ \mathrm{GeV}$.

As a general remark, we can see that bounds on the mixing between heavy (sterile) neutrinos and light ones $|V_{l1}|^{2}$ can be as large as $10^{-7}$ for $m_{1} \in [0.01,1] \ \mathrm{GeV}$, being less stringent for lighter and heavier states. In the case of the mixing with $\nu_{e}$, the most stringent bounds come from neutrinoless double beta decay, applying only if $N_{1}$ is a Majorana fermion, though.

\chapter{TeV-scale right-handed bulk neutrino in RS}\label{Chapter 6}

The hierarchy between the right-handed neutrinos generated in the 3+1+2 seesaw model, implemented by warped compactification, make us wonder if some configurations of the 5-dimensional fermion mass parameters can simultaneously address the smallness of neutrino masses and somehow give us an acceptable DM candidate.

In order to study this possibility, our purpose is to systematically explore the parameter space, restricting it to have a WIMP-like DM candidate, taking advantage of the results of Refs. \cite{Folgado2020a,Folgado2020b} (outlined in Chap. \ref{Chapter 4}). The model has many free parameters and, therefore, the scan of the parameter space is a difficult task.

\section{Fermionic DM as Majorana Fermions}\label{Section 6.1}

Since our DM candidate would be one of the right-handed neutrinos, the Feynman rules of the model should be properly modified, in order to take into account the Majorana nature of this particle. Majorana fields give rise to different Wick contractions from those of Dirac fields, as they are self-conjugate.

In a field theory containing Majorana fermions, fermion number flow is violated, and therefore it is not useful to indicate the orientation of fermion lines. A solution proposed in \cite{Denner1992} consist on taking arbitrary orientations for each fermion chain, what is called a continuous fermion flow. This way, one encounters new diagrams contributing to the amplitude of the processes. In the case of DM annihilation cross-sections, the results obtained following this method are given in App. \ref{Appendix A}. These results expand those presented in Refs. \cite{Folgado2020,Folgado2020b} for Dirac fermions in extra-dimensional scenarios.

\section{Pushing DM into the bulk}

\begin{figure}[t]
	\centering
	\includegraphics[width=0.9\linewidth]{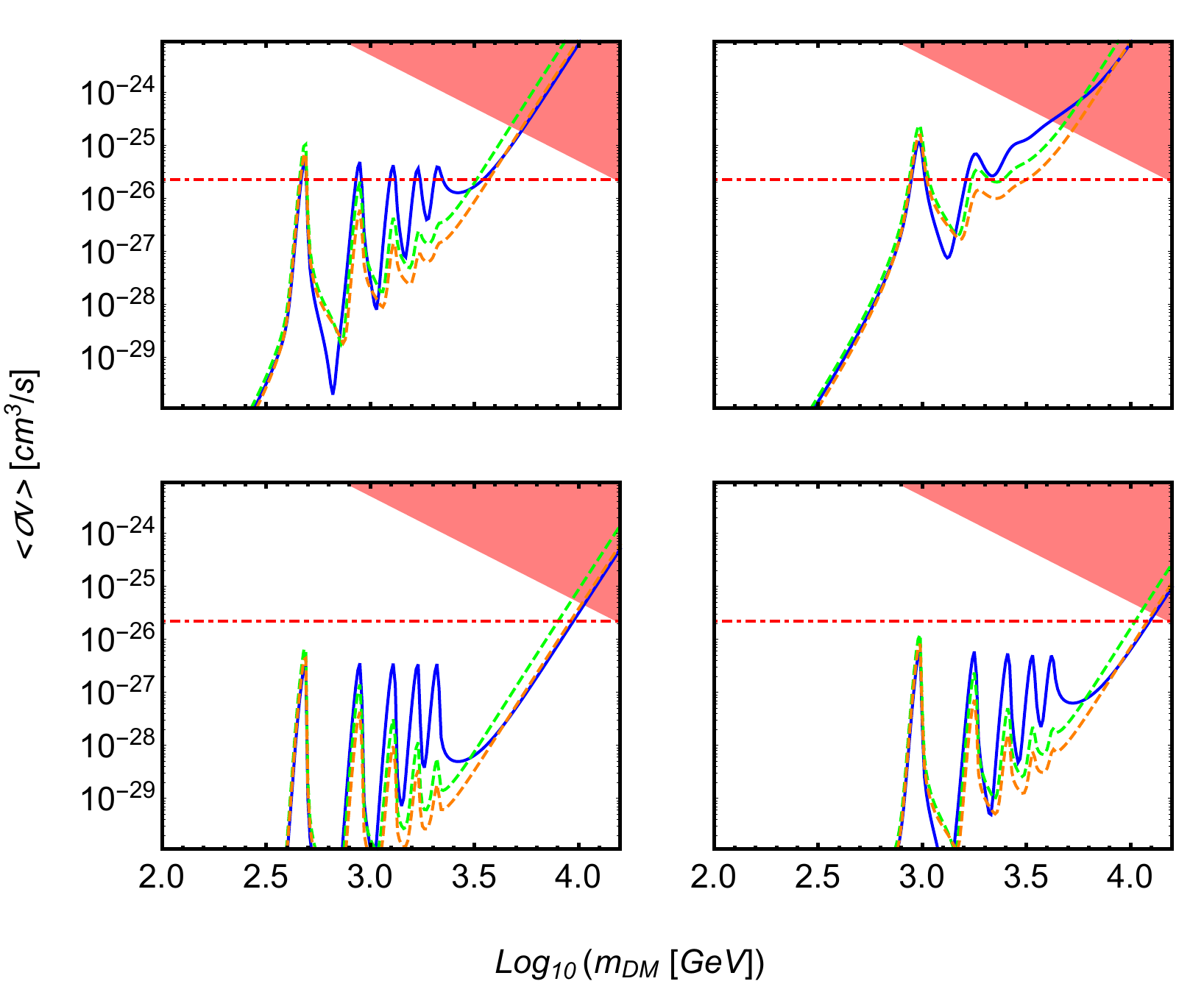}
	\caption{\textit{Thermally-averaged fermionic DM annihilation cross-section via virtual graviton KK-modes and production of two KK-gravitons, as a function of the DM mass. In all panels, the solid blue line correspond to the case where the DM particle is Dirac and it is localized in the IR-brane. On the other hand, the dashed green and orange lines represent Majorana DM particles placed in the bulk, with bulk parameters $\nu=1$ and $\nu=1/2$ respectively. Top left panel: $\Lambda=10 \ \mathrm{TeV}$, $m_{G_1}=1 \ \mathrm{TeV}$; top right panel: $\Lambda=5 \ \mathrm{TeV}$, $m_{G_1}=2 \ \mathrm{TeV}$; bottom left panel: $\Lambda=40 \ \mathrm{TeV}$, $m_{G_1}=1 \ \mathrm{TeV}$; bottom right panel: $\Lambda=30 \ \mathrm{TeV}$, $m_{G_1}=2 \ \mathrm{TeV}$.}}
	\label{fig:bulkcross}
\end{figure}

In Sec. \ref{Section3.2} we performed the KK-reduction of the action of a bulk fermion field. Placing the field content of a theory in the bulk carries some phenomenological implications: in particular, the couplings between graviton KK-modes and the fields we have placed in the bulk change in the 4-dimensional effective field theory.

In general, the couplings to the n-th KK-graviton mode are of the form \cite{Fitzpatrick2007}:
\begin{equation}\label{6.1}
C_{n}^{XX}\int \diff x^4 h^{(n)}_{\mu\nu} T^{\mu\nu}_{XX},
\end{equation}
being $T^{\mu\nu}_{XX}$ the 4-dimensional effective energy-momentum tensor of the $X$ field. The coefficients $C_{n}$ arise from the dimensional reduction performed from the full 5-dimensional theory to the 4-dimensional effective one, and represent the overlap of wavefunctions of the field species in the bulk. In the case of a bulk fermion, for its zero-mode we can write the coefficients as \cite{Davoudiasl2001}:
\begin{equation}\label{6.2}
C_{00n}^{\chi\chi}=\mathrm{e}^{\pi k r_c} \frac{1+2\nu}{1-\mathrm{e}^{-(1+2\nu)\pi k r_c}} \int_{\mathrm{e}^{-\pi k r_c}}^{1} \diff y \ y^{2+2\nu} \frac{J_2(x_{n}^{G}y)}{|J_2(x_{n}^{G})|},
\end{equation}
and for one zero-mode and a generic $l$-mode we have:
\begin{equation}\label{6.3}
C_{0ln}^{\chi\chi}=\mathrm{e}^{\pi k r_c} \left[\frac{2(1+2\nu)}{1-\mathrm{e}^{-(1+2\nu)\pi k r_c}}\right]^{1/2} \int_{\mathrm{e}^{-\pi k r_c}}^{1} \diff y \ y^{5/2+\nu} \frac{J_2(x_{n}^{G}y)}{|J_2(x_{n}^{G})|} \frac{J_f(x_{l}^{L}y)}{J_f(x_{l}^{L})},
\end{equation}
where $x_{l}^{L}$ is the $l$-th zero of Eq. (\ref{3.12a}) and the function $f$ is defined as:
\begin{equation}\label{6.4}
f=\left\{\begin{array}{ccc}\nu-1/2&;&\nu>-1/2,\\-\nu+1/2&;&\nu<-1/2.\end{array}\right.
\end{equation}

When we push the DM out of the IR-brane, into the bulk, we have to take into account the effects of these coefficients. Thus, we have to modify the expressions for the decay width of gravitons into DM particles and the ones corresponding to DM annihilation cross section. In App. \ref{Appendix B} we give the modified expressions after considering this effect and the one mentioned in Sec. \ref{Section 6.1}.

\begin{figure}[t]
	\centering
	\includegraphics[width=1\linewidth]{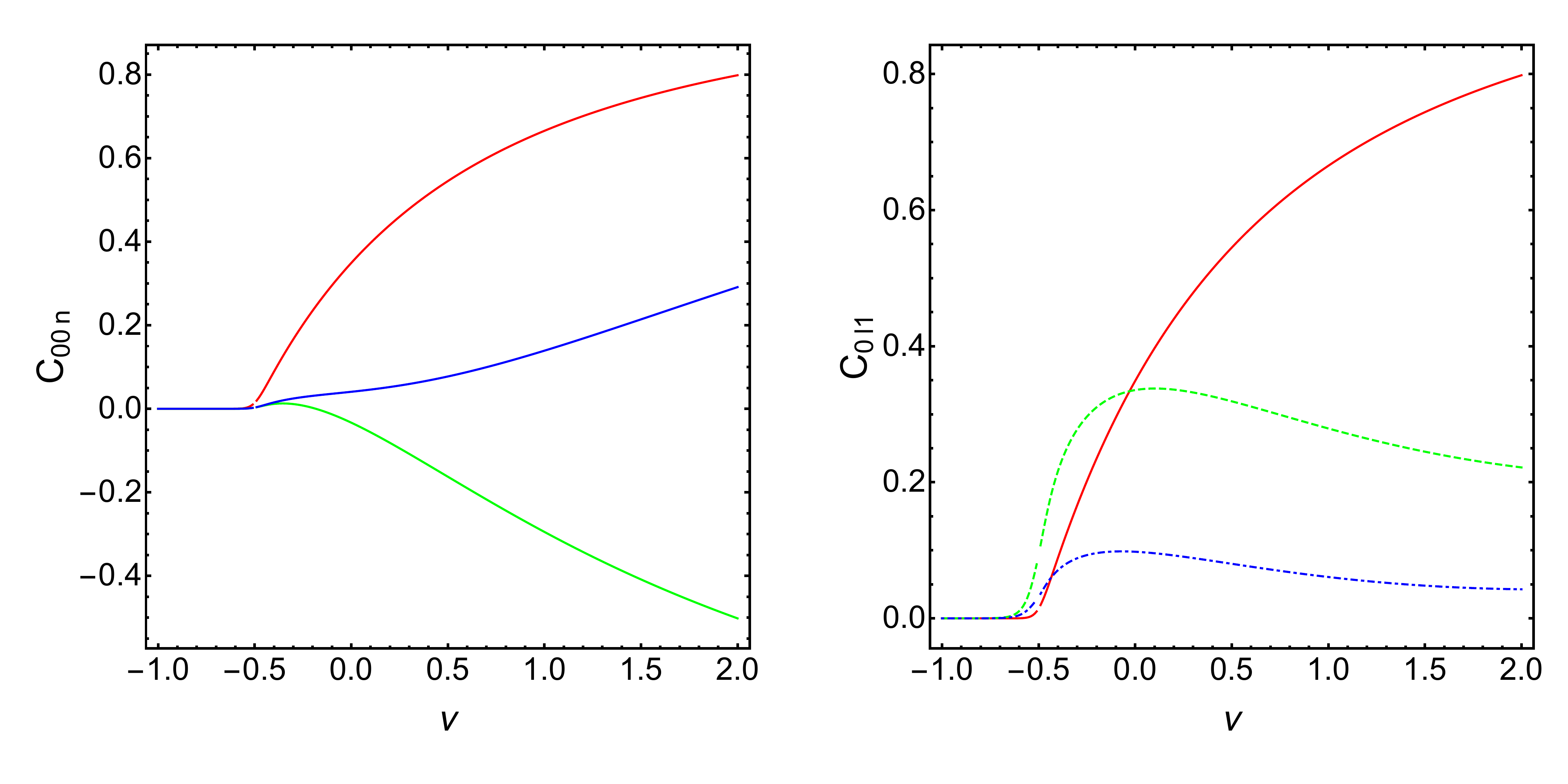}
	\caption{\textit{Left panel: Values of the bulk coefficients $C_{00n}$ as a function of $\nu$, for a fixed value $\Lambda=10 \ \mathrm{TeV}$. The red, green and blue solid lines correspond to the choices $n=1$, $n=2$ and $n=3$ respectively. Right panel: Values of the bulk coefficients $C_{0l1}$ as functions of $\nu$, for the choices $l=0, \ 1, \ 2$ (solid red, dashed green and dash-dotted blue lines respectively) and $\Lambda=10 \ \mathrm{TeV}$.}}
	\label{fig:bulkpar}
\end{figure}

The impact of these modifications is shown in Fig. \ref{fig:bulkcross}, where we compare the thermal-averaged total DM cross-section in the case of Dirac DM particles in the brane and Majorana DM particles in the bulk, for various choices of $m_{G_1}$ and $\Lambda$. The solid blue line correspond to the known case where the DM particle is a Dirac fermion and it is localized in the IR-brane, whereas the green and orange dashed lines correspond to DM particles being Majorana fermions in the bulk, with dimensionless bulk parameters $\nu=1$ and $\nu=1/2$ respectively. In all panels, the horizontal red dash-dotted line correspond to the value of the thermal-averaged cross-section for which the correct DM relic abundance is achieved $\expval{\sigma_{\mathrm{FO}} v}=2.2\times 10^{-26} \ \mathrm{cm^3/s}$. It can be seen that in all three cases the first graviton resonance remains almost the same, slightly enhanced in the Majorana cases. On the contrary, the following resonance peaks are severely suppressed for the bulk fermions. This suppression makes quite difficult achieving the relic abundance through KK-graviton exchange, in the Majorana bulk fermion case. In some configurations, the DM relic abundance can be achieved by the resonant virtual first KK-graviton exchange channel, but a significant amount of fine-tuning in the DM mass is needed. For sufficiently large values of the DM particle mass where the two on-shell KK-graviton production channel is allowed, it quickly dominates. Through this channel the relic abundance is achievable, without entering the red-shaded untrustable region $\expval{\sigma v} \geq 1/s$. In most configurations $\left(\Lambda,m_{G_1}\right)$, the Majorana case is slightly favoured with respect to Dirac, especially for larger values of $\nu$.

The variations in the couplings between DM and different KK-graviton modes can be understood directly from the dependence on $n$ of Eq. (\ref{6.2}). In Fig. \ref{fig:bulkpar} (left panel) we show the values of the $C_{00n}$ coefficients for increasing $n$, as a function of the bulk parameter $\nu$, for a fixed value $\Lambda=10 \ \mathrm{TeV}$. The one representing $n=1$ (solid red line) dominates among the rest, $n=2$ and $n=3$ (solid green and blue lines respectively). It can be noticed that, in spite of not being constant, the difference between the $n=2$ and $n=3$ lines is smaller than the one between $n=1$ and $n=2$. When $\nu \rightarrow \infty$, all the coefficients go to $1$ (in modulus), meaning that the fermion field is located at the IR-brane. This mathematical sense is somehow incompatible with the physical meaning of $\nu$, which is not expected to be much larger than $1$. Nevertheless, one can check that for $\nu\gtrsim 1$ the squared-modulus of the fermion wavefunction generically shows a peak near $\phi=\pi$, meaning that in this range the fermion field is essentially localized at the IR-brane.

In Fig. \ref{fig:bulkpar} (right panel) we represent the $C_{0ln}$ coefficients for the coupling between the KK-graviton zero-mode, one fermion zero-mode and a fermion $l$-mode, for $\Lambda=10 \ \mathrm{TeV}$. From this, we can see the dependence of the coupling on the specific fermionic mode. The solid red line represents the coefficient $C_{001}$, whereas the dashed green and the dash-dotted blue lines represent $C_{011}$ and $C_{021}$, respectively.

\begin{figure}[t]
	\centering
	\includegraphics[width=0.7\linewidth]{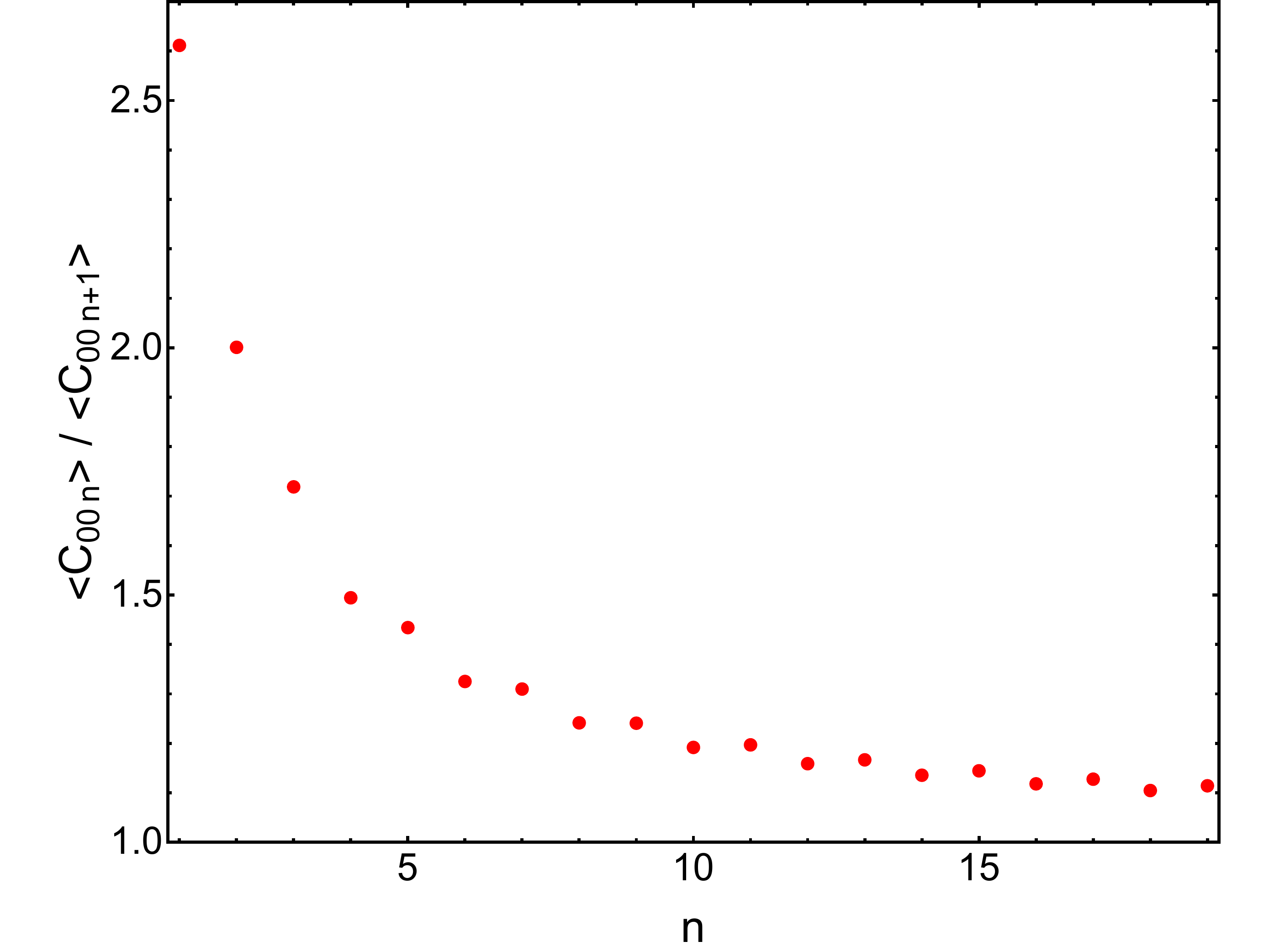}
	\caption{\textit{Ratio of the averages of two consecutive values of the bulk coefficients $C_{00n}$, for a fixed $\Lambda=10 \ \mathrm{TeV}$.}}
	\label{fig:bulkpar2}
\end{figure}

In Fig. \ref{fig:bulkpar2} we represent the ratio between the integral averages of the coefficients of two consecutive values of $n$, with $\Lambda=10 \ \mathrm{TeV}$. This average must not get mixed up with the previously used thermal average. We compute it as:
\begin{equation}\label{6.5}
\expval{C_{00n}}\equiv\frac{\int \diff \nu \ C_{00n}}{\int \diff \nu}.
\end{equation}
We can see that this ratio tends to stabilize for large values of $n$.

\section{Exploring the neutrino parameter space}

The parameter space of this particular model is quite large. Assuming that the two Yukawa coupling matrices, $M$ and $m_{D}$, are real, from these we have $18$ parameters. Then, we will have $3$ from the values of $\left\{\nu_1,\nu_2,\nu_3\right\}$ and $2$ more from the specific values of $\Lambda$ and $\tilde{v}_{\mathrm{BL}}$. We end up with $23$ free parameters.

Our goal is to generate light masses for the left-handed neutrino sector, have a right-handed neutrino that can play the role of a sterile neutrino, another one in the appropriate mass range to be a thermal DM candidate and eventually a third one sufficiently heavy to decouple. To achieve this we must ensure $m_{\nu}^{4\times4}\lesssim1 \ \mathrm{eV}$. In the light of Eq. (\ref{3.3}), this implies (assuming no ad-hoc cancellations due to some underlying texture of the mass matrices):
\begin{equation}\label{6.6}
\left\{\begin{array}{ccc}M_{\nu}^{4\times4}&\lesssim &1 \ \mathrm{eV},\\m_{D}^{4\times2}\left(M^{2\times2}\right)^{-1}\left(m_{D}^{4\times2}\right)^{T}&\lesssim &1 \ \mathrm{eV}.\\\end{array}\right.
\end{equation}

In order to successfully fulfill the first condition, we can set $M_{11}\simeq m_{\alpha1}$ (for any $\alpha$). Following Eq. (\ref{3.27}) this implies:
\begin{equation}\label{6.7}
\omega_1\simeq\frac{\bar{M}_{P}}{\Lambda} \frac{v_{\mathrm{EW}}}{\tilde{v}_{\mathrm{BL}}},
\end{equation}
assuming the corresponding Yukawa couplings to be of the same order of magnitude. Since $\tilde{v}_{\mathrm{BL}}$ is approximately fixed to be Planck scale order, this quantity would depend noticeably on the hierarchy between $\Lambda$ and $v_{\mathrm{EW}}$. Therefore the order of the Yukawa couplings should be fixed according to the value of $\Lambda$, to maintain these entries sub-eV. Using again Eq. (\ref{3.27}) together with the condition of Eq. (\ref{6.7}), one gets the following bound:
\begin{equation}\label{6.8}
\frac{v_{\mathrm{EW}}^2}{\tilde{v}_{\mathrm{BL}}}\frac{\bar{M}_{Pl}}{\Lambda} \ \lambda_{11}\lesssim 1 \ \mathrm{eV}.
\end{equation}

The second condition in Eq. (\ref{6.6}) is directly affected by Eq. (\ref{6.7}). It automatically sets $M_{1j}\simeq m_{\alpha j}$; $j=1, \ 2$, in the case that its Yukawa couplings are of the same order. If one takes $\omega_j$ ($j=1, \ 2$) to be $\mathcal{O}(1)$, the values of these entries will depend on the order of the Yukawa couplings. Assuming that the Yukawa couplings of the heavy sector are approximately of the same order as the ones of the coupling sector, the condition reads:
\begin{equation}\label{6.9}
\frac{v_{\mathrm{EW}}^2}{\tilde{v}_{\mathrm{BL}}}\frac{\bar{M}_{Pl}}{\Lambda} \ \lambda_{ij}\lesssim 1 \ \mathrm{eV}.
\end{equation}

For the sake of simplicity, we can take all the mentioned Yukawa couplings to be of the same order. Thus, the conditions expressed in Eqs. (\ref{6.8}) and (\ref{6.9}) merge into a unique condition involving just one coupling $\lambda$. The value of this parameter can be fixed simply as:
\begin{equation}\label{6.10}
\lambda\lesssim\frac{\tilde{v}_{\mathrm{BL}}\Lambda}{v_{\mathrm{EW}}^2\bar{M}_{Pl}} \ 1 \ \mathrm{eV}.
\end{equation}

If we set the parameters as discussed until now, the diagonalization of the heavy sector matrix $M^{2\times2}$ may rise some troubles. The symmetric layout of this matrix can produce a vanishing eigenvalue. This can be avoided if we introduce some fluctuations in their Yukawa couplings, one safe choice is, for example, $\lambda_{22} \sim \lambda_{33} = 1 + \delta_{\lambda}$ with $\delta_{\lambda} \ll 1$.

For the time being, we are going to assume the equality in Eq. (\ref{6.10}), as it represent the minimal choice for the value of the parameter $\lambda$. Moreover, we set $\lambda_{22}, \ \lambda_{33} \sim \mathcal{O}(1)$ to avoid problems involving the heavy sector. From this point, we can explore different choices of the neutrino mass parameters $\left\{\nu_2,\nu_3\right\}$, as $\nu_1$ is fixed in virtue of Eq. (\ref{6.7}), noticing its dependence on both $\Lambda$ and $\tilde{v}_{\mathrm{BL}}$.

\begin{figure}[t]
	\centering
	\includegraphics[width=1\linewidth]{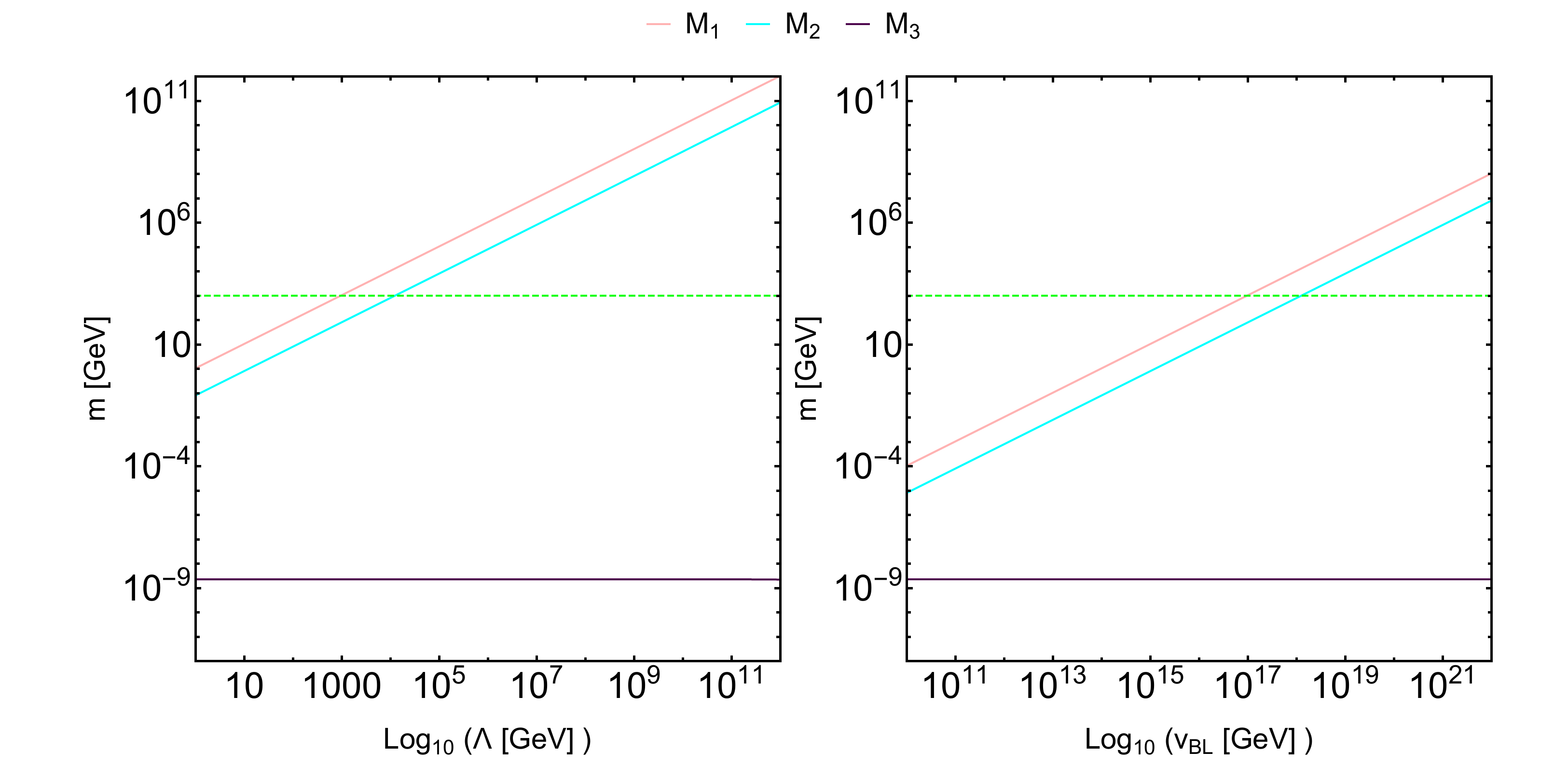}
	\caption{\textit{Masses of the three right-handed neutrinos (light red, light blue and purple lines respectively) for the choice $\left\{\nu_{2}=-0.4,\nu_{3}=0.8\right\}$ as functions of $\Lambda$ (Left panel) and $\tilde{v}_{\mathrm{BL}}$ (Right panel). The green dashed line represents the reference value $1 \ \mathrm{TeV}$.}}
	\label{fig:ExampleNu}
\end{figure}

\begin{figure}[h!]
	\centering
	\includegraphics[width=0.9\linewidth]{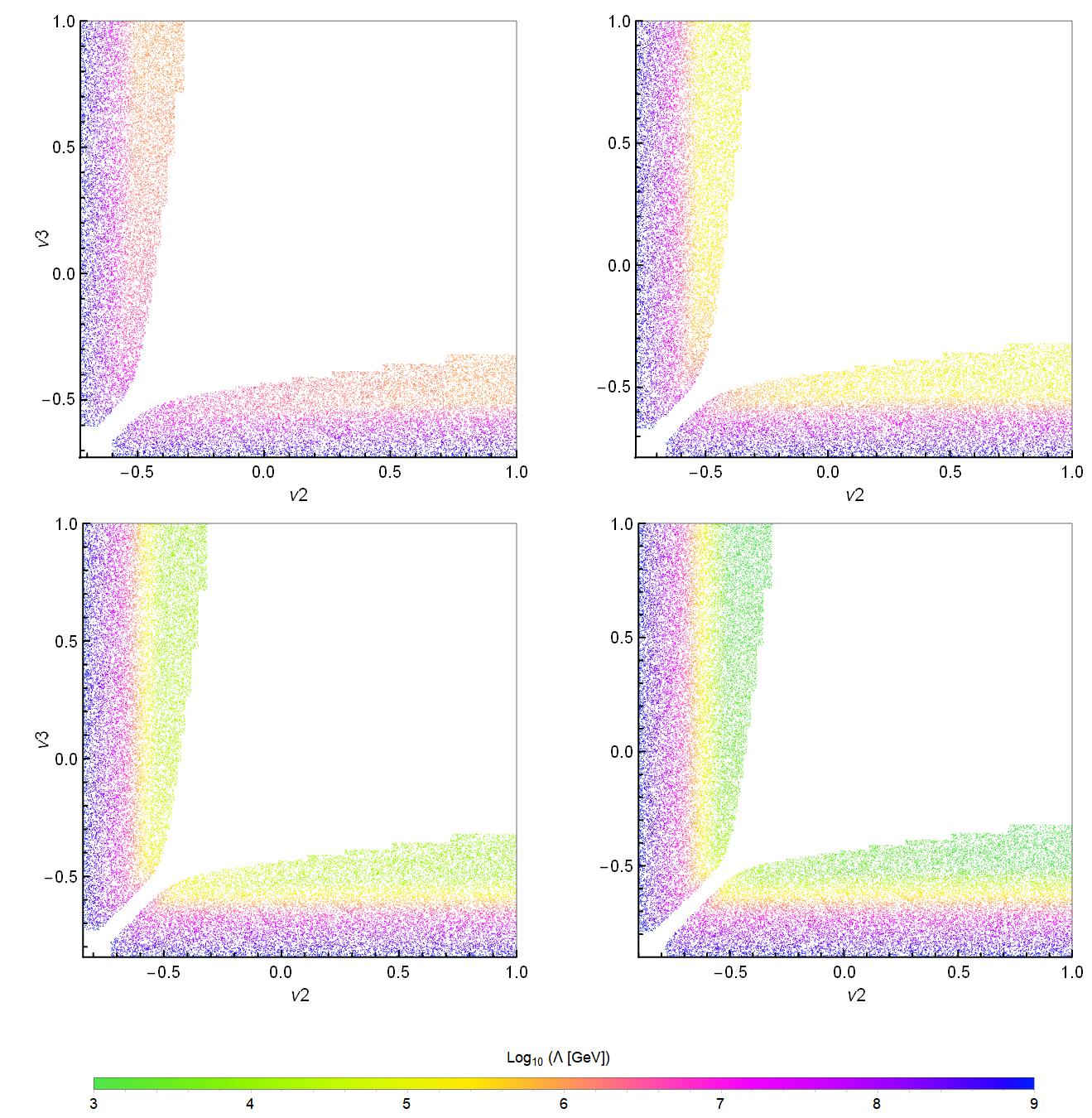}
	\caption{\textit{Values of $\Lambda$ for which the second right-handed neutrino can be a DM candidate ($M_2\sim1$ TeV),the first one is decoupled ($M_1\gtrsim1000$ TeV) and the third is relatively light ($M_3 \lesssim 1$ GeV) in the $\nu_2$, $\nu_3$ plane, with $\nu_1$ fixed in virtue of Eq. (\ref{6.7}). Top Left panel: $\tilde{v}_{\mathrm{BL}}=10^{16}$ GeV; Top Right panel: $\tilde{v}_{\mathrm{BL}}=10^{17}$ GeV; Bottom Left panel: $\tilde{v}_{\mathrm{BL}}=10^{18}$ GeV; Bottom Right panel: $\tilde{v}_{\mathrm{BL}}=10^{19}$ GeV. The required $\Lambda$ ranges from $1$ TeV to $10^{6}$ TeV, as shown by the color legend.}}
	\label{fig:plotnu1nu23}
\end{figure}

In Fig. \ref{fig:ExampleNu} (left panel) we plot the masses of the three right-handed neutrinos as functions of $\Lambda$, for the fixed values $\left\{\nu_{2}=-0.4,\nu_{3}=0.8\right\}$ and $\tilde{v}_{\mathrm{BL}}=10^{19} \ \mathrm{GeV}$. One can see that the mass of the lightest right-handed neutrino, $M_3$ (solid purple line), is fixed in the considered interval, as a consequence of our choice of parameters. The masses of the heavy species, $M_1$ and $M_2$ (light red and blue solid lines respectively), grow at an almost constant rate. For a value $\Lambda\approx10 \ \mathrm{TeV}$ the masses of both heavy neutrinos are above the reference value of $1 \ \mathrm{TeV}$ (dashed green line). If we want the second right-handed neutrino to be the DM, it must have the correct mass to achieve the relic abundance, which lies in the range $\left[1,15\right] \ \mathrm{TeV}$ in the present case. This shape may give one configuration of parameters that reproduce the appropriate $m_{\mathrm{DM}}$.

In Fig. \ref{fig:ExampleNu} (right panel) the masses of the right-handed neutrinos are represented as functions of $\tilde{v}_{\mathrm{BL}}$, for the same fixed values of the mass parameters  $\left\{\nu_{2}=-0.4,\nu_{3}=0.8\right\}$ and $\Lambda=10^{4} \ \mathrm{GeV}$. We can appreciate the same pattern as before, reaching the reference line at $\tilde{v}_{\mathrm{BL}}\approx10^{18} \ \mathrm{GeV}$. This behaviour is expected from the structure of the mass matrix. In the considered regime $\Lambda\ll M_{P}$, and $\omega_i$ from Eq. (\ref{3.15}) reduces to:
\begin{equation}
\omega_i\approx\sqrt{1+2\nu_i},
\end{equation}
for $\nu_i>1/2$. Therefore, the leading contribution of $\Lambda$ comes from the rescaling of the 5-dimensional $\tilde{v}_{\mathrm{BL}}$:
\begin{equation}
v_{\mathrm{BL}}=\mathrm{e}^{-\pi k r_c} \tilde{v}_{\mathrm{BL}} = \frac{\Lambda}{\bar{M}_{Pl}} \tilde{v}_{\mathrm{BL}}.
\end{equation}

As a consequence, the distribution of the masses as functions of $\Lambda$ for a fixed value of $\tilde{v}_{\mathrm{BL}}$ results quite similar to the one where $\tilde{v}_{\mathrm{BL}}$ is the independent variable and $\Lambda$ is a fixed parameter.

Now that we have taken a look to the model for a specific choice of $\nu_2$, $\nu_3$, a more generic search of the parameter space is needed in order to understand the possibilities that we can exploit. Since we are interested in a particular hierarchy between the right-handed neutrinos, that is one has to be sufficiently heavy ($M_1\geq 10^{3} \ \mathrm{TeV}$), another one must be in the interesting range for DM ($1 \ \mathrm{TeV}\lesssim M_2 \lesssim 15 \ \mathrm{TeV}$) and the last should be light ($M_3 \sim 1 \ \mathrm{eV}$), we can focus our search to obtain results within this specific choice.

In Fig. \ref{fig:plotnu1nu23} we show the values of $\Lambda$ for which the masses of the right-handed neutrinos lie within the specified bounds, in the $\left(\nu_2,\nu_3\right)$ plane for several values of $\tilde{v}_{\mathrm{BL}}$. We notice the presence of an excluded region in the upper-right margin of every panel, where our requirements are not fulfilled. The basic structure of all four plots are quite similar. All of them present a symmetric layout, with the interior boundaries where the required values of $\Lambda$ are smaller. The minimum value of $\Lambda$ appearing in each plot decreases as one increase the value of $\tilde{v}_{\mathrm{BL}}$.

As our main motivation is to achieve the DM relic abundance with our neutrino candidate, we must focus on the $\Lambda$ range not ruled out by theoretical and experimental bounds represented in Fig. \ref{fig:DMBounds}. Therefore, as we are interested in $\Lambda \in \left[10^{4},10^{9}\right] \ \mathrm{GeV}$, we will just consider  $\tilde{v}_{\mathrm{BL}}\sim10^{17}-10^{19} \ \mathrm{GeV}$ (since it can not be larger than the Planck scale).

\begin{figure}[t]
	\centering
	\begin{tabular}{cc}
			\includegraphics[width=0.5\linewidth]{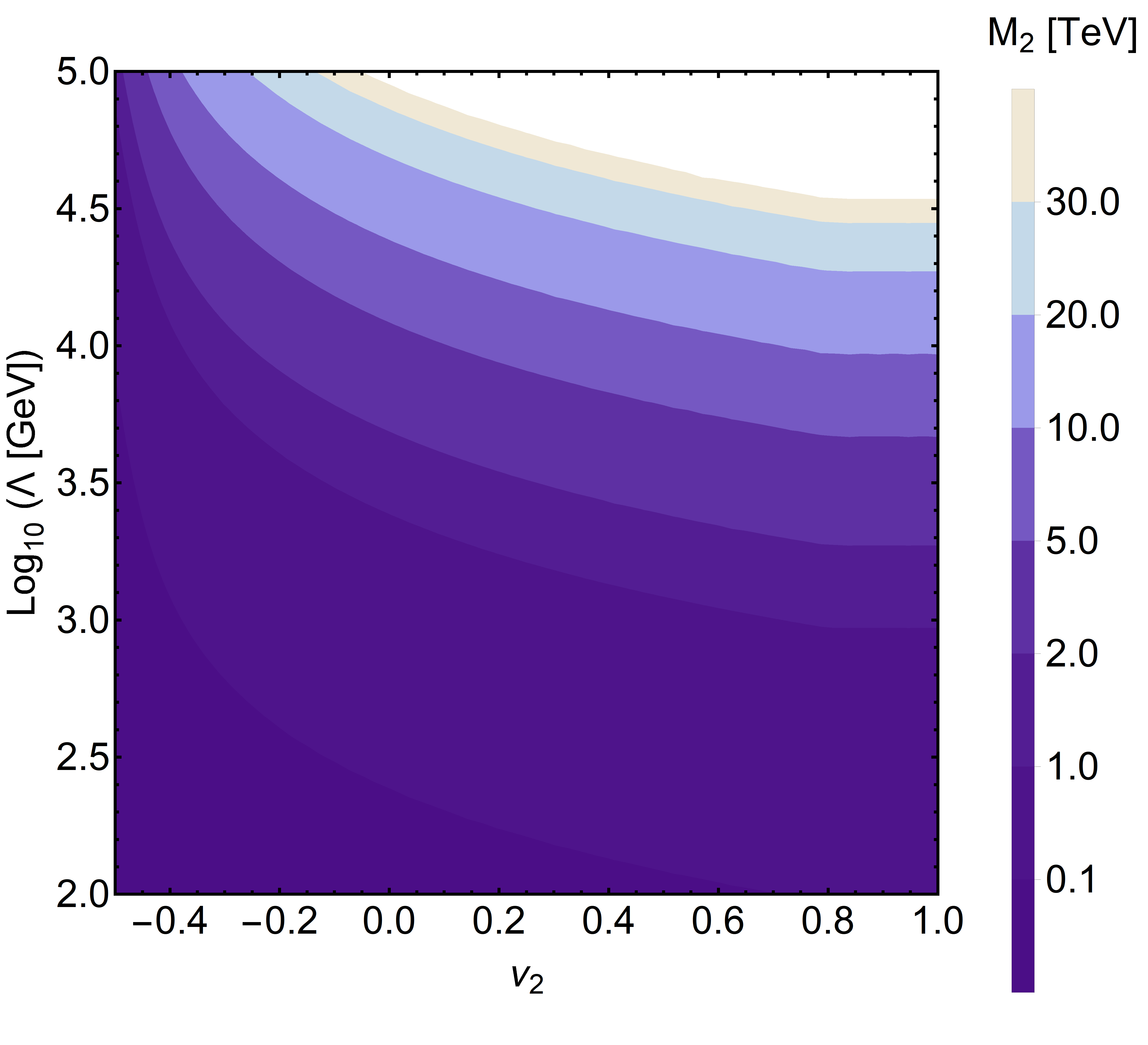}&
				\includegraphics[width=0.5\linewidth]{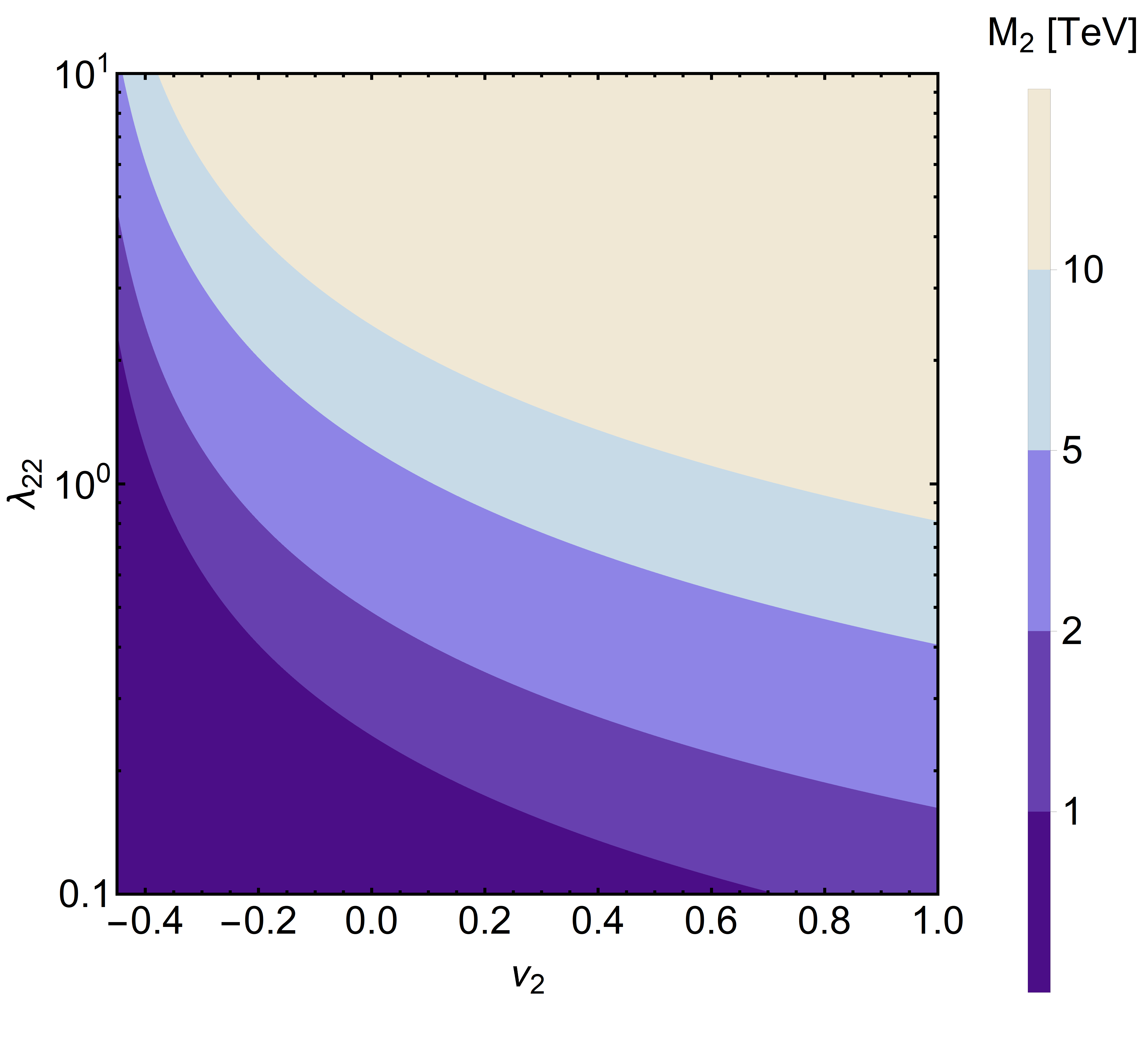}
	\end{tabular}
	\caption{\textit{Values of the mass of the second right-handed neutrino as a function of $\nu_{2}$ and $\Lambda$ (left panel) and in the $\left(\nu_{2},\lambda_{22}\right)$ plane with $\Lambda=10^{4} \ \mathrm{GeV}$ (right panel). Both panels assumes $\tilde{v}_{\mathrm{BL}}=10^{18} \ \mathrm{GeV}$.}}
	\label{fig:masses_contour}
\end{figure}

In Fig. \ref{fig:masses_contour} (left panel), we represent the mass of the second right-handed neutrino, our DM candidate, in the $\left(\nu_{2},\Lambda\right)$ plane, for $\tilde{v}_{\mathrm{BL}}=10^{18} \ \mathrm{GeV}$. We notice the strong dependence on $\nu_{2}$ in the range $\nu_{2}\in\left[-0.5,0\right]$, which precisely turns out to be the allowed range shown in Fig. \ref{fig:plotnu1nu23}. For larger values of $\nu_{2}$, the value of $M_{2}$ begins to stabilize for constant $\Lambda$. A similar behaviour can be observed in Fig. \ref{fig:masses_contour} (right panel), where we show the values of the mass as functions of $\nu_{2}$ and the Yukawa coupling $\lambda_{22}$ (which previously we set to be $\mathcal{O}(1)$). We do not consider the effect on $M_{2}$ of varying the value of the Yukawa coupling $\lambda_{33}$, as it has little effect on it when diagonalizing the heavy sector mass matrix.

\section{Neutrinos meet DM}

Before, we have searched for the values of $\Lambda$ for which the heavy neutrino masses lie within a region that seemed reasonable to us. But if we are to obtain the relic abundance, at least the mass of the DM candidate is univocally fixed for given values of $\Lambda$ and $m_{G_1}$. As we are considering the case of a Majorana fermion in the bulk, we also need to take into account the value of the dimensionless mass parameter $\nu_2$. Therefore, for each choice of $\left(\Lambda,m_{G_1}\right)$, we will have values of $m_{\mathrm{DM}}$ as functions of $\nu_2$.

\begin{figure}
	\centering
	\includegraphics[width=1\linewidth]{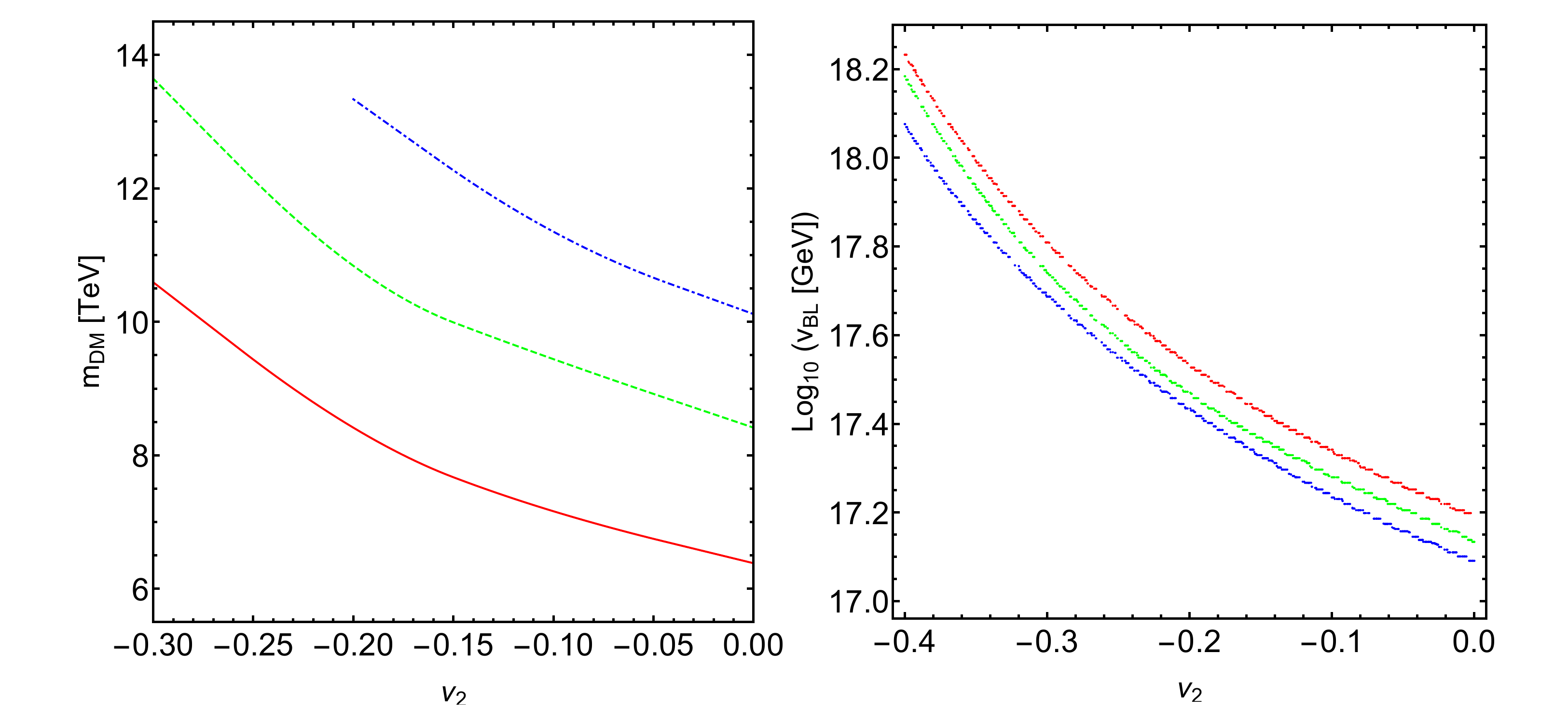}
	\caption{\textit{Left panel: DM masses for which the correct relic abundance is obtained, as a function of the dimensionless parameter $\nu_2$ for several values of $\Lambda$ and $m_{G_1}=0.1 \ \mathrm{TeV}$. Right panel: points on the plane $(\nu_2,\tilde{v}_{\mathrm{BL}})$ for which the required DM mass to obtain the relic abundance is reproduce, depending on the value of $\nu_2$, for several $\Lambda$. In both cases $\Lambda=100, \ 150, \ 200 \ \mathrm{TeV}$ (red, green and blue lines respectively).}}
	\label{fig:relicnu}
\end{figure}

Using the modified cross-sections of App. \ref{Appendix B}, given a particular choice of $\left(\Lambda,m_{G_1}\right)$, we have to compute the thermal-average of the DM annihilation cross-section for different bulk parameters. In Fig. \ref{fig:relicnu} (left panel) we have plotted the required values of $m_{\mathrm{DM}}$ to obtain the relic abundance, for $m_{G_1}=0.1 \ \mathrm{TeV}$ and $\Lambda=100,150,200 \ \mathrm{TeV}$ (solid red, dashed green and dash-dotted blue lines respectively), as functions of $\nu_{2}$. From this one can see that generally greater values of $\Lambda$ imply larger masses at a fixed $\nu_2$. Moreover, notice that in every case the required mass decreases as the dimensionless parameter grows larger.

In order to adjust the mass of the second right-handed neutrino to the correct value required to achieve the DM relic abundance, there are several parameters that can be put into play. Following the parametrization of Eq. (\ref{6.10}), an interesting possibility can be study the values of $\tilde{v}_{\mathrm{BL}}$ needed to reach $m_{\mathrm{DM}}$. In Fig. \ref{fig:relicnu} (right panel) we represent the values of $\tilde{v}_{\mathrm{BL}}$ versus the dimensionless parameter $\nu_{2}$ for which we recover the appropriate masses. Notice that the range of $\tilde{v}_{\mathrm{BL}}$ is consistent with what we expected, as it must be of order $M_{P}$.

\chapter{Towards a realistic implementation of the model}\label{Chapter 7}

In the previous chapter, we found significant regions of the parameter space where the idea of one of the right-handed neutrinos being the DM was feasible. Nevertheless, further considerations are required if we want our model to be acceptable from a phenomenological point of view. Thus, in the present chapter we review the difficulties that our model has to face.

We are going to focus on two main issues: the consistency of our light-neutrino sector with experimental data and the possibly dangerous decay channels of the heavy species. Once we have identified these, we propose different ways to settle them in future work.

\section{Light neutrino sector}

Despite the fact that we have achieved sub-$\mathrm{eV}$ masses for the light neutrino sector, our model does not give us the observed mass splittings nor the flavour mixing. This problem can be addressed using the Casas-Ibarra parametrization \cite{Casas2001} or its extension to all orders in the seesaw expansion \cite{Donini2012}. To first order in the seesaw expansion, the neutrino mass matrix takes the form:
\begin{equation}\label{7.1}
\left(\begin{array}{cc}U_{PMNS}&i U_{PMNS} m_{\nu}^{1/2} R^{\dagger} M_{H}^{-1/2}\\i U_{PMNS} M_{H}^{-1/2} R m_{\nu}^{1/2}&I\end{array}\right),
\end{equation}
where $m_{\nu}$ and $M_{H}$ are the diagonal mass matrices of the light and heavy neutrino sectors respectively, $U_{PMNS}$ the standard $3\nu$ mixing matrix and $R$ a generic $3 \times 3$ orthogonal matrix. This way, the parameter space of the model would be substantially reduced. For example, the Dirac Yukawa coupling would move from $9$ free parameters (assuming that the matrix is real) to only $3$ mixing angles. The other $6$ parameters are fixed, $3$ related to the masses of the light neutrinos and the remaining $3$ to the mixing between them. Moreover, this parametrization will put stringent bounds to the regions tested in this study. This task is beyond the scope of this thesis, and will be explored elsewhere.

\section{Lepton-number violating decays}

For the present study to be consistent, we must check that our DM candidate is sufficiently stable in order to maintain the correct relic abundance after the freeze-out. Therefore, its lifetime must be greater than the time elapsed from the freeze-out era to nowadays. The dangerous decays of the heavy Majorana neutrinos come from their mixing with the light neutrinos (that, after electroweak spontaneous symmetry breaking, give the Dirac mass matrix $m_D$), thus coupling the $N_i$ with the SM Higgs. After diagonalization of the neutrino mass matrix,
eventually, the right-handed neutrinos also couple to the $W$ and $Z$ gauge bosons through charged and neutral currents interactions. As $M_{2}>m_{W}$, the dominant decay modes of the heavy Majorana neutrino are expected to be the ones to heavy species, such as the Higgs boson, a $t \bar{t}$ pair, and the massive gauge bosons. The partial width of the heavy neutrino at leading order in the mixing, are given below. The Lagrangian, from which the corresponding Feynman rules for the interaction vertices can be obtained, is given in App. \ref{Appendix B}.
\begin{equation}\label{7.2}
\begin{split}
&\Gamma(N_{2} \rightarrow \nu_{\ell} H)=\frac{\left|V_{\ell 2}\right|^2}{16 \pi v_{\mathrm{EW}}^2} M_{2}^{3} \left(1-\frac{m_{H}^2}{M_{2}^2}\right)^2\\
&\Gamma(N_{2} \rightarrow \ell^{\pm} W^{\mp})=\frac{\left|V_{\ell 2}\right|^{2}}{16 \pi v_{\mathrm{EW}}^2} M_{2}^{3} \left(1+\frac{2 m_{W}^{2}}{M_{2}^{2}}\right) \left(1-\frac{m_{W}^2}{M_{2}^2}\right)^2,\\
&\Gamma(N_{2} \rightarrow \nu_{\ell} Z)=\frac{\left|V_{\ell 2}\right|^{2}}{16 \pi v_{\mathrm{EW}}^2} M_{2}^{3} \left(1+\frac{2 m_{Z}^{2}}{M_{2}^{2}}\right) \left(1-\frac{m_{Z}^2}{M_{2}^2}\right)^2,\\
&\Gamma(N_{2} \rightarrow \nu_{\ell} \bar{t} t)=\frac{m_{t}^2}{1024 \pi^3 v_{\mathrm{EW}}^4} M_{2}^{3} \left|V_{\ell 2}\right|^2.
\end{split}
\end{equation}

The listed decay modes contribute to the total decay width of the heavy Majorana neutrino, which is given by:
\begin{equation}\label{7.3}
\Gamma_{N_{2}}=\sum_{\ell} \left[\Gamma(N_{2} \rightarrow \nu_{\ell} H) + 2 \Gamma(N_{2} \rightarrow \ell^{\pm} W^{\mp}) + \Gamma(N_{2} \rightarrow \nu_{\ell} Z) + 3 \Gamma(N_{2} \rightarrow \nu_{\ell} \bar{t} t) \right],
\end{equation}
where we introduced a $2$ factor in order to take into account the processes $N_{2} \rightarrow \ell^{+} W^{-}$ and their charge conjugates $N_{2} \rightarrow \ell^{-} W^{+}$ and a $3$ factor due to the three colour configurations of the quark pair.

\begin{figure}
	\centering
	\includegraphics[width=1\linewidth]{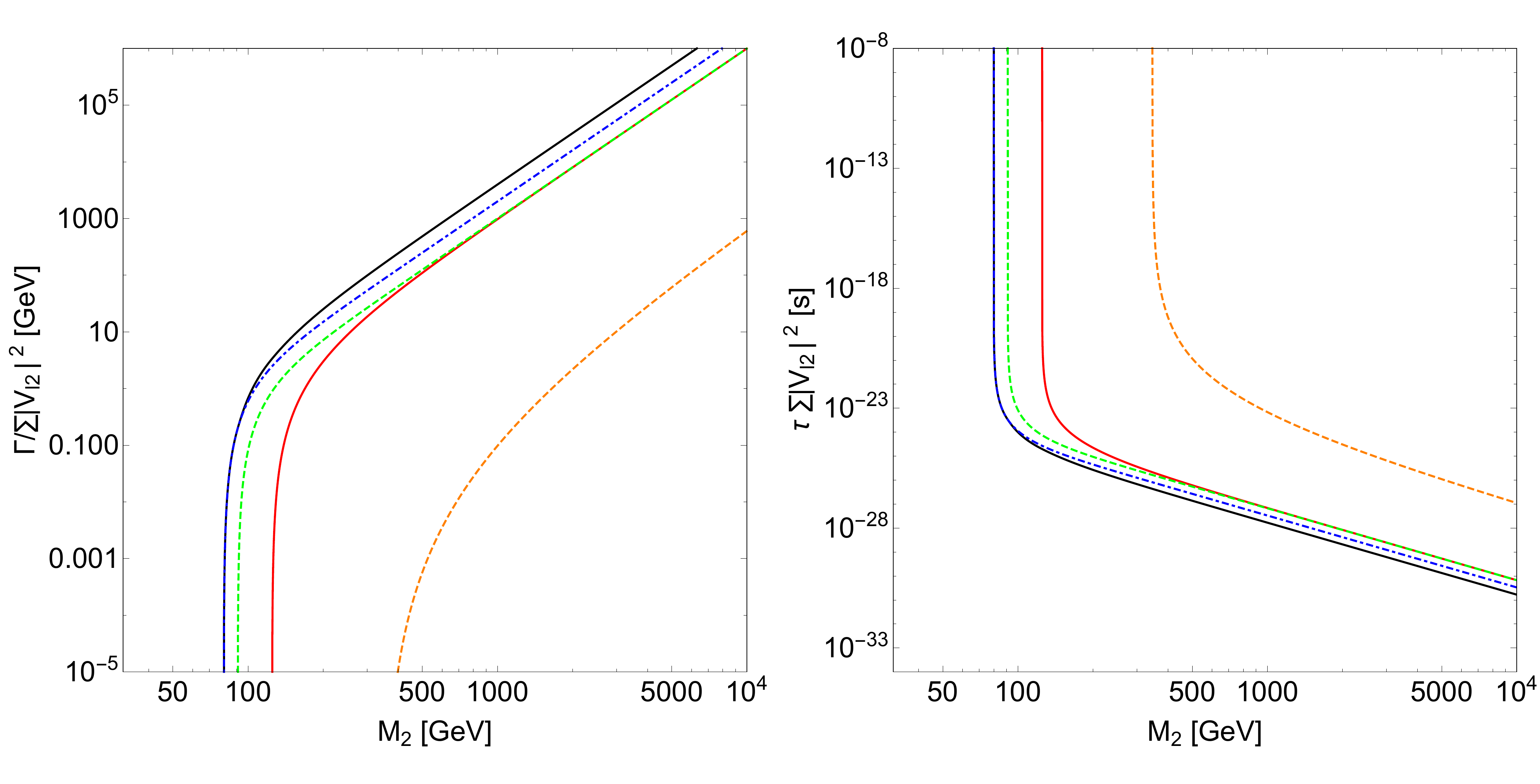}
	\caption{\textit{Decay width (left panel) and mean lifetime (right panel) of the heavy Majorana neutrino normalized by $\sum_{\ell} \left|V_{\ell 2}\right|$, as a function of its mass $M_{2}$. In both figures, the total quantity is represented by the solid black line. The dashed orange, solid red, dash-dotted blue and dashed green lines correspond to the decays to $\bar{t}t$, $H$, $W^{\pm}$ and $Z$ respectively.}}
	\label{fig:decay_widths}
\end{figure}

In Fig. \ref{fig:decay_widths} we show the partial and total decay width (left panel) and mean lifetime (right panel) versus the mass of the heavy neutrino. We have normalized both quantities by a factor $\sum_{\ell} \left|V_{\ell 2}\right|^{2}$. One can see that the contribution of the $N_{2} \rightarrow \nu_{\ell} \bar{t} t$ channel (dashed orange line) is almost negligible compared to the other three. Moreover, the contribution of the $H$ and $Z$ production channels (solid red and dashed green lines) are almost identical in the considered region. The predominant channel is found to be the corresponding of the $W^{\pm}$ production (dash-dotted blue line). For $M_2>m_{H}$, one can see that the total width grows as $M_{2}^3$, as expected.

According to the expected value of the thermally-averaged annihilation cross-section at freeze-out, one finds that the freeze-out point is approximately $x\equiv m_{\mathrm{DM}}/T_{\mathrm{FO}}\approx25$. For the considered mass range $m_{\mathrm{DM}}\in\left[0.1,10\right] \ \mathrm{TeV}$, that translates into a range of temperatures of $T_{\mathrm{FO}}\in\left[10^{13},10^{15}\right] \ \mathrm{K}$. That places freeze-out near the cosmological quark epoch, at around $10^{-12} \ \mathrm{s}$ to $10^{-6} \ \mathrm{s}$ after Big Bang. Since the estimated age of the universe is $4.35 \times 10^{17} \ \mathrm{s}$ at $68\%$ C.L. \cite{Ade2016}, the effects of $T_{\mathrm{FO}}$ can be neglected, thus the lifetime of the heavy Majorana neutrino should be of the same order of magnitude as the age of the universe. Therefore, we can estimate that it must be of order:
\begin{equation}\label{7.4}
\sum_{\ell} \left|V_{\ell 2}\right|^{2}\approx 10^{-42}.
\end{equation}

This value is far below the scaling imposed by the seesaw mechanism:
\begin{equation}\label{7.5}
\left|V_{\ell N}\right|^{2}\sim\frac{m_{\nu}}{M_{N}},
\end{equation}
which in the case of our DM candidate is usually of order $\sim 10^{-14}$. From this we learn that in order to make our model consistent, we need to implement a mechanism that forbids or at least suppresses these decay channels. The origin of the problem is the Yukawa coupling between the light and heavy neutrino species, thus it should be modified in some way. We are currently working in this direction, considering some minimal extensions of the model that can guarantee this requisite. Some of the possibilities involve extensions of the scalar sector of the model, radiative origin of the Dirac Yukawa coupling, etc.

\section{Type-I Seesaw in a Two Higgs Doublet Model}

A possible solution to the previous issue entails the implementation of a two Higgs doublet model (2HDM). The inclusion of a second Higgs doublet leads to a Yukawa sector of the form:
\begin{equation}\label{7.6}
\begin{split}
-\lag_{Y}&=y_{\mathrm{I}}^{d} \bar{Q}_{L} \Phi_{\mathrm{I}} d_{R}+y_{\mathrm{I}}^{u} \bar{Q}_{L} \tilde{\Phi}_{\mathrm{I}} u_{R}+y_{\mathrm{I}}^{\ell} \bar{L}_{L} \Phi_{\mathrm{I}} \ell_{R}+y_{\mathrm{I}}^{\nu} \bar{L}_{L} \tilde{\Phi}_{\mathrm{I}} N_{R}\\
&+y_{\mathrm{II}}^{d} \bar{Q}_{L} \Phi_{\mathrm{II}} d_{R}+y_{\mathrm{II}}^{u} \bar{Q}_{L} \tilde{\Phi}_{\mathrm{II}} u_{R}+y_{\mathrm{II}}^{\ell} \bar{L}_{L} \Phi_{\mathrm{II}} \ell_{R}+y_{\mathrm{II}}^{\nu} \bar{L}_{L} \tilde{\Phi}_{\mathrm{II}} N_{R}\\
&+\lambda \bar{N}^{c}_{L} \Phi_{\mathrm{BL}} N_{R} + \mathrm{h.c.},
\end{split}
\end{equation}

Extending our model with an Abelian gauge symmetry $U(1)_{X}$ we can avoid the flavour changing neutral currents (FCNC) that can arise from these Yukawa interactions. Our motivation to introduce two Higgs doublets is the possibility that neutrinos couple to a different Higgs than the rest of SM fermions. With these specifications we can fix the charges of the different fields under this new symmetry:
\begin{equation}\label{7.7}
\begin{array}{cc}Q_{1}=\frac{u-d}{2},&Q_{2}=(n-e)-\frac{u-d}{2},\\q=\frac{u+d}{2},&l=e+\frac{u-d}{2},\end{array}
\end{equation}
where $q$ is the $U(1)_{X}$ charge of the quark doublet, $l$ the charge of the lepton doublet, $u$ and $d$ the charges of the right-handed quark singlets, $e$ and $n$ the charges of the right-handed lepton singlets and $Q_i$ the charge of the $i$-th scalar doublet. In addition to these relations, we must ensure $Q_{1}\neq Q_{2}$, in order to avoid FCNC. Another one of the charges above, for example $n$, can be constrained by the cancellation of the $\left[U(1)_{X}\right]^{3}$ gauge anomaly. This condition reads:
\begin{equation}\label{7.8}
\sum_{L} Q_{L}^{3}-\sum_{R} Q_{R}^{3}=0.
\end{equation}

\begin{figure}
	\centering
	\includegraphics[width=0.6\linewidth]{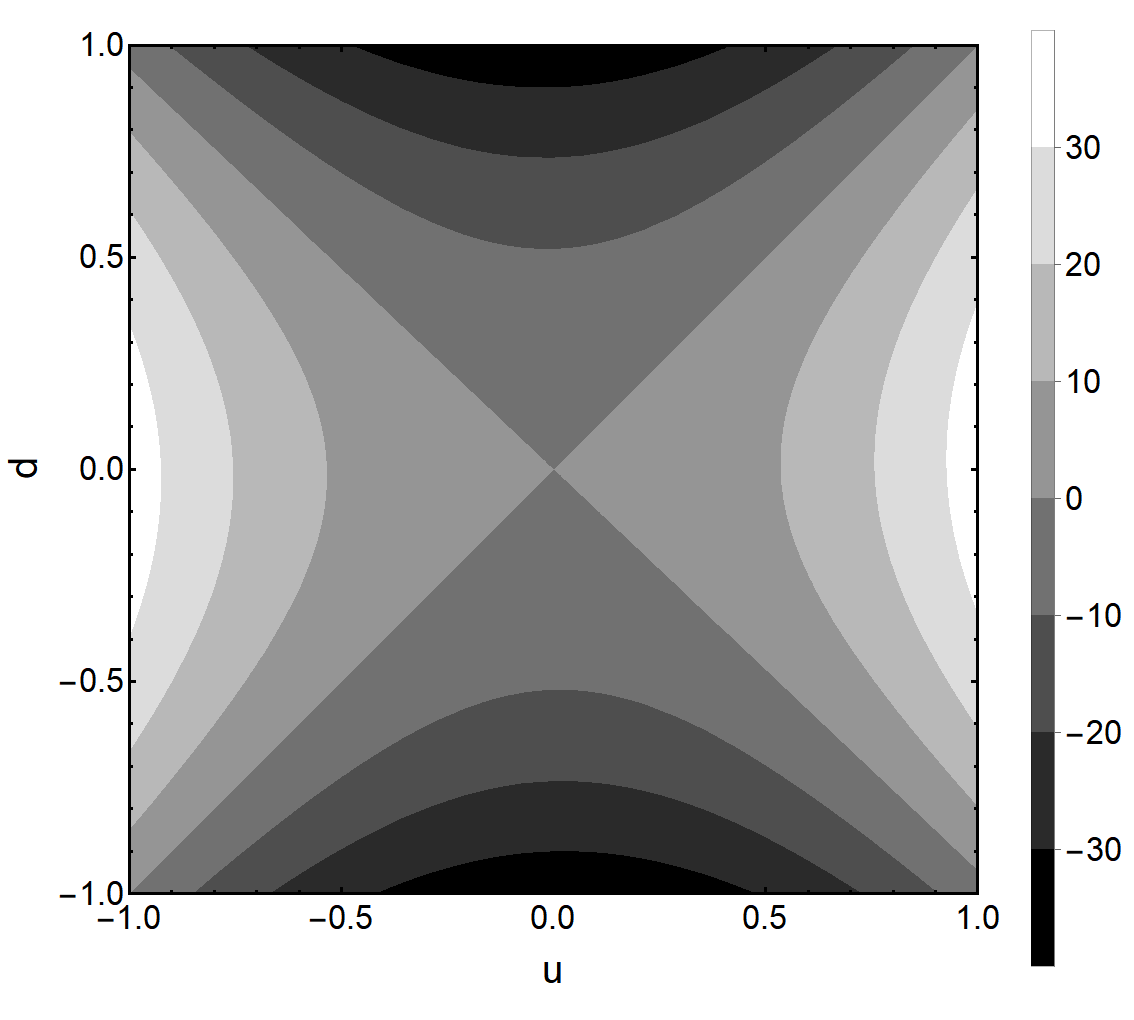}
	\caption{\textit{Value of the discriminant of Eq. (\ref{7.9}) as a polynomial in $n$, as a function of the charges of the right-handed quark singlets $u$ and $d$.}}
	\label{fig:discriminant}
\end{figure}

If one assumes the simple relation $Q_{1}=-Q_{2}$ for the scalar doublet charges, we have $e = n$ and Eq. (\ref{7.8}) reduces to:
\begin{equation}\label{7.9}
\left(d-u\right)\left[5d^2+6n^2-4u^2+3nu-3dn-du\right]=0.
\end{equation}
In general, this polynomial would have two roots. In order to ensure that $n\in \mathbb{R}$, its discriminant must be positive. In Fig. \ref{fig:discriminant} we show the value of the discriminant as a function of the charges $u$ and $d$. It is clear that we can achieve a real value for $n$ in a wide area of parameter space. If we choose the discriminant to be zero, we automatically fix the values of $d$ and $n$:
\begin{equation}\label{7.10}
\begin{array}{cc}d=-\frac{35}{37} u,&n=-\frac{18}{37} u.\end{array}
\end{equation}

Therefore, all charges are fixed once we fix a specific value for $u$. The Yukawa interaction in Eq. (\ref{7.6}) reduces, once $U(1)_{X}$ charges are assigned as we have sketched above, to:
\begin{equation}\label{7.11}
\begin{split}
-\lag_{Y}&=y_{\mathrm{I}}^{d} \bar{Q}_{L} \Phi_{\mathrm{I}} d_{R}+y_{\mathrm{I}}^{u} \bar{Q}_{L} \tilde{\Phi}_{\mathrm{I}} u_{R}+y_{\mathrm{I}}^{\ell} \bar{L}_{L} \Phi_{\mathrm{I}} \ell_{R}+y_{\mathrm{II}}^{\nu} \bar{L}_{L} \tilde{\Phi}_{\mathrm{II}} N_{R}\\
&+\lambda \bar{N}^{c}_{L} \Phi_{\mathrm{BL}} N_{R} + \mathrm{h.c.},
\end{split}
\end{equation}
so that $\Phi_{\mathrm{I}}$ is "neutrinophobic", whereas $\Phi_{\mathrm{II}}$ is "neutrinophilic". The most general scalar potential invariant under $U(1)_{X}$ is:
\begin{equation}\label{7.12}
\begin{split}
V&=m_{11}^2 \Phi_{\mathrm{I}}^{\dagger}\Phi_{\mathrm{I}}+m_{22}^2 \Phi_{\mathrm{II}}^{\dagger}\Phi_{\mathrm{II}}+\frac{\lambda_{1}}{2}\left(\Phi_{\mathrm{I}}^{\dagger}\Phi_{\mathrm{I}}\right)^2+\frac{\lambda_{2}}{2}\left(\Phi_{\mathrm{II}}^{\dagger}\Phi_{\mathrm{II}}\right)^2\\
&+\lambda_{3}  \Phi_{\mathrm{I}}^{\dagger}\Phi_{\mathrm{I}} \Phi_{\mathrm{II}}^{\dagger}\Phi_{\mathrm{II}}+\lambda_{4}  \Phi_{\mathrm{I}}^{\dagger}\Phi_{\mathrm{II}} \Phi_{\mathrm{II}}^{\dagger}\Phi_{\mathrm{I}},
\end{split}
\end{equation}
where we have adopted the notation of Ref. \cite{Branco2011}. As we have two scalar $SU(2)$ doublets, we end up with eight fields:
\begin{equation}\label{7.13}
\Phi_{i}=\left(\begin{array}{c}\phi^{+}_{i}\\\frac{v_{i}+\rho_{i}+i \eta_{i}}{\sqrt{2}}\end{array}\right).
\end{equation}

Three of these fields give mass to the massive gauge bosons, thus there are only five physical scalar fields. As a consequence of our $U(1)_{X}$ charges choice, the pseudoscalar boson results massless.

For the present study, the relevant mixing is the one between the neutral species. The mass term of these scalars results:
\begin{equation}\label{7.14}
\lag_{\rho}=-\left(\begin{array}{cc}\rho_{\mathrm{I}}&\rho_{\mathrm{II}}\end{array}\right)\left(\begin{array}{cc}\lambda_{1}v_{1}^2&\lambda_{34}v_{1}v_{2}\\\lambda_{34}v_{1}v_{2}&\lambda_{2}v_{2}^2\end{array}\right)\left(\begin{array}{c}\rho_{\mathrm{I}}\\\rho_{\mathrm{II}}\end{array}\right),
\end{equation}
where $\lambda_{34}=\lambda_{3}+\lambda_{4}$. Assuming that this mixing between scalars is sufficiently small, we can identify the first one with the SM Higgs boson, $H$. The second would be a heavier boson that only couples to the neutrino sector, $\mathcal{H}$. If $M_{\mathcal{H}}>M_{2}$, the decay to this scalar on-shell $N_{2}\rightarrow\nu_{\ell}\mathcal{H}$ is forbidden, and the ones mediated by its virtual exchange are highly suppressed. The Higgs-gauge interaction sector will be given by:
\begin{equation}
\lag_{V} = D_{\mu} \Phi_{\mathrm{I}}^{\dagger} D^{\mu} \Phi_{\mathrm{I}} + D_{\mu} \Phi_{\mathrm{II}}^{\dagger} D^{\mu} \Phi_{\mathrm{II}} - \frac{1}{4} W^{i}_{\mu \nu} W^{i\mu\nu} - \frac{1}{4} X_{\mu\nu} X^{\mu\nu},
\end{equation}
with $D_{\mu} = \partial_{\mu} - i g t^{i} W^{i}_{\mu} - i g^{\prime} Y/2 B_{\mu} - i   g^{\prime\prime} Q X_{\mu}$.

Eventually, the two unavoidably dangerous decay channels are $N_{2} \rightarrow \ell^{\pm} W^{\mp},\nu_{\ell} Z$. After SSB occurs, breaking $SU(2)_{L} \times U(1)_{Y} \times U(1)_{X}$ to $U(1)_{Q}$, the Yukawa Lagrangian in Eq. (\ref{7.11}) gives the SM fermion masses plus the neutrino mass matrix of Eq. (\ref{3.19}). After diagonalization, we get the charged and neutral currents neutrino interactions given in Eq. (\ref{B.7}) of App. \ref{Appendix B}. After $U(1)_{X}$ spontaneous symmetry breaking, we can still have a remnant global $Z_{2}$ symmetry. This new parity can be used to distinguish between SM particles (including left-handed neutrinos) and the right-handed singlet states, much as R-parity is used in SUSY-models to differentiate SM particles from their superpartners. It could be possible that using this parity we may suppress the undesirable $N_{2} \rightarrow \ell^{\pm} W^{\mp},\nu_{\ell} Z$ decays, whilst maintaining the features of the model described in Chap. \ref{Chapter 6}. Further work is needed in order to have a consistent and complete picture of this implementation. This specific scenario will be studied elsewhere.

\chapter{Conclusion and Outlook}\label{Chapter 8}

In this thesis we have studied the possibility of identify one of the right-handed neutrinos entering the type-I seesaw mechanism as a DM candidate, within the warped extra-dimensions paradigm. We focused on generate a specific hierarchy between the right-handed neutrinos, that allows a correct realization of the seesaw while having an eV-scale sterile neutrino, another one lying in the appropriate mass range for the DM and the last one being super heavy. For every configuration of parameters, the DM must have a specific mass in order to achieve the correct relic abundance. Therefore, we correlated these with the parameters of the seesaw, encountering a large region of parameter space where both ideas are compatible. Moreover, these results are not excluded by the bounds from DM and heavy neutrino searches, which are also discussed.

Even though, the model has some issues that compromise its integrity: it does not account for the measured masses and mixing of the light neutrino sector and there are some dangerous decay channels that prevents the DM candidate to be sufficiently stable. We discussed both problems and propose some reasonable solutions. The first one can be easily addressed using a specific parametrization that fixes those outputs, therefore constraining the parameter space. This is expected to be done in future work. The second problem is still under analysis. In the thesis we proposed a feasible extension of the model involving a new Higgs doublet, and briefly study its implementation. However, further research is needed. Another considered possibility is a radiative origin of the Yukawa coupling between the heavy and light species. A full study of these scenarios will be done elsewhere.

\begin{appendices}
\chapter{Annihilation DM cross-section}\label{Appendix A}

In the present section, we show the DM annihilation cross-sections in the case where the DM particles are Majorana fermions. To do so, we have made use of the Feynman rules given in Ref. \cite{Folgado2020}. First, we will consider the production of SM model particles from DM annihilation, through virtual exchange of KK-graviton modes. Then, we focus on the production of on-shell KK-gravitons.

In this analysis, we are going to make use of the so-called \textit{velocity expansion} for the DM particles:
\begin{equation}
s\approx m_{\mathrm{DM}}^2(4+V_{rel}^2),
\end{equation}
where $V_{rel}$ is its relative velocity. In this approximation, further simplification arise from the different scalar products of the particles momenta:
\begin{equation}
\left\{\begin{array}{ccccc}p_{1}p_{4}&=&p_{2}p_{3}&\approx&m_{\mathrm{DM}}^2+\frac{1}{2} m_{\mathrm{DM}}^2 \sqrt{1-\frac{m_{out}^2}{m_{\mathrm{DM}}^2}} \cos{\theta}V_{rel} + \frac{1}{4} m_{\mathrm{DM}}^2 V_{rel}^2\\p_{1}p_{3}&=&p_{2}p_{4}&\approx&m_{\mathrm{DM}}^2-\frac{1}{2} m_{\mathrm{DM}}^2 \sqrt{1-\frac{m_{out}^2}{m_{\mathrm{DM}}^2}} \cos{\theta}V_{rel} + \frac{1}{4} m_{\mathrm{DM}}^2 V_{rel}^2\\\end{array}\right.,
\end{equation}
where $p_{1}$, $p_{2}$ are the momenta of the two DM ingoing particles and $p_{3}$, $p_{4}$ the corresponding momenta of the outgoing particles.

\section{DM annihilation via KK-graviton exchange}

\begin{figure}[t]
	\centering
	\subfigure{
		\begin{fmfgraph*}(96,64)
			\fmfleft{i1,i2}
			\fmfright{o1,o2}
			\fmf{plain}{v1,i1}
			\fmf{plain}{i2,v1}
			\fmf{dbl_wiggly}{v2,v1}
			\fmf{dashes}{v2,o1}
			\fmf{dashes}{v2,o2}
		\end{fmfgraph*}
	}
	\subfigure{
		\begin{fmfgraph*}(96,64)
			\fmfleft{i1,i2}
			\fmfright{o1,o2}
			\fmf{plain}{v1,i1}
			\fmf{plain}{i2,v1}
			\fmf{dbl_wiggly}{v2,v1}
			\fmf{fermion}{o2,v2}
			\fmf{fermion}{v2,o1}
		\end{fmfgraph*}
	}
	\subfigure{
		\begin{fmfgraph*}(96,64)
			\fmfleft{i1,i2}
			\fmfright{o1,o2}
			\fmf{plain}{v1,i1}
			\fmf{plain}{i2,v1}
			\fmf{dbl_wiggly}{v2,v1}
			\fmf{wiggly}{v2,o1}
			\fmf{wiggly}{v2,o2}
		\end{fmfgraph*}
	}
	\caption{\textit{Feynman diagrams corresponding to the different amplitudes that contribute to fermionic DM annihilation into SM particles via KK-graviton exchange.}}
	\label{fig:SM_decay}
\end{figure}
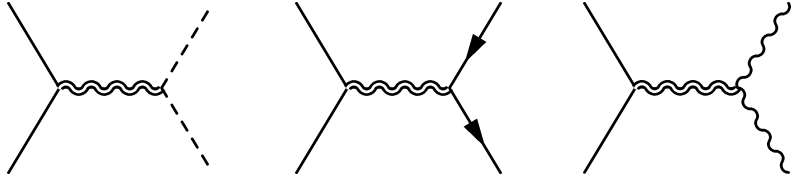

In the first place, two DM particles can annihilate into a pair of SM Higgs bosons. The $s$-channel amplitude of this process will be given by:
\begin{equation}
\mathcal{M}_s= \bar{v}^{s'}(p_2) \Gamma_{\mu\nu}^{F}(p_1,p_2) u^{s}(p_1) \Delta_{\mu\nu\alpha\beta}(p_1+p_2) \Gamma_{\alpha\beta}^{S}(p_3,p_4),
\end{equation}
where $\Gamma_{\mu\nu}^{F}$ is the vertex that involves two fermions and a KK-graviton, $\Gamma_{\mu\nu}^{S}$ the vertex between two scalars and a KK-graviton and $\Delta_{\mu\nu\alpha\beta}$ the graviton propagator. Using the velocity expansion, the cross-section at leading order in the relative velocity results:
\begin{equation}
\sigma\left(\chi\chi\rightarrow HH\right)\approx V_{rel} \left|S_{KK}\right|^2 \frac{m_{\mathrm{DM}}^6}{36 \pi} \left(1-\frac{m_{H}^2}{m_{\mathrm{DM}}^2}\right)^{5/2}.
\end{equation}

The annihilation amplitude of two DM particles producing two SM fermions through virtual KK-graviton exchange reads:
\begin{equation}
\mathcal{M}_s= \bar{v}^{s'}(p_2) \Gamma_{\mu\nu}^{F}(p_1,p_2) u^{s}(p_1) \Delta_{\mu\nu\alpha\beta}(p_1+p_2) \bar{u}^{r}(p_3) \Gamma_{\alpha\beta}^{F}(p_3,p_4) v^{r'}(p_4).
\end{equation}

The total cross-section for such a process results:
\begin{equation}
\sigma\left(\chi\chi\rightarrow \psi \bar{\psi}\right)\approx V_{rel} \left|S_{KK}\right|^2 \frac{m_{\mathrm{DM}}^6}{6 \pi} \left(1-\frac{m_{\psi}^2}{m_{\mathrm{DM}}^2}\right)^{3/2}\left(1+\frac{2m_{\psi}^2}{3m_{\mathrm{DM}}^2}\right).
\end{equation}

In the end, two DM particles can annihilate giving rise to a pair of gauge bosons, with an $s$-channel amplitude:
\begin{equation}
\mathcal{M}_s= \bar{v}^{s'}(p_2) \Gamma_{\mu\nu}^{F}(p_1,p_2) u^{s}(p_1) \Delta_{\mu\nu\alpha\beta}(p_1+p_2) \Gamma_{\alpha\beta}^{V}(p_3,p_4) e^{\sigma}_{\alpha}(p_3)e^{\lambda}_{\beta}(p_4),
\end{equation}
where $\Gamma_{\alpha\beta}^{V}$ is the vertex corresponding to one KK-graviton and two vectors. In the case of the massive gauge bosons, one gets:
\begin{equation}
\left\{\begin{array}{ccc}\sigma\left(\chi\chi\rightarrow W^{+} W^{-}\right)&\approx&V_{rel} \left|S_{KK}\right|^2 \frac{13 m_{\mathrm{DM}}^6}{18 \pi} \left(1-\frac{m_{W}^2}{m_{\mathrm{DM}}^2}\right)^{1/2}\left(1+\frac{14m_{W}^2}{13m_{\mathrm{DM}}^2}+\frac{3m_{W}^4}{13m_{\mathrm{DM}}^4}\right),\\\sigma\left(\chi\chi\rightarrow Z Z\right)&\approx&V_{rel} \left|S_{KK}\right|^2 \frac{36 m_{\mathrm{DM}}^6}{18 \pi} \left(1-\frac{m_{Z}^2}{m_{\mathrm{DM}}^2}\right)^{1/2}\left(1+\frac{14m_{Z}^2}{13m_{\mathrm{DM}}^2}+\frac{3m_{Z}^4}{13m_{\mathrm{DM}}^4}\right),\end{array}\right.
\end{equation}
whereas for the case of the massless gauge bosons we have:
\begin{equation}
\left\{\begin{array}{ccc}\sigma\left(\chi\chi\rightarrow \gamma \gamma\right)&\approx&V_{rel} \left|S_{KK}\right|^2 \frac{m_{\mathrm{DM}}^6}{3 \pi},\\\sigma\left(\chi\chi\rightarrow g g\right)&\approx&V_{rel} \left|S_{KK}\right|^2 \frac{8 m_{\mathrm{DM}}^6}{3 \pi}.\end{array}\right.
\end{equation}

\section{On-shell KK-gravitons production via DM annihilation}

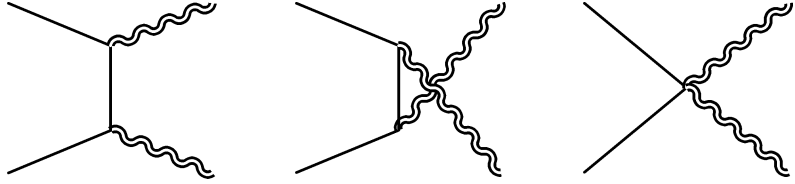
\begin{figure}[t]
	\centering
	\subfigure{
		\begin{fmfgraph*}(96,64)
			\fmfleft{i1,i2}
			\fmfright{o1,o2}
			\fmf{plain}{v1,i1}
			\fmf{plain}{i2,v2}
			\fmf{plain}{v2,v1}
			\fmf{dbl_wiggly}{v1,o1}
			\fmf{dbl_wiggly}{v2,o2}
		\end{fmfgraph*}
	}
	\subfigure{
		\begin{fmfgraph*}(96,64)
			\fmfleft{i1,i2}
			\fmfright{o1,o2}
			\fmf{plain}{v1,i1}
			\fmf{phantom}{v1,o1} 
			\fmf{plain}{i2,v2}
			\fmf{phantom}{v2,o2} 
			\fmf{plain}{v2,v1}
			\fmf{dbl_wiggly,tension=0}{v1,o2}
			\fmf{dbl_wiggly,tension=0}{v2,o1}
		\end{fmfgraph*}
	}
	\subfigure{
		\begin{fmfgraph*}(96,64)
			\fmfleft{i1,i2}
			\fmfright{o1,o2}
			\fmf{plain}{v1,i1}
			\fmf{plain}{i2,v2}
			\fmf{plain}{v1,v2}
			\fmf{dbl_wiggly}{v1,o2}
			\fmf{dbl_wiggly}{v2,o1}
		\end{fmfgraph*}
	}
	\caption{\textit{Feynman diagrams corresponding to the different amplitudes that contribute to fermionic DM annihilation into two on-shell KK-gravitons.}}
	\label{fig:KK_onshell}
\end{figure}

When $m_{DM}>m_{G_{1}}$, the channel $\chi\bar{\chi}\rightarrow G_{n} G_{m}$ is open. The contribution of the 4-point function vanishes \cite{Folgado2020}. The amplitude of the \textit{t}- and \textit{u}-channel result:
\begin{equation}
\begin{split}
\mathcal{M}_t&=e^{\lambda}_{\mu \nu}(p_4) \bar{v}^{r}(p_2) \Gamma_{\mu\nu}^{F}(p_2,k) S(k) \Gamma_{\alpha\beta}^{F}(k,p_1) u^{s}(p_1) e^{\sigma}_{\alpha \beta}(p_3),\\
\mathcal{M}_u&=e^{\sigma}_{\alpha \beta}(p_3) \bar{v}^{r}(p_2) \Gamma^{F}_{\alpha\beta}(p_2,k) S(k) \Gamma^{F}_{\mu\nu}(k,p_1) u^{s}(p_1) e^{\lambda}_{\mu\nu}(p_4),\\
\end{split}
\end{equation}
being $S(k)$ the fermion propagator. We can write the total amplitude as:
\begin{equation}
\begin{split}
\mathcal{M}&=\mathcal{M}_t+\mathcal{M}_u\\
&=e^{\sigma}_{\alpha \beta}(p_3)e^{\lambda}_{\mu\nu}(p_4) \bar{v}^{r}(p_2)\left[\Gamma^{F}_{\mu\nu}(p_2,k) S(k) \Gamma^{F}_{\alpha\beta}(k,p_1)+\Gamma^{F}_{\alpha\beta}(p_2,k) S(k) \Gamma^{F}_{\mu\nu}(k,p_1)\right]u^{s}(p_1)\\
&=e^{\sigma}_{\alpha \beta}(p_3)e^{\lambda}_{\mu\nu}(p_4) \ A^{sr}_{\mu\nu\alpha\beta}(p_1,p_2,k).
\end{split}
\end{equation}

To calculate the cross section, we must first compute the averaged modulus squared of this amplitude:
\begin{equation}
\begin{split}
\langle|\mathcal{M}|^2\rangle&=\frac{1}{4} \sum_{s,r,\lambda,\sigma} \mathcal{M} \mathcal{M}^*\\
&=\frac{1}{4} \sum_{s,r,\lambda,\sigma} e^{\sigma}_{\alpha \beta}(p_3) e^{*\sigma}_{\alpha' \beta'}(p_3) e^{\lambda}_{\mu\nu}(p_4) e^{*\lambda}_{\mu'\nu'}(p_4) \ A^{sr}_{\mu\nu\alpha\beta}(p_1,p_2,k) A^{*sr}_{\mu'\nu'\alpha'\beta'}(p_1,p_2,k)\\
&=\frac{1}{4} \sum_{s,r} P_{\alpha \beta \alpha' \beta'}(p_3) P_{\mu\nu \mu'\nu'}(p_4) \ A^{sr}_{\mu\nu\alpha\beta}(p_1,p_2,k) A^{*sr}_{\mu'\nu'\alpha'\beta'}(p_1,p_2,k).\\
\end{split}
\end{equation}

The differential cross-section then will be given by:
\begin{equation}
\frac{\diff \sigma}{\diff \Omega}=\frac{1}{64\pi^2 s} \ \frac{|\vec{p}_f|}{|\vec{p}_i|} \ \langle|\mathcal{M}|^2\rangle
\end{equation}

After applying the velocity expansion, taking only the leading order term we obtain:

\begin{equation}
\begin{split}
\sigma\left(\chi\bar{\chi}\rightarrow G_{n} G_{m}\right)&\approx V_{rel}^{-1}C_{\psi }^4 \frac{\left(\left(m_n^2-4 m_{\text{DM}}^2\right)^2-2 m_m^2 \left(4 m_{\text{DM}}^2+m_n^2\right)+m_m^4\right)^3}{16384 \pi\left(m_n^2+m_m^2-4 m_{\text{DM}}^2\right)^2}\\
&\times \frac{1}{{\Lambda^{4} m_m^2 m_{\text{DM}}^3 m_n^2}} \sqrt{\frac{\left(4 m_{\text{DM}}^2+m_n^2-m_m^2\right){}^2}{16 m_{\text{DM}}^2}-m_n^2}
\end{split}
\end{equation}
\chapter{Lepton mixing formalism}\label{Appendix B}

After spontaneous symmetry breaking of both $SU(2)_{L}\times U(1)_{Y}$ and $U(1)_{\mathrm{B-L}}$, the diagonalization of the neutrino mass term of Eq. (\ref{3.19}) gives:
\begin{equation}\label{B.1}
\lag_{\nu}=-\frac{1}{\sqrt{2}} m_{\alpha} \left(\bar{\nu}_{\alpha}\right)_{m} \left(\nu_{\alpha}^{c}\right)_{m}-M_{i} \left(\bar{N}_{i}^{c}\right)_{m} \left(N_{i}\right)_{m} + \mathrm{h.c.},
\end{equation}
where the subscript $m$ points out that we are talking about mass eigenstates. The relation between flavour and mass eigenstates reads:
\begin{equation}\label{B.2}
\left\{\begin{array}{ccc}\nu_{\beta}&=&U_{\beta\alpha} \left(\nu_{\alpha}\right)_{m} + V_{\beta i} \left(N_{i}^{c}\right)_{m},\\
N_{j}^{c}&=&X_{j\alpha} \left(\nu_{\alpha}\right)_{m} + Y_{j i} \left(N_{i}^{c}\right)_{m}.\end{array}\right.
\end{equation}

The matrices appearing in the mixing are built from the eigenvectors of the neutrino mass matrix, being these the blocks that form its modal matrix:
\begin{equation}\label{B.3}
L=\left(\begin{array}{cc}U&V\\X&Y\end{array}\right),
\end{equation}
a $6\times 6$ unitary matrix parametrized such that:
\begin{equation}\label{B.4}
L^{\dagger}\left(\begin{array}{cc}0&m_D\\ m_D^{T}&M\end{array}\right)L=\left(\begin{array}{cc}m_{\nu}&0\\0&M_{N}\end{array}\right).
\end{equation}

Along with these previous expansions, the Dirac Yukawa term of Eq. (\ref{3.17}) gives us the interaction between the right-handed neutrinos and the SM Higgs boson:
\begin{equation}\label{B.5}
\lag_{H}\supset-\frac{1}{\sqrt{2}} \frac{H}{v} M_{i} V_{\ell i}^{*} \bar{N}^{c}_{i} \nu^{c}_{\ell},
\end{equation}
where $\nu_{\ell}$ are the left handed neutrinos rotated in the same way as the charged lepton flavours:
\begin{equation}\label{B.6}
\nu_{\alpha}=\left(O_{L}\right)_{\alpha\ell}\nu_{\ell}, \ V_{\alpha i}=\left(O_{L}\right)_{\alpha \ell}V_{\ell i}.
\end{equation}

In this basis, the part of the gauge Lagrangian involving neutrinos can be written as:
\begin{equation}\label{B.7}
\begin{split}
\lag_{V}&\supset-\frac{g}{\sqrt{2}} W_{\mu}^{+} \left[\left(O_{L}^{\dagger}U\right)_{\ell \alpha}^{*} \left(\bar{\nu}_{\alpha}\right)_{m} \gamma^{\mu} P_{L} \ell + \left(O_{L}^{\dagger}V\right)_{\ell i}^{*} \left(\bar{N}^{c}_{i}\right)_{m} \gamma^{\mu} P_{L} \ell\right]+\mathrm{h.c.}\\
&-\frac{g}{2 \cos{\theta_{W}}} Z_{\mu} \left[\left(O_{L}^{\dagger}U\right)_{\ell \alpha}^{*} \left(\bar{\nu}_{\alpha}\right)_{m} \gamma^{\mu} P_{L} \nu_{\ell} + \left(O_{L}^{\dagger}V\right)_{\ell i}^{*} \left(\bar{N}^{c}_{i}\right)_{m} \gamma^{\mu} P_{L} \nu_{\ell}\right]+\mathrm{h.c.}.
\end{split}
\end{equation}
\end{appendices}

\bibliographystyle{jhepmod}

\setcounter{page}{45}

\chapter*{Bibliography}
\addcontentsline{toc}{chapter}{Bibliography}
\markboth{Bibliography}{Bibliography}

\begin{multicols}{2}\setlength{\columnseprule}{0pt}
	
\interlinepenalty=10000

\bibliography{./TeX_files/tfmbib}

\end{multicols}

\end{fmffile}
\end{document}